\documentclass[twocolumn, linenumbers,usenames]{aastex631}

\newcommand\N[1]{{\color{blue} \bf #1}} 
\newcommand\R[1]{{\color{red} \bf #1}}
\usepackage{soul}

\shorttitle{{\sl HST} Snapshots of Supernovae}
\shortauthors{Baer-Way et al.}

\graphicspath{{./}{figures/}}

\begin{document}
\nolinenumbers
\title{A Snapshot Survey of Nearby Supernovae with the {\sl Hubble Space Telescope}}

\author[0009-0004-7268-7283]{Raphael Baer-Way}
\affiliation{Department of Astronomy, University of California, Berkeley, CA 94720-3411, USA}
\affiliation{Department of Astronomy, University of Virginia, 530 McCormick Road, Charlottesville, VA, 22904}

\author[0009-0001-2794-8278]{Asia DeGraw}
\affiliation{Department of Astronomy, University of California, Berkeley, CA 94720-3411, USA}

\author[0000-0002-2636-6508]{WeiKang Zheng}
\affiliation{Department of Astronomy, University of California, Berkeley, CA 94720-3411, USA}
\affiliation{Eustace Specialist in Astronomy}

\author[0000-0001-9038-9950]{Schuyler D.~Van Dyk}
\affiliation{Caltech/IPAC, Mailcode 100-22, Pasadena, CA 91125, USA}

\author[0000-0003-3460-0103]{Alexei~V.~Filippenko}
\affiliation{Department of Astronomy, University of California, Berkeley, CA 94720-3411, USA}

\author[0000-0003-2238-1572]{Ori D.~Fox}
\affiliation{Space Telescope Science Institute, 3700 San Martin Drive, Baltimore, MD 21218, USA}

\author[0000-0001-5955-2502]{Thomas G.~Brink}
\affiliation{Department of Astronomy, University of California, Berkeley, CA 94720-3411, USA}
\affiliation{Wood Specialist in Astronomy}
\author[0000-0003-3142-997X]{Patrick L.~Kelly}
\affiliation{University of Minnesota, School of Physics and Astronomy, 116 Church St. SE, Minneapolis, MN 55455, USA}

\author[0000-0001-5510-2424]{Nathan Smith}
\affiliation{Steward Observatory, University of Arizona, 933 North Cherry Avenue, Tucson, AZ 85721, USA}

\author[0000-0002-4951-8762]{Sergiy~S.Vasylyev}
\affiliation{Department of Astronomy, University of California, Berkeley, CA 94720-3411, USA}
\affiliation{Steven Nelson Graduate Fellow}

\author[0000-0001-6069-1139]{Thomas de Jaeger}
\affiliation{Institute for Astronomy, University of Hawai`i, 2680 Woodlawn Dr., Honolulu, HI 96822, USA}
\affiliation{CNRS/IN2P3 (Sorbonne Université, Université Paris Cité), Laboratoire de Physique Nucléaire et de Hautes Énergies, 75005, Paris,
France}
\author{Keto Zhang}
\affiliation{Department of Astronomy, University of California, Berkeley, CA 94720-3411, USA}
\affiliation{Caltech/IPAC, Mailcode 100-22, Pasadena, CA 91125, USA}
\author{Samantha Stegman}
\affiliation{Department of Astronomy, University of California, Berkeley, CA 94720-3411, USA}
\affiliation{Department of Chemistry, University of Wisconsin, Madison, WI 53706, USA}
\author[0000-0003-3975-8110]{Timothy Ross}
\affiliation{Department of Astronomy, University of California, Berkeley, CA 94720-3411, USA}
\affiliation{Lumentum, 1001 Ridder Park Dr.,
San Jose, CA 95131, USA}
\author[0009-0001-0135-8472]{Sameen Yunus}
\affiliation{Department of Astronomy, University of California, Berkeley, CA 94720-3411, USA}
\affiliation{Department of Physics, University of California, Merced, 5200 Lake Rd., Merced, CA 95343, USA}
\begin{abstract}
Over recent decades, robotic (or highly automated) searches for supernovae (SNe) have discovered several thousand events, many of them in quite nearby galaxies (distances $< 30$ Mpc). Most of these SNe, including some of the best-studied events to date, were found before maximum brightness and have associated with them extensive follow-up photometry and spectroscopy. Some of these discoveries are so-called ``SN impostors,'' thought to be superoutbursts of luminous blue variable stars, although possibly a new, weak class of massive-star explosions. We conducted a Snapshot program with the {\sl Hubble Space Telescope\/} ({\sl HST}) and obtained images of the sites of 31 SNe and four impostors, to acquire late-time photometry through two filters. The primary aim of this project was to reveal the origin of any lingering energy for each event, whether it is the result of radioactive decay or, in some cases, ongoing late-time interaction of the SN shock with pre-existing circumstellar matter, or the presence of a light echo. Alternatively, lingering faint light at the SN position may arise from an underlying stellar population (e.g., a host star cluster, companion star, or a chance alignment).  
The results from this study complement and extend those from Snapshot programs by various investigators in previous {\sl HST\/} cycles.
\end{abstract}


\section{Introduction} \label{sec:intro}

Supernovae (SNe) represent the final, explosive stage in the evolution of certain varieties of stars (e.g., \citealt{Woosley1986,Wheeler1990,Filippenko1997,GalYam2017}). Studies of SNe, both observational and theoretical, are closely tied with the physics of stellar evolution, explosion mechanisms and nucleosynthesis, the chemical evolution of galaxies and the Universe, the formation of neutron stars and black holes, and gamma-ray bursts. SNe~Ia are also exceedingly useful cosmological tools, revealing the accelerating expansion of the Universe. 

Despite the well-sampled early-time light curves of relatively nearby SNe, observations are quite sparse at late times ($t \gtrsim 6$ months), primarily because the SNe are extremely faint or their ground-based photometry is contaminated by neighboring stars within the seeing disk. Thus, high-spatial-resolution observations, such as with the {\sl Hubble Space Telescope\/} ({\sl HST}), are required to obtain accurate photometry. Following the multifilter light-curve shapes of these SNe over their long evolution provides important information on their progenitor systems and on the underlying physics leading to the lingering light, and can reveal ``SN impostors'' --- events which are not genuine SNe involving a terminal explosion, but instead are powerful stellar outbursts which occasionally approach the peak luminosity of some kinds of 
true SNe. 

There have been a number of {\sl HST\/} Snapshot surveys of the sites of SNe. Various studies have conducted detailed analyses of the late-time emission of these SNe and of their immediate environments (e.g., \citealt{Li2002,Fransson2002,VanDyk2003,Sun2023}). In each case it has been shown that {\sl HST\/} can effectively resolve the faint SNe at late times from their immediate environments. In some cases, more than one epoch of {\sl HST\/} observations was obtained, enabling the measurements of late-time decline rates and providing important information on the nebular evolution of SNe. Observed differences in late-time decline rates, particularly for those significantly diverging from power due to $^{56}$Co radioactive decay, motivate the need for larger samples of light curves to be collected.

Additional sources of late-time luminosity can originate via contribution from light echoes or from interaction of the SN shock with circumstellar matter (CSM). For instance, light echoes observed with {\sl HST\/} have been spatially resolved around four nearby SNe~Ia: SN 1991T \citep{Sparks1999}, SN 1998bu \citep{Cappellaro2001}, SN 2006X \citep{Wang2008}, and SN 2014J \citep{Crotts2015,Yang2017}. The presence of light echoes is not limited to SNe~Ia --- resolved echoes have been detected around a number of core-collapse SNe as well, such as SN 1987A (e.g., \citealt{Bond1990}), SN 1993J \citep{Sugerman2002,Liu2003}, SN 2003gd \citep{Sugerman2005,VanDyk2006}, SN 2008bk \citep{VanDyk2013c}, SN 2012aw \citep{VanDyk2015}, and SN 2016adj \citep{Stritzinger2022}. Recently, a light echo has also been detected around SN 1987A as detailed by \citet{Ding2021} and \citet{Cikota2023}. Snapshot programs, in particular, can provide statistics on the frequency of light echoes around various types of SNe.

The sustained late-time luminosity of some SNe~II can be explained by interaction of the SN shock with large amounts of CSM set up by the pre-SN wind (e.g., \citealt{Fox2013,smith14araa,smith17hsn,Smith2017}). The sustained optical emission in this case likely arises from a radiatively-cooled shell.
{\sl HST\/} Snapshot programs can also help reveal the nature of the SN impostors, events similar to Type~IIn SNe (with relatively narrow H emission lines in their spectra), but which are 
subluminous compared with core-collapse SNe ($M_V \approx -14$~mag) near maximum brightness \citep{Smith2011b,VanDyk2012}.

In this paper, we present the results of {\sl HST\/} Snapshot program GO-16179 (PI A.~Filippenko), along with some data from previous Snapshot programs such as GO-14668 and GO-15166 (PI A.~Filippenko). The primary goal of this study is fairly simple: to determine whether the SNe at late times are essentially following the exponential light-curve decline, as a result of reprocessing of $\gamma$-rays and positrons from radioactive $^{56}$Co decay, or whether an additional power source is at work. In Section~\ref{observations} we provide the details of the {\sl HST\/} observations, and Section~\ref{analysis} describes our analysis, including the data reduction and results. The results as they pertain to all of the individual objects are discussed in Section~\ref{sec:individual}. Section~\ref{conclusions}  provides our summary and conclusions.

\section{Observations}\label{observations}

The program was executed during {\sl HST\/} Cycle 28 from 2020 November 11 through September 24 2021 (UTC dates are used throughout), with the Wide Field Camera 3 (WFC3) UVIS. The original observing request was for 55 visits, consisting of 9 SNe~Ia, 27 SNe~Ib/c, 12 SNe~II, and 7 SN impostors (for  reviews of SN spectral classification, see \citet{Filippenko1997}, ,\citet{GalYam2017}), of which 38 visits ($\sim 70$\%) were actually executed. 
However, one visit, of the target SN IIn 2005ip, failed outright (all data lost), and two other visits, of the SN~Ia 2018hfp and SN~Ia 2019cth, experienced loss of guiding and were  rendered useless. The final observed sample of 35 targets (3 SNe~Ia, 9 SNe~Ib/c, 19 SNe~II, and 4 SN impostors) is summarized in Table~\ref{table:properties}. All of the data presented in this article were obtained from the Mikulski Archive for Space Telescopes (MAST) at the Space Telescope Science Institute. The specific observations analyzed can be accessed via
the associated \dataset[DOI]{http://dx.doi.org/10.17909/qz85-k192}.

The observing scheme for the program was to obtain all of the executed visits within an optimum ``Snapshot'' orbit of $\sim 38$--40~min, which was intended to increase the likelihood that a visit would be scheduled. To accomplish this, each visit orbit consisted of observations in two bands, with a total exposure time of 710~s in one band and 780~s in the other. Typically, we observed in the F555W band with the shorter exposures and in F814W with the longer ones, but varied the actual filter combination used depending on the specific science goal for each visit. The observations in each band per visit were split into two exposures of equal duration, employing a line dither between exposures to mitigate as best as possible against cosmic-ray hits and detector cosmetic defects.

All of the data from the program had no exclusive access period and were publicly available from the Mikulski Archive for Space Telescopes (MAST)\footnote{https://mast.stsci.edu/search/ui/\#/hst} as soon as they were processed through the Space Telescope Science Institute (STScI) {\sl HST\/} standard pipeline. We note that at least one study not focused on SNe or related transients, but on the optical counterparts of extragalactic ultraluminous X-ray sources \citep{Allak2022}, has already made use of our data.

\begin{deluxetable*}{llcclcccc}
\tablewidth{0pt}
\tablecolumns{9}
\tablecaption{Properties of the Targeted Events and Their Hosts}
\tablehead{
\colhead{SN} & \colhead{Type} & \colhead{$\alpha$ (h,m,s)} & \colhead{$\delta$ ($^\circ,',''$)} & \colhead{Discovery} & \colhead{Host galaxy} & \colhead{$A_V ({\rm MW})$} & \colhead{Distance} & \colhead{$v_{{\rm hel}}$(host)} \\
\colhead{} & \colhead{} & \colhead{(J2000)} & \colhead{(J2000)} & \colhead{Date (UTC)} & \colhead{} & \colhead{(mag)} & \colhead{(Mpc)} & \colhead{(km\,s$^{-1}$)}
}
\startdata
SN 1988Z          & IIn    & 10:51:50.10 & +16:00:01.01                    & 1988-12-12 & MCG +03-28-022                             & 0.07 &  70.7(11)       &     6748(3) \\
SN 1993J          & IIb     & 09:55:25.00 & +69:01:13.01                    & 1993-03-28 & NGC 3031                                       & 0.22 &   3.6(0.3) &    $-$39(3) \\
SN 2000ch        & Imp.   & 10:52:41.40 & +36:40:08.51                   & 2000-05-03 & NGC 3432                                       & 0.04 & 11.7(4.2) &       613(4) \\
SN 2010jl          & IIn      & 09:42:53.33 & +09:29:41.78                   & 2010-11-03 & UGC 5189A                                     & 0.08 & \nodata   &     3207(37) \\
SN 2010mc       & IIn      & 17:21:30.67 & +48:07:47.39                   & 2010-08-20 & GALEXASCJ172130.92+480747.6 & 0.05 & \nodata   &   10493      \\ 
SN 2011dh        & IIb      & 13:30:05.12 & +47:10:10.81                   & 2011-06-01 & NGC 5194                                        & 0.10 &   7.2(2.1) &     463(3) \\
SN 2012A          & II-P      & 10:25:07.39 & +17:09:14.62                  & 2012-01-07 & NGC 3239                                        & 0.09 &   9.7(1.6) &     710(1) \\
SN 2012aw        & II-P     & 10:43:53.76 & +11:40:17.90                   & 2012-03-16 & NGC 3351                                        & 0.08 &   9.9(1.1) &   1128(24) \\
SN 2013df          & IIb     & 12:26:29.33 & +31:13:38.32                   & 2013-06-07 & NGC 4414                                        & 0.05 & 18.1(3.1) &     708(2)  \\
SN 2013ej          & II-P/II-L & 01:36:48.16 & +15:45:31.00                  & 2013-07-25 & NGC 628                                          & 0.19 &   7.5(3.1) &     657(1)  \\
SN 2014C          & Ib       & 22:37:05.60 & +34:24:31.90                  & 2014-01-05 & NGC 7331                                        & 0.25 & 13.4(2.7) &      816(1) \\
SN 2015cp         & Ia       & 03:09:12.76 & +27:31:16.95                  & 2015-12-28 & WISEAJ030912.10+273106.9         & 0.74 & \nodata    &  \nodata  \\
SN 2016G          & Ic-BL  & 03:03:57.70 & +43:24:03.60                  & 2016-01-09 & NGC 1171                                        & 0.43 & 26.6(6.2) &   2742(5) \\
SN 2016adj        & Ic?  & 13:25:24.11 & $-$43:00:57.50                & 2016-02-08 & NGC 5128                                       & 0.32 &   3.8(0.8) &     547(5) \\
SN 2016bkv       & II-P      & 10:18:19.31 & +41:25:39.30                  & 2016-03-21 & NGC 3184                                        & 0.05 & 12.3(2.2) &    582(1) \\
AT 2016blu & Imp.   & 12:35:52.30 & +27:55:55.9\phantom{0} & 2021-01-11 & NGC 4559                                        & 0.05 &   8.9(0.2) &     814(1) \\
SN 2016coi        & Ic-BL  & 21:59:04.14 & +18:11:10.46                  & 2016-05-27 & UGC 11868                                      & 0.23 & 17.2(7)         &  1093(5) \\
SN 2016coj        & Ia       & 12:08:06.80 & +65:10:37.80                  & 2016-05-28 & NGC 4125                                        & 0.05 & 22.8(7.6) &  1281(14) \\
SN 2016gkg       & IIb      & 01:34:14.46 & $-$29:26:25.00               & 2016-09-20 & NGC 613                                          & 0.05 & 20.9(5.7) &  1481(5) \\
AT 2016jbu         & Imp.? & 07:36:25.96 & $-$69:32:55.25               & 2016-12-01 & NGC 2442                                        & 0.56 & 20.1(0.5) &  1466(5) \\
SN 2017cfd        & Ia       & 08:40:49.09 & +73:29:15.11                  & 2017-03-16 & IC 511                                               & 0.06 & \nodata    &  3623(49) \\
SN 2017eaw      & II-P      & 20:34:44.24 & +60:11:35.84                  & 2017-05-14 & NGC 6946                                        & 0.94 &   7.3(1.5)  &      40(2) \\
SN 2017gax       & Ib/c    & 04:45:49.50 & $-$59:14:42.50               & 2017-08-14 & NGC 1672                                        & 0.06 & 11.8(1.4)  &  1331(3) \\
SN 2017gkk       & IIb      & 09:13:44.57 & +76:28:44.54                  & 2017-08-31 & NGC 2748                                        & 0.07 & 19.2(2.8)  &  1476(2) \\
SN 2017ixv        & Ic-BL  & 19:21:31.24 & +61:08:51.76                  & 2017-12-17 & NGC 6796                                        & 0.19 & 36.2(2.8)  &  2189(6) \\
SN 2018aoq      & II-P      & 12:10:38.22 & +39:23:47.87                  & 2018-04-01 & NGC 4151                                        & 0.08 & 15.8(0.4)  &    997(2) \\
SN 2018gj         & II-P      & 16:32:02.30 & +78:12:40.93                  & 2018-01-14 & NGC 6217                                        & 0.12 &  20.6(7.3)  &  1361(3) \\
AT 2018cow      & Ic-BL  & 16:16:00.22 & +22:16:04.83                  & 2018-06-16 & CGCG 137$-$068                            & 0.24 & \nodata      &  4241(39) \\
SN 2018ivc       & II-pec  & 02:42:41.28 & $-$00:00:31.92               & 2018-11-24 & NGC 1068                                        & 0.09 & 10.6(3.0)   &  1137(3) \\
SN 2018zd        & II-P      & 06:18:03.19 & +78:22:01.16                  & 2018-03-02 & NGC 2146                                        & 0.26 & 19.6(8.2)   &    892(4) \\
SN 2019ehk      & Ib/gap tr. & 12:22:56.15 & +15:49:34.03                 & 2019-04-29 & NGC 4321                                        & 0.09 & 16.2(3.1)   &  1571(1)  \\
AT 2019krl         & Imp.    & 01:36:49.65 & +15:46:46.21                 & 2019-07-06 & NGC 628                                          & 0.19 &  7.5(3.1)    &    657(1) \\
SN 2020dpw      & II-P      & 20:37:10.55 & +66:06:10.66                 & 2020-02-26 & NGC 6951                                        & 1.02 & 23.1(3.5)   &  1424(1) \\
SN 2020hvp       & Ib       & 16:21:45.39 & $-$02:17:21.37              & 2020-04-21 & NGC 6118                                        & 0.43 & 20.8(3.9)   &  1573(1) \\
SN 2020jfo         & II-P      & 12:21:50.48 & +04:28:54.05                 & 2020-05-06 & NGC 4303                                        & 0.06 & 14.6(7.3)   &  1566(2) \\
\enddata
\tablecomments{SN positions and discovery dates are adopted from the TNS. Foreground Milky Way visual extinction, $A_V ({\rm MW})$, is adopted in each case from the NASA/IPAC Extragalactic Database (NED; \citealt{Schlafly2011}). Distances and heliocentric velocities ($v_{{\rm hel}}$) are also obtained from NED.}
\label{table:properties}
\end{deluxetable*}

\begin{deluxetable*}{cccccccc}
\tablecaption{{\sl HST\/} Photometry of the Observed Supernovae}
\tablecolumns{8}
\tablenum{2}
\tablehead{\colhead{Object} & \colhead{Obs.~Date} & \colhead{MJD} & \colhead{Age} & \colhead{Bands} & \colhead{Exp.} & \colhead{Brightness} & \colhead{Exponential}\\
\colhead{} & \colhead{(UTC)} & \colhead{} & \colhead{(days)} & \colhead{} & \colhead{Time (s)} & \colhead{(Vegamag)} & \colhead{Decline?}}
\startdata
SN 1988Z  & 2021-02-19 & 59264.8 & 11757.8 & F625W, F814W & 710, 780 & 24.83(03), 24.98(10) & No \\
SN 1993J  & 2020-12-14 & 59197.2 & 10123.2 & F336W, F814W & 710, 780 & 23.38(03), 23.25(03) & No \\
SN 2000ch & 2020-12-13 & 59196.8 &  7529.1 & F555W, F814W & 710, 780 & 20.90(03), 19.01(01) & N/A \\
SN 2010jl & 2020-12-29 & 59213.0 &  3699.9 & F336W, F814W & 710, 780 &  $>$24.8, $>$25.6 & No? \\
SN 2010mc & 2021-09-24 & 59481.3 &  4053.3 & F555W, F814W & 710, 780 & 24.56(04), 25.44(12) & No \\    
SN 2011dh & 2020-12-10 & 59193.1 &  3480.1 & F555W, F814W & 710, 780 & 23.15(02), 23.22(03) & No \\
SN 2012A  & 2021-02-16 & 59261.8 &  3328.8 & F606W, F814W & 710, 780 & $>$26.9, $>$25.8     & Yes? \\
SN 2012aw & 2021-02-17 & 59262.9 &  3260.9 & F555W, F814W & 710, 780 & 25.87(10), 25.17(13)    & No \\
SN 2013df & 2021-02-15 & 59260.9 &  2810.9 & F336W, F555W & 780, 710 & 23.62(03), 23.21(03) & No \\
SN 2013ej & 2021-08-19 & 59446.0 &  2948.0 & F555W, F814W & 710, 780 & 24.37(03), 23.41(03) & No \\
SN 2014C  & 2021-08-20 & 59446.7 &  2785.7 & F555W, F814W & 710, 780 & 22.18(04), 20.88(01) & No \\
SN 2015cp  & 2020-11-30 & 59183.0 & 1799.0 & F275W, F625W & 710, 780 & $>$25.1, $>$27.2 & Yes? \\
SN 2016G   & 2020-12-20 & 59203.4 & 1807.4 & F606W, F814W & 710, 780 & $>$25.8, $>$25.1     & Yes? \\
SN 2016adj & 2021-07-28 & 59423.6 & 1997.6 & F438W, F555W & 710, 780 & $>$27.4, $>$26.5      & Yes? \\ 
SN 2016bkv & 2020-12-13 & 59196.0 & 1721.0 & F555W, F814W & 710, 780 & 23.39(02), 23.33(04) & No \\
AT 2016blu & 2021-02-17 & 59262.9 & 1779.0 & F606W, F814W & 710, 780 & 19.51(00),19.32(00) & N/A \\
SN 2016coi & 2020-12-06 & 59189.7 & 1654.7 & F336W, F814W & 710, 780 & $>$26.1, $>$26.0 & Yes? \\
SN 2016coj & 2020-12-09 & 59192.2 & 1656.2 & F555W, F814W & 710, 780 & $>$26.8, $>$25.5     & Yes? \\
SN 2016gkg & 2021-08-19 & 59445.8 & 1794.8 & F438W, F606W & 710, 780 & $>$26.1, 24.88(04) & No \\
AT 2016jbu & 2021-08-21 & 59447.9 & 1724.9 & F555W, F814W & 710, 780 & 26.63(04), 25.69(05) & N/A \\
SN 2017cfd  & 2021-09-16 & 59473.5 & 1645.4 & F555W, F814W & 710, 780 & $>$26.9, $>$26.0    & Yes? \\
SN 2017eaw & 2020-11-11 & 59164.8 & 1277.8 & F555W, F814W & 710, 780 & 23.66(02), 23.00(05) & No \\
SN 2017gax & 2020-11-27 & 59180.2 & 1201.2 & F336W, F814W & 710, 780 & $>$26.3, $>$26.0 & Yes? \\
SN 2017gkk & 2021-09-24 & 59481.0 & 1485.1 & F555W, F814W & 710, 780 & 24.69(05), 23.91(07) & No \\
SN 2017ixv & 2021-01-11 & 59225.8 & 1121.8 & F555W, F814W & 710, 780 & $>$26.4, $>$24.9   & Yes? \\
SN 2018gj  & 2021-01-27 & 59241.4 & 1109.3 & F555W, F814W & 710, 780 & 24.77(04), 23.32(06) & No\\
SN 2018zd  & 2021-02-07 & 59252.6 & 1073.7 & F555W, F814W & 710, 780 & $>$27.0, $>$26.1   & Yes? \\
SN 2018aoq & 2020-12-05 & 59188.8 &  983.8 & F555W, F814W & 710, 780 & $>$26.9, $>$25.9& Yes? \\
AT 2018cow  & 2021-07-25 & 59420.8 & 1135.8  & F555W, F814W & 710, 780 & 25.69(09), 26.21(03) & No \\
SN 2018ivc  & 2020-12-02 & 59185.0 & 739.0 & F555W, F814W & 710, 780 & 22.71(02), 21.97(03) & No \\
SN 2019ehk & 2021-02-21 & 59266.9 & 664.9  & F438W, F625W & 710, 780 & $>$27.2, $>$26.6  & Yes? \\
AT 2019krl & 2021-02-15 & 59260.6 & 590.8  & F438W, F625W & 710, 780 & 23.85(04), 23.77(03) & N/A \\
SN 2020dpw & 2020-12-13 & 59196.9 & 291.9  & F555W, F814W & 710, 780 & 20.33(00), 18.54(00) & Yes? \\
SN 2020hvp & 2021-05-22 & 59356.8 & 396.8  & F555W, F814W & 710, 780 & 22.81(01), 22.20(02) & Yes \\
SN 2020jfo & 2021-07-28 & 59423.5 & 448.5  & F555W, F814W & 710, 780 & 22.23(01), 21.34(01) & Yes \\
\enddata
\tablecomments{Obs.~Date is the Snapshot observation date. Modified Julian Date (MJD) is Julian date (JD) $-$ 2,400,000.5. Age is days since discovery date. Exposure times (``Exp.") are the total time in each {\sl HST\/} band.}
\label{table:phot}
\end{deluxetable*}

\section{Analysis}\label{analysis}

\subsection{Data Reduction}

We ran the suite of STScI {\tt Drizzlepac} routines \citep{STSCI2012} on the data from each visit, to construct a drizzled image mosaic in each band. To locate the general sites of SNe in each of the image mosaics, coordinates from the Transient Name Server (TNS)\footnote{https://www.wis-tns.org/} were first used. To more precisely isolate the SN site, either we directly compared the new data with previous {\sl HST\/} images containing the SN from previous epochs (in many of the cases, from previous Snapshot programs) when the SN was brighter, or, if no prior {\sl HST\/} data were available, astrometrically aligned the {\sl HST\/} Snapshot mosaics with ground-based images of the SN (in many cases, obtained by us with the 0.76~m Katzman Automatic Imaging Telescope \citep[KAIT;][]{Filippenko2003} or at the Nickel 1~m telescope, both at Lick Observatory).
In the former cases, we could simply blink the Snapshot images with the previous {\sl HST\/} images and visually identify the SN in the new data. In the latter cases, stars were found in common between the {\sl HST\/} and ground-based images, and astrometric registration between the two datasets was performed. The SN position in the ground-based data was then transformed to the {\sl HST\/} reference frame. The typical astrometric uncertainty was in the range of $\sim 0.1''$, and in no case where we performed this alignment were there any other sources in the error circle.  In most cases, it was then readily apparent which object in the Snapshot data was the SN in question.

To obtain photometry from the data in both bands for each visit, we ran the individual frames for the entire visit through the {\tt Dolphot} package \citep{Dolphin2016}, using one of the mosaics as the reference image for source detection. Generally, the recommended WFC3/UVIS parameters from \citet{Dolphin2016} were used. 

\subsection{Results}

Once an SN location was isolated, its image coordinates were matched to the output from {\tt Dolphot} to retrieve photometric information for the SN, along with any potential error/quality flags. The photometric results for all of the SNe are given in Table~\ref{table:phot}. The brightness measurements for the SNe are in Vega magnitudes for the {\sl HST\/} flight system bands, as indicated in the table. In a number of cases, nothing was detected by {\tt Dolphot} at the SN position, and subsequently we estimated upper limits to detection and provide these in the table. The upper limits were based on the formal estimations of signal-to-noise ratio ($S/N$) from {\tt Dolphot}, and we have set the significance at $S/N=5$. The major caveat 
is that the uncertainties, particularly at low flux levels and in crowded environments, are underestimated by {\tt Dolphot} \citep{Williams2014}, so the significance levels of the nondetections are likely overestimates; see \citet{VanDyk2023} for a discussion of this issue.
 
For each event we also indicate in Table~\ref{table:phot} our assessment of whether the late-time light curves appear to be powered by radioactive decay (the $^{56}$Co decay rate is shown for comparison in the light-curve figures; see Section~\ref{sec:individual}), with either a ``yes'' or ``no.'' In some cases we were unable to make this assessment, since the event is either an SN impostor (and therefore still likely undergoing a superoutburst) or early-time photometry did not exist; these are listed as ``N/A,'' for ``not applicable.'' In a number of cases we were unable to confidently ascribe the decline to radioactive decay, primarily because we could only place an upper-limit constraint on the late-time emission, and these are indicated by a question mark.

In the next section 
we discuss each of the individual objects separately. 

\section{Individual Objects}\label{sec:individual}

\subsection{SN 1988Z}\label{88Z}

SN 1988Z was recognized early as an unusual SN II \citep{Stathakis1991}. From the luminous radio \citep{vanDyk1993,Williams2002} and X-ray \citep{Fabian1996,Schlegel2006} emission detected from the SN, along with the characteristics of its optical photometric and spectroscopic emission \citep{Turatto1993,Aretxaga1999}, it was posited that long-lived interaction of the SN shock with a pre-existing dense CSM was the likely source of this radiation. In fact, recent spectra show that SN~1988Z is still strongly interacting with dense CSM even three decades after explosion \citep{Smith2017}. SN 1988Z is generally considered to be a Type IIn SN, even a prototype of this subclass.

\begin{figure*}[htb]
\gridline{\fig{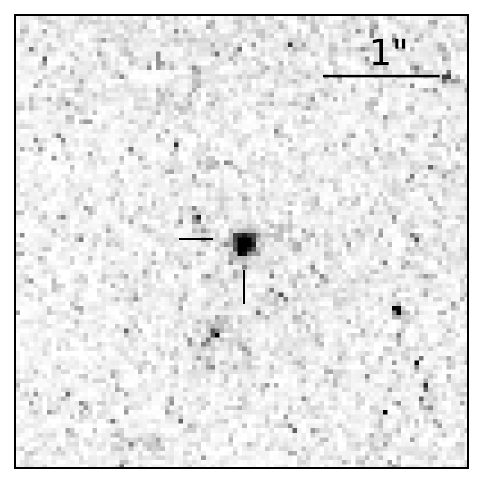}{0.25\textwidth}{(a) F625W}
          \fig{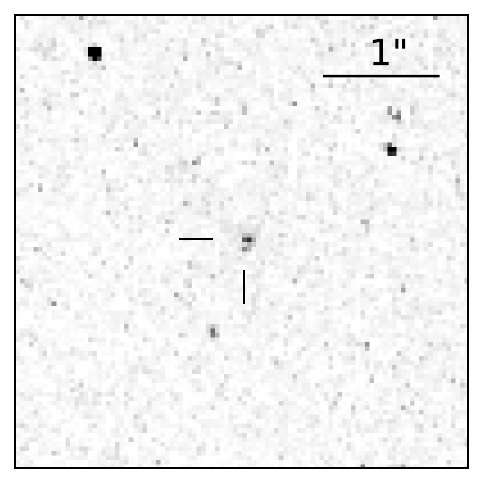}{0.25\textwidth}{(b) F814W}
          \fig{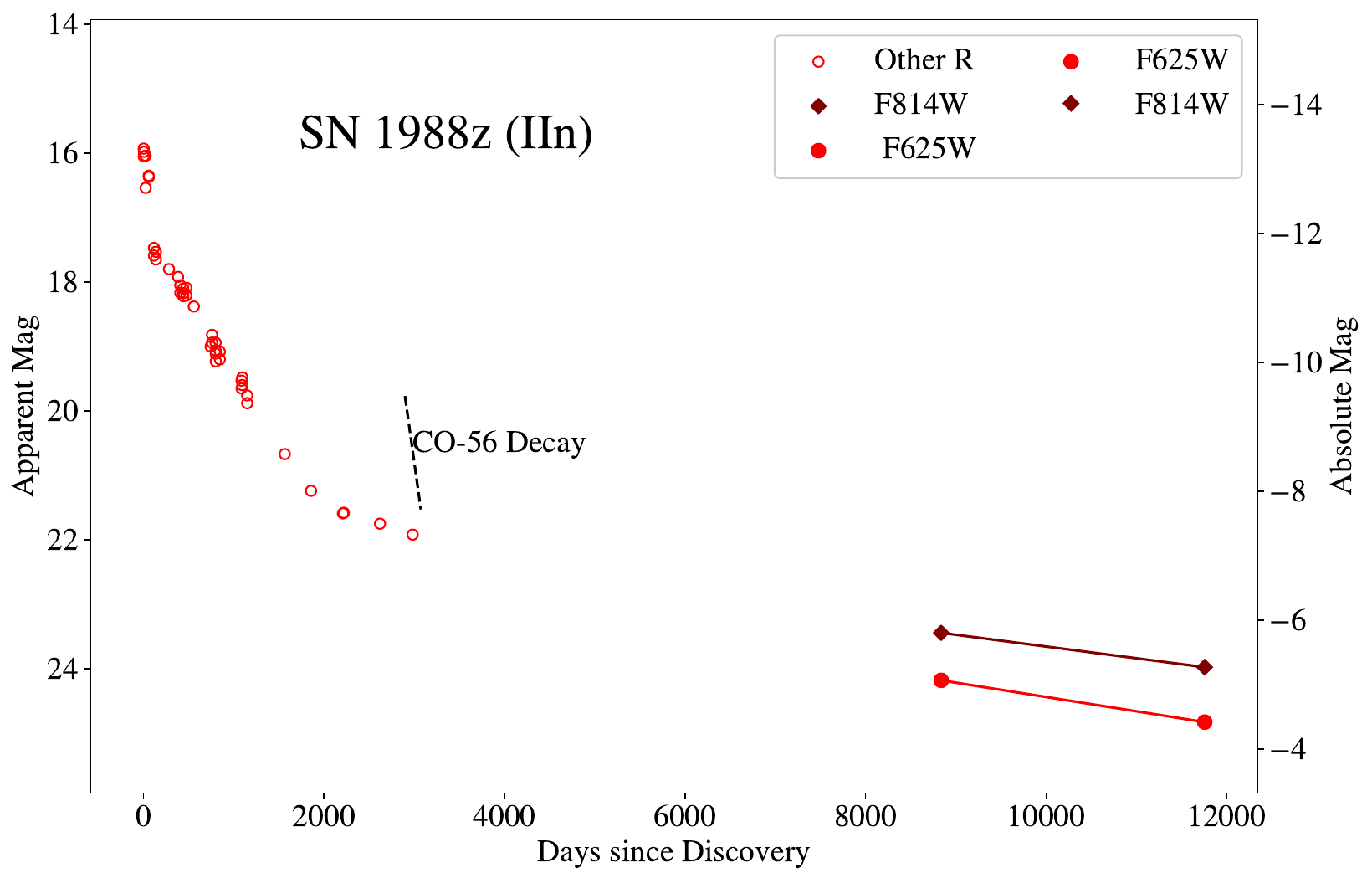}{0.4\textwidth}{(c) light curve}}
\caption{A portion of the WFC3 image mosaic containing SN 1988Z, from observations on 2021 February 19, in (a) F625W and (b) F814W. Here, and in all other figures showing {\sl HST\/} images in this paper, north is up and east is to the left; also, whenever the SN is visible, it is indicated by tick marks. Also shown are the (``Other'') $R$ (c) light curves from \citet{Aretxaga1999} and \citet{Turatto1993}, together with the Snapshot detections from programs GO-13029 and GO-16179.}
\label{fig:88Z}
\end{figure*}

The Snapshot observations were obtained on 2021 February 19 in F625W ($\sim R$) and F814W ($\sim I$). As can be seen in Figure~\ref{fig:88Z}, amazingly the SN was still detectable in the {\sl HST\/} images 11,758~d (32.2~yr) after discovery, at $m_{\rm F625W}=24.83 \pm 0.03$ and $m_{\rm F814W}=24.98 \pm 0.10$ mag. (We had intentionally used the F625W band, sensitive to any remaining H$\alpha$ line emission, and the F814W band, potentially sensitive to hot dust, to increase the probability of detection at such late times.) We pinpointed the location of the faint SN in these images by employing previously unpublished imaging in 2013 February from our Snapshot program in Cycle 20 (GO-13029, PI A.~Filippenko), when the SN was brighter ($m_{\rm F625W}=24.08 \pm 0.04$ and $m_{\rm F814W}=24.49 \pm 0.07$~mag).

We have included the two sets of Snapshot data together with the ground-based, earlier-time $R$ light curves from \citet{Aretxaga1999} and \citet{Turatto1993}. The light curve clearly does not follow the trend for radioactive-decay power, and the very late-time points from {\sl HST\/} appear to follow the break in the light curve that began at $\sim 2000$~d.

\subsection{SN 1993J}\label{sec:sn1993J}

SN 1993J is one of the best-studied and historically prominent SNe ever discovered. It remains a benchmark SN~IIb to which more recent discoveries are often compared, with a rich array of multiwavelength observations collected over the last 30~yr. The proximity of its host galaxy facilitated detection and characterization of its progenitor as a K-type supergiant, even from the ground \citep{1993J_Fillipenko,Aldering1994,Cohen1995,VanDyk2002}, and excess flux in the blue and near-ultraviolet (UV) bands suggested the presence of a binary companion, consistent with models of supergiant mass loss onto the secondary (e.g., \citealt{Podsiadlowski1993,Maund2004,Fox2014}). The optical light curve of the SN has been powered by ongoing interaction with the CSM in a relatively slow decline. The SN had faded enough by 2004 that it was evident that the supergiant progenitor had vanished \citep{Maund2009}. The SN had remained too luminous, however, for a binary companion to be isolated via imaging, until \citet{Fox2014} were able to claim detection of UV excess emission indicative of such a star in {\sl HST\/} data from 2011 and 2012.  Up to the present epoch, the visual-wavelength spectrum of SN~1993J is still dominated by strong ongoing CSM interaction, with prominent broad emission lines of H$\alpha$, [O~{\sc i}], [O~{\sc ii}], and [O~{\sc iii}] \citep{Smith2017}.

\begin{figure*}[htb]
\gridline{\fig{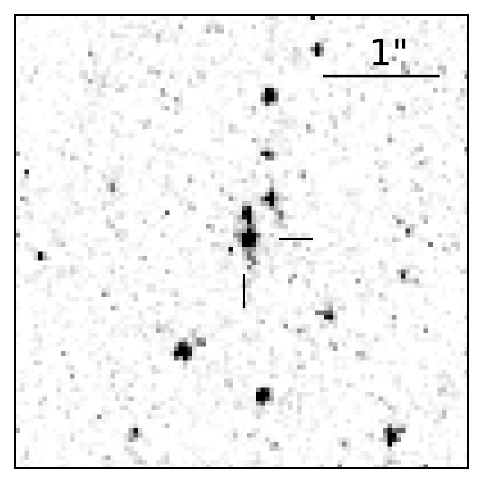}{0.25\textwidth}{(a) F336W}
          \fig{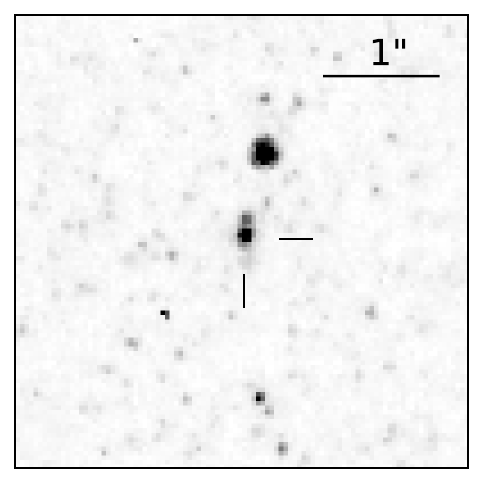}{0.25\textwidth}{(b) F814W}
          \fig{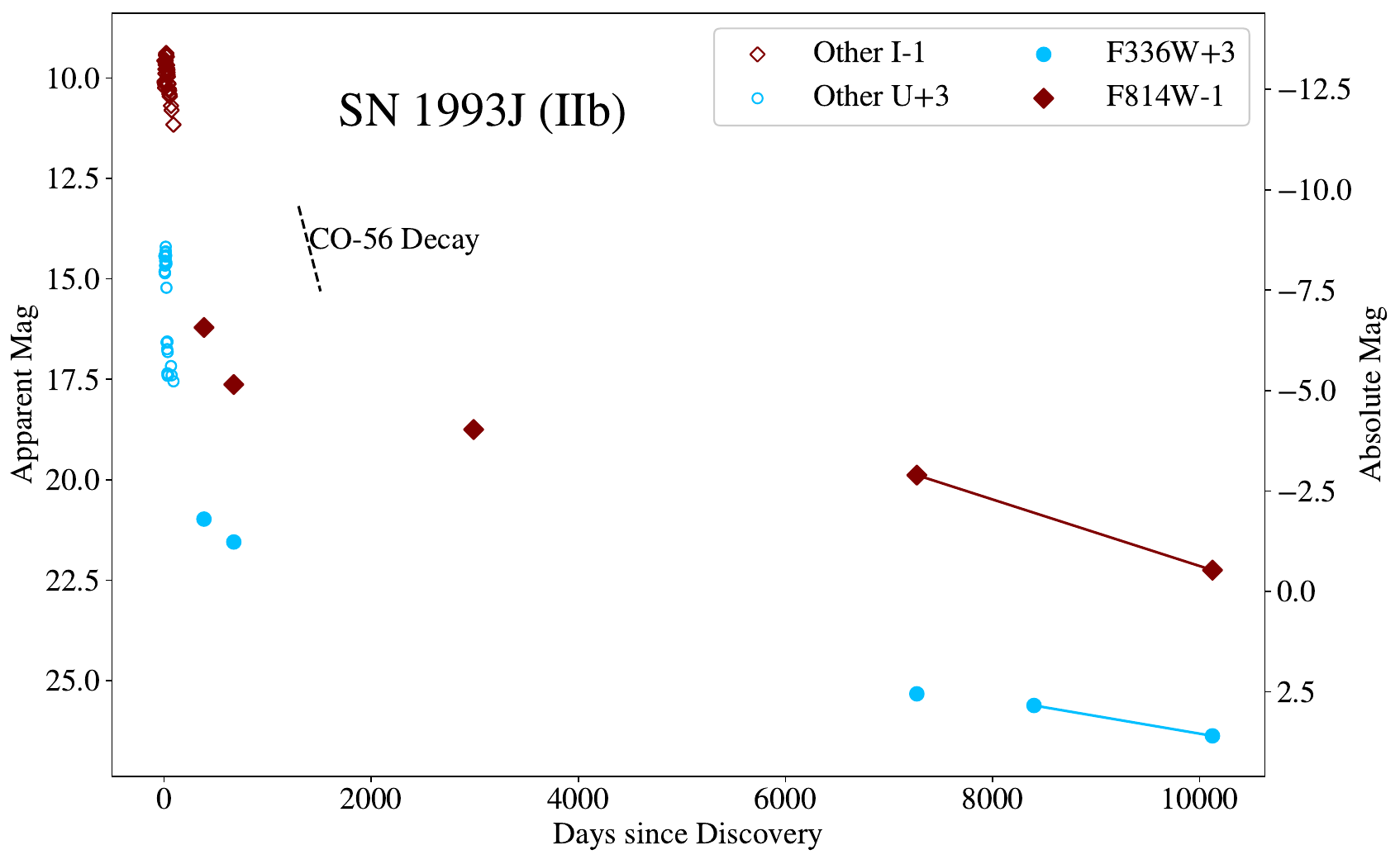}{0.4\textwidth}{(c) light curve}}
\caption{A portion of the WFC3 image mosaic containing SN 1993J, from observations on 2020 December 14, in (a) F336W and (b) F814W. 
Also shown are the (``Other'') $U$ and $I$ (c) light curves from \citet{Richmond1996}, together with prior {\sl HST\/} data from \citet{VanDyk2002}, \citet{Fox2014}, and previously-unpublished data from GO-13648 (PI O.~Fox), as well as our Snapshots.}
\label{fig:93j}
\end{figure*}

We located the SN site in our Snapshot data in F336W ($\sim U$) and F814W from 2020 December 14, when the SN was at 10,123~d (27.7~yr), by consulting \citet[][ their Figure 1]{Fox2014} and also comparing with the data from 2012 February from program GO-12531 (PI A.~Filippenko), when the SN was brighter in F336W and F814W ($22.33 \pm 0.02$ and $20.87 \pm 0.01$~mag, respectively). We also compared to previously-unpublished data from 2015 March at F336W from GO-13648 (PI O.~Fox; the SN was at $22.62 \pm 0.05$~mag); see Figure~\ref{fig:93j}. SN 1993J clearly is not following the radioactive-decay trend, which has been the case for most of its late-time ($\gtrsim 500$~d) history. However, the SN appears to have faded more rapidly, compared to the more gradual decline up to (and possibly beyond) 2015. This could be indicating that the SN shock was encountering a less dense circumstellar environment than previously, consistent with the results of modeling of the declines in both the radio and X-ray emission \citep{Kundu2019}.

We can now clearly detect in F814W (see Figure~\ref{fig:93j}) a star immediately to the northwest of the SN, the presence of which was just hinted at by \citet{Fox2014}. Following the labeling scheme from \citet{Fox2014}, this star ``O'' has $m_{\rm F814W}=22.88$~mag. 
We have reprocessed the 2011 data from \citet{Fox2014}, adopting the same {\tt Dolphot} parameters that we used here (which differ somewhat from those used in that previous study); our results are presented in Table~\ref{table:93J_comparison}. We also include our measurements for F336W data (640~s) from  GO-13648 (PI O. Fox). Furthermore, we present our results for these same stars from our Snapshot data. On average, the stars have essentially the same measured brightnesses in F814W, with the Snapshot values being slightly fainter (by $\sim 0.06$~mag), whereas the F336W Snapshot measurements appear to differ by substantially more, $\sim 0.46$~mag fainter, than our remeasurements of the \citet{Fox2014} data. We can potentially account for this large difference in that the 2011 total exposures (3000~s) in F336W were a factor of $\sim 4.2$ deeper than the 710~s total Snapshot exposure, so the $S/N$ was substantially higher for the former than the latter.

\begin{deluxetable*}{cccccc}
 \tablecolumns{6}
 \tablewidth{0pt}
 \tablenum{3}
 \tablecaption{SN 1993J Field Photometry Comparisons}
  \tablehead{
  \colhead{Star} & \multicolumn2c{Fox et al. (2014)} & \colhead{GO-13648} & \multicolumn2c{This paper} \\
  \colhead{} & \colhead{F336W} & \colhead{F814W} & \colhead{F336W} & \colhead{F336W} & \colhead{F814W} \\
  \colhead{} & \colhead{(mag)} & \colhead{(mag)} & \colhead{(mag)} & \colhead{(mag)} & \colhead{(mag)}}
\startdata
SN & 22.33 (0.02) & 20.87 (0.01) & 22.62 (0.05) & 23.85 (0.10) & 22.01 (0.02) \\
A  & 23.39 (0.03) & 20.60 (0.01) & 23.65 (0.09) & 24.79 (0.21) & 20.47 (0.01) \\
B  & 22.66 (0.02) & 22.89 (0.03) & 22.93 (0.05) & 23.06 (0.05) & $>$25.6      \\
C  & 22.55 (0.03) & 23.71 (0.05) & 22.82 (0.05) & 22.93 (0.05) & 23.56 (0.04) \\
D  & 22.44 (0.02) & 24.19 (0.06) & 22.63 (0.05) & 22.73 (0.06) & 24.04 (0.06) \\
E  & 23.29 (0.03) & 24.21 (0.07) & 23.54 (0.09) & 23.65 (0.07) & 24.22 (0.07) \\
F  & 23.90 (0.04) & 24.98 (0.12) & 24.07 (0.12) & 24.04 (0.16) & $>$25.6      \\
G  & 23.21 (0.03) & $>$25.7      & 23.53 (0.09) & 23.65 (0.07) & $>$25.6      \\
H  & 24.14 (0.04) & 24.65 (0.10) & 24.10 (0.13) & 24.49 (0.12) & 24.38 (0.08) \\
I  & 24.51 (0.06) & $>$25.7     & 24.44 (0.17) & $>$25.5      & $>$25.6      \\
J  & $>$26.4      & 23.84 (0.05) & $>$26.1      & $>$25.5      & 23.93 (0.06) \\
K  & 23.58 (0.03) & 24.43 (0.07) & 23.86 (0.10) & 24.04 (0.09) & 24.77 (0.11) \\
L  & 25.84 (0.13) & $>$25.7      & $>$26.1      & $>$25.5      & $>$25.6      \\
M  & 23.54 (0.03) & 24.73 (0.10) & 23.42 (0.08) & 23.96 (0.09) & 25.56 (0.21) \\
N  & 24.35 (0.06) & 24.72 (0.10) & 25.08 (0.27) & 24.72 (0.14) & 24.66 (0.10) \\
O  & \nodata      & \nodata      & $>$26.1      & $>$25.5      & 22.88 (0.03) \\
\enddata
\tablenotetext{}{Uncertainties ($1\sigma$) are in parentheses.}
\label{table:93J_comparison}
\end{deluxetable*}

\subsection{SN 2000ch}

SN 2000ch, discovered with KAIT at magnitude 17.4, was suspected early on to be an unusual and very luminous variable in NGC 3432. \citet{Wagner2004} initially described its erratic behavior, and \citet{Pastorello2010} later detailed the occurrence of multiple outbursts from the star. \citet{Smith2011b} compared SN 2000ch to a host of other objects considered to be luminous blue variables or SN impostors, which may survive their eruptive outbursts. SN 2000ch has continued to experience 
brief, regularly recurring outbursts \citep{Aghakhanloo2022a}, and can fool transient hunters as being a new event (e.g., \citealt{VanDyk2013a}). 

\begin{figure*}[htb]
\gridline{\fig{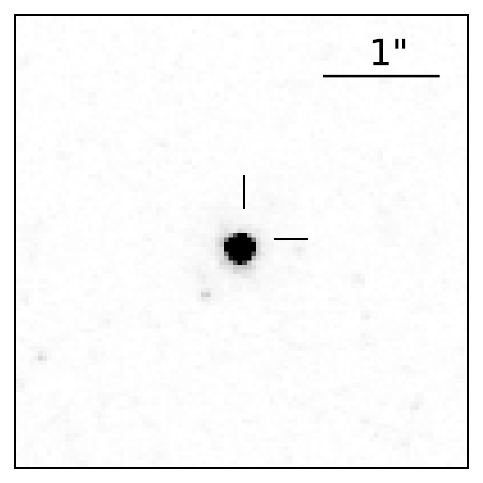}{0.25\textwidth}{(a) F555W}
          \fig{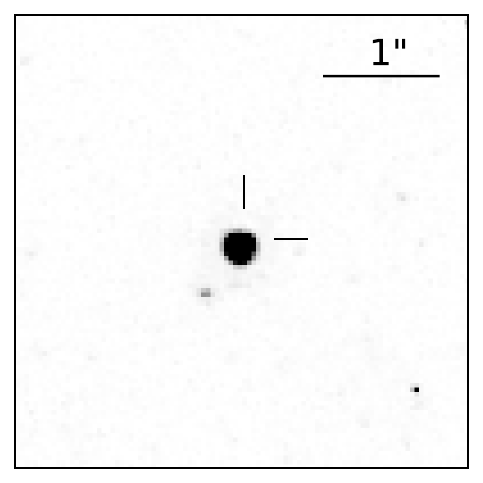}{0.25\textwidth}{(b) F814W}
          \fig{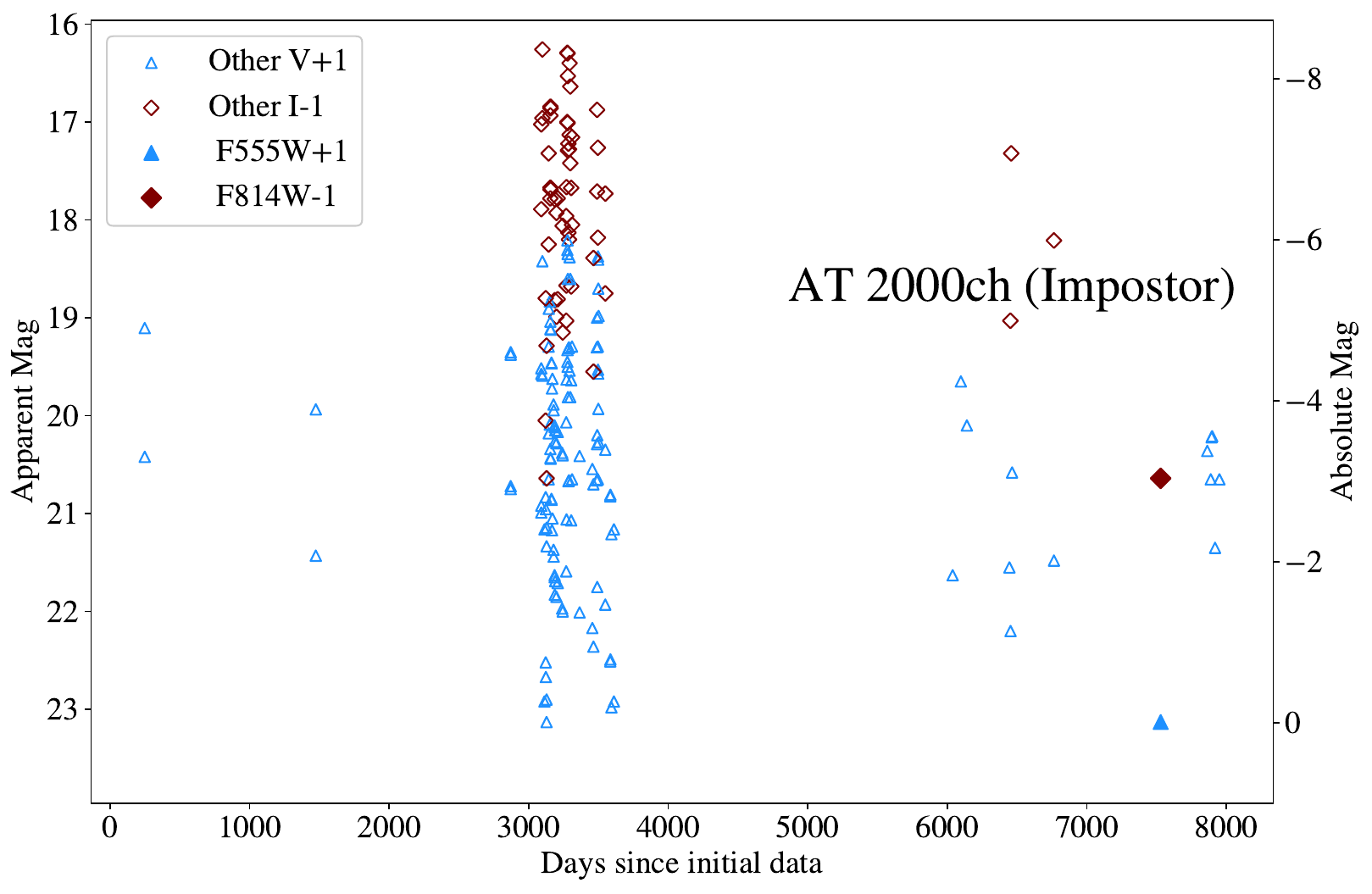}{0.4\textwidth}{(c) light curve}}
\caption{A portion of the WFC3 image mosaic containing SN 2000ch, from observations on 2020 December 13, in (a) F555W and (b) F814W. 
Also shown are the (``Other'') $V$ and $I$ (c) light curves from \citet{Pastorello2010} and \citet{Aghakhanloo2022a}, together with the Snapshot detections.}
\label{fig:00ch}
\end{figure*}

Our Snapshot observations were obtained in F555W ($\sim V$) and F814W on 2020 December 13. As one can see in Figure~\ref{fig:00ch}, the object is easily detectable in the {\sl HST\/} images, and its light curve appears very much unlike that of a typical SN. \citet{Aghakhanloo2022a} have recently analyzed the continued photometric evolution of SN 2000ch, finding periodicity to the cycle of repeating outbursts, which suggests a binary nature for the transient. The object, since it is a likely SN impostor, is not powered at late times by radioactive decay.

\subsection{SN 2010jl}

SN 2010jl was classified as a luminous Type IIn SN, and \citet{Stoll2011} presented early-time light curves and spectra. \citet{Smith2011a} identified a luminous blue ($M_{\rm F300W} \approx -12.0$~mag) point source at the SN location that they identified as a candidate progenitor, although \citet{Fox2017} have since demonstrated that this is less likely. 
\citet{Smith2011a} noted that even if the blue source was a nearby star cluster, its young age suggested a high initial mass of $> 30~M_{\odot}$ for the progenitor of SN~2010jl. Further optical and near-infrared (IR) monitoring of SN 2010jl has been presented by \citet{Zhang2012}, \citet{Ofek2014}, \citet{Borish2015}, \citet{Jencson2016}, and others. As an SN~IIn, similar to the case of SN 1988Z (Section~\ref{88Z}), there has long been multiwavelength evidence for strong circumstellar interaction (e.g., \citealt{Smith2011a,smith12,Fransson2014,Chandra2015}). Most notable is the observational and analytical indications for the presence of dust associated with the SN (e.g., \citealt{Andrews2011,smith12,Gall2014,Sarangi2018,Bevan2020}).

\begin{figure*}[htb]
\gridline{\fig{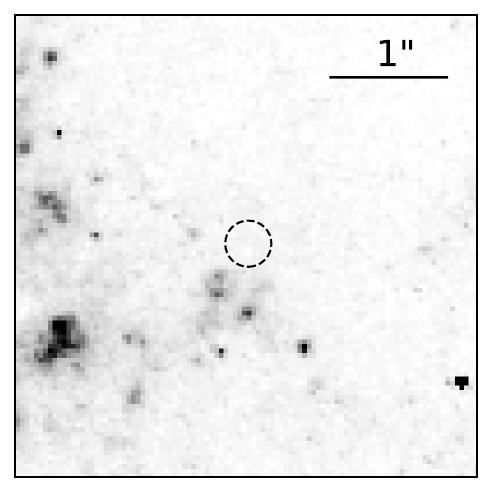}{0.25\textwidth}{(a) F336W}
          \fig{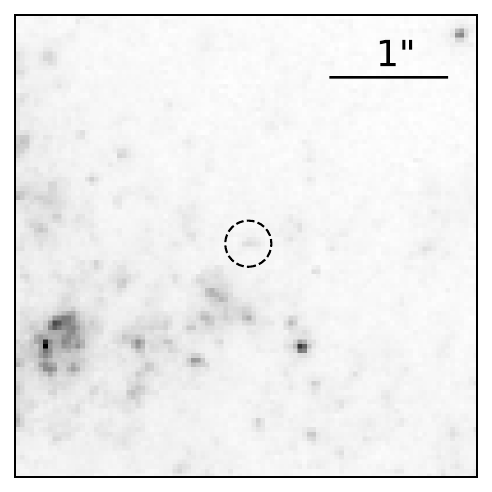}{0.25\textwidth}{(b) F814W}
          \fig{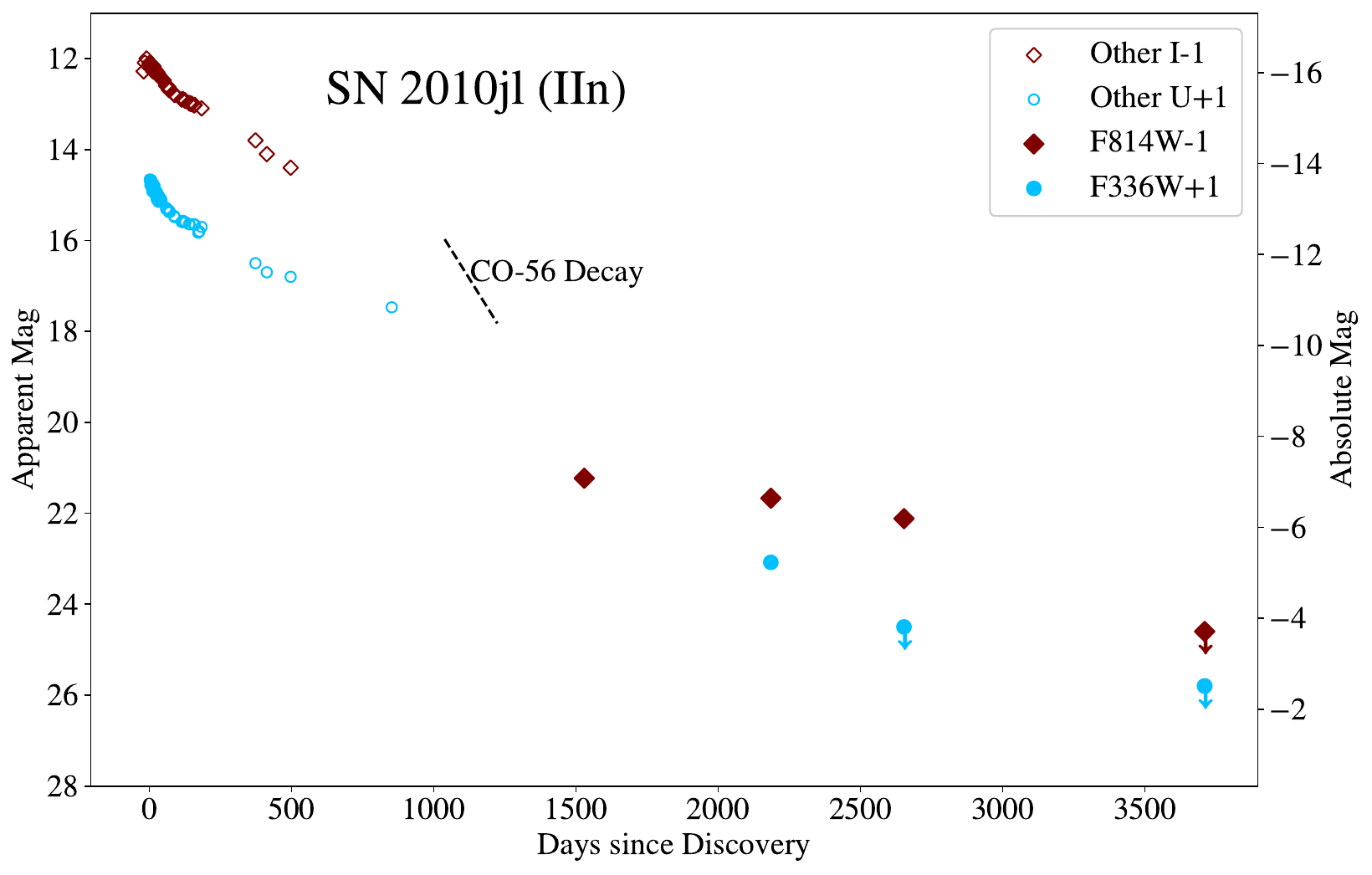}{0.4\textwidth}{(c) light curve}}
\caption{A portion of the WFC3 image mosaic containing SN 2010jl, from observations on 2020 December 29, in (a) F336W and (b) F814W.  As the SN is not detected in either band, its location is indicated by the dashed circle. Also shown are the (``Other'') $U$ and $I$ (c) light curves from \citet{Ofek2014} and \citet{Jencson2016}, and F336W and F814W measurements from \citet{Fox2017}, together with {\sl HST\/} points from GO-15166 (PI A.~Filippenko) and the Snapshot detections.}
\label{fig:10jl}
\end{figure*}

\citet{Fox2017} detected SN 2010jl in {\sl HST\/} images up to 1618~d (4.4~yr) after discovery. Our Snapshots in F336W and F814W from 2020 December 29 are when the SN is significantly older, at 4118~d (11.3~yr); see Figure \ref{fig:10jl}. We located the SN 2010jl site via comparison with prior {\sl HST\/} images obtained in 2015 October by GO-14149 (PI 
A.~Filippenko), when the SN was at $m_{\rm F336W}=21.77 \pm 0.04$ and $m_{\rm F814W}=22.23 \pm 0.02$~mag, and October 2016 by GO-14668 (PI A.~Filippenko), at $m_{\rm F336W}=22.08 \pm 0.03$ and $m_{\rm F814W}=22.67 \pm 0.03$~mag (see \citealt{Fox2017}). Analysis of a set of {\sl HST\/} images from February 7 2018(nearly 3~yr prior to our data), previously-unpublished data from GO-15166 (PI A.~Filippenko), shows that SN 2010jl was still detectable at F814W, with $m_{\rm F814W}=23.11 \pm 0.04$~mag, but had already become undetectable in F336W (with an upper limit of 23.5~mag). Given the late-time brightness of the SN, radioactive decay could not have been the object's primary source of power. We conclude from analysis of our Snapshot images that the SN has now vanished, with upper limits of 24.8 mag in F336W and 25.6 mag in F814W. 

\subsection{SN 2010mc}

\citet{Ofek2012} discovered SN 2010mc during the course of the PTF survey. \citet{Howell2012} classified it subsequently as an SN~IIn at redshift $z=0.035$. Looking back in the PTF data, \citet{Ofek2013} discovered an astounding outburst event $\sim 40$~d prior to the apparent SN.  \citet{smith2013} pointed out that SN~2010mc was a near twin of the remarkable event SN~2009ip, and \citet{Smith2014} 
proposed that both events were the terminal SN~IIn explosions arising from eruptive blue supergiant progenitors. 
A terminal SN explosion has since been confirmed for SN~2009ip \citep{smith22}. 
The last published spectrum of SN 2010mc was from day 1024 \citep{Smith2014}, 
which at that time showed strong shock-broadened H$\alpha$ emission indicative of ongoing CSM interaction.

\begin{figure*}[htb]
\gridline{\fig{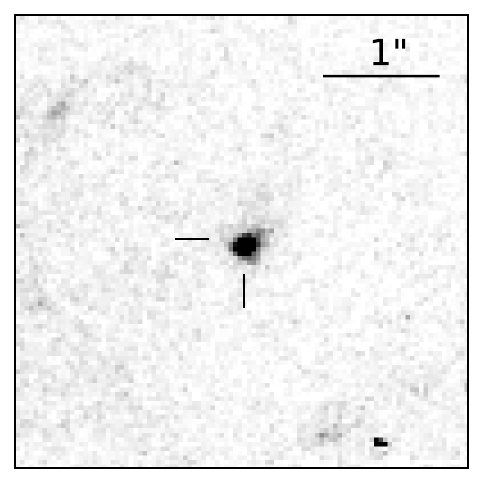}{0.25\textwidth}{(a) F555W}
          \fig{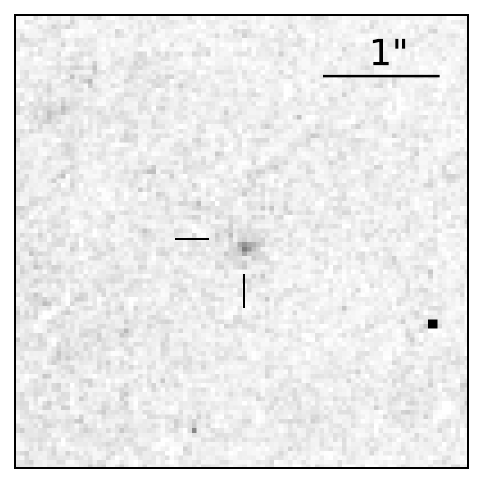}{0.25\textwidth}{(b) F814W}
          \fig{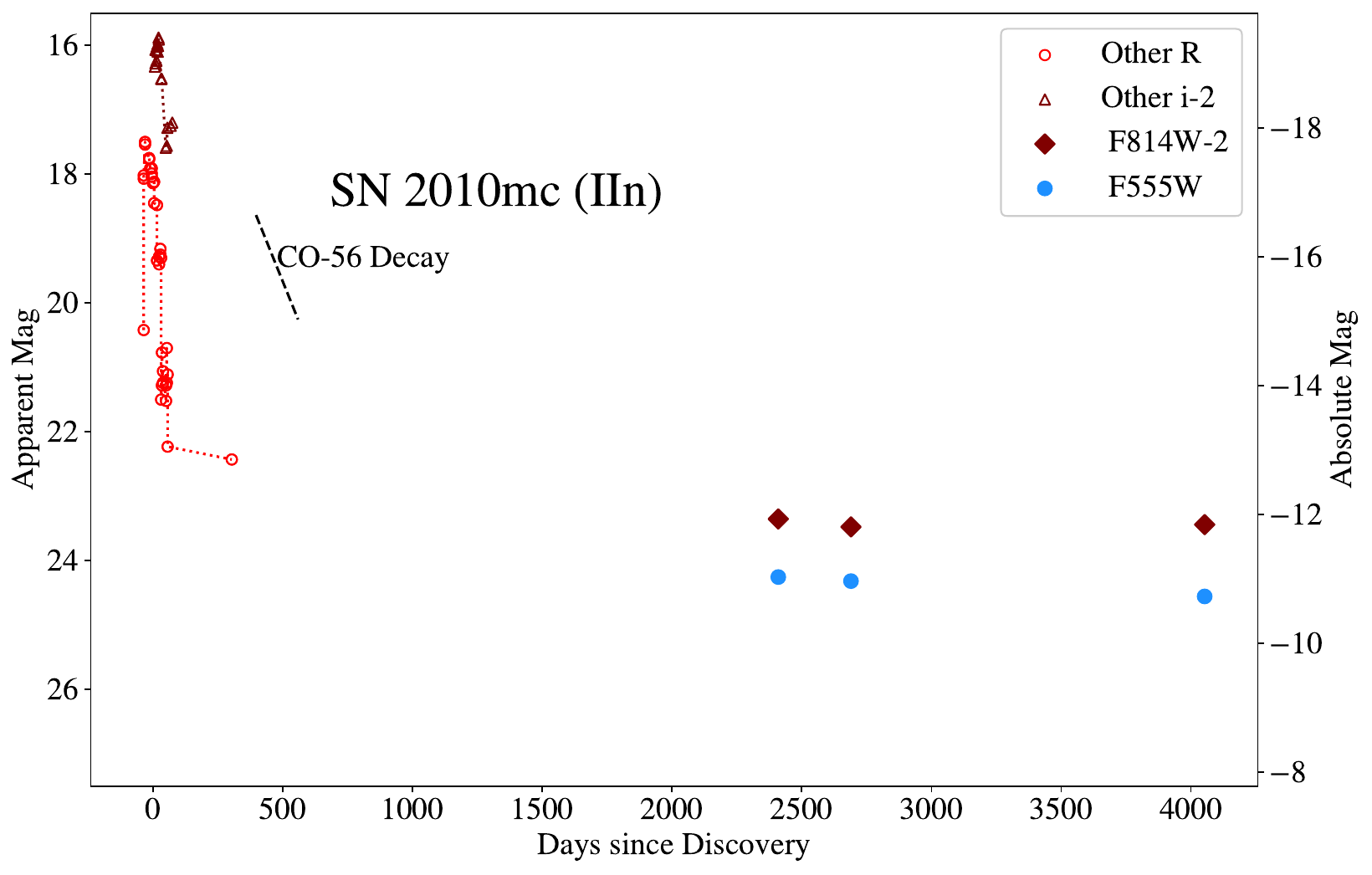}{0.4\textwidth}{(c) light curve}}
\caption{A portion of the WFC3 image mosaic containing SN 2010mc, from observations on 2021 September 24, in (a) F555W and (b) F814W. 
Also shown are the (``Other'') $R$ and $i$ (c) light curves from \citet{Ofek2013} and \citet{Smith2014}, with the Snapshot data at the two {\sl HST\/} bands for comparison.}
\label{fig:10mc}
\end{figure*}

We detected SN 2010mc in both our F555W and F814W Snapshots from 2021 September 24, 4053~d (11.1~yr) after discovery; see Figure \ref{fig:10mc}. We had isolated the site of the SN using {\sl HST\/} data from 2017 March 26 obtained by program GO-14668 (PI A.~Filippenko), when the SN was at $m_{\rm F555W}=24.26 \pm 0.04$ and $m_{\rm F814W}=25.35 \pm 0.10$~mag. One will note that, both in 2017 and 2021, the SN is significantly brighter in F555W than in F814W, which we speculate must be the result of sustained luminous H$\alpha$ emission from the SN within the F555W bandpass, with much less luminous continuum emission in F814W. SN 2010mc has diverged from radioactive-decay power since day $\sim 400$ and continues to do so, most likely as a result of sustained CSM interaction. However, it is possible that some of the light is contributed by a star cluster coincident with the SN, as was the case for SN 2009ip; deeper and higher-resolution observations are needed to identify such a cluster.

\subsection{SN 2011dh}\label{sec:sn2011dh}

The nearby SN 2011dh in M51 has become, along with SN 1993J, one of the best-studied SNe~IIb, if not one of the best-studied SNe of any type thus far. Extensive UV, optical, and near-IR follow-up observations were carried out not long after discovery by \citet{Arcavi2011}, \citet{Sahu2013}, \citet{Shivvers2013}, \citet{Ergon2014,Ergon2015}, \citet{Mauerhan2015}, \citet{Marion2014}, and others. Multiwavelength observations, including X-ray and radio, were indicative of circumstellar interaction; see, for example, \citet{Marti2011}, \citet{Krauss2012}, \citet{Horesh2013}, \citet{Maeda2014}, \citet{deWitt2016}, and \citet{Kundu2019}. Both \citet{Maund2011} and \citet{VanDyk2011} independently identified the SN's progenitor. \citet{Soderberg2012}, followed by \citet{VanDyk2011}, argued that the progenitor was compact, while \citet{Maund2011} pointed to the detected yellow supergiant as the star that exploded, and this was supported by the modeling by \citet{Bersten2012} and subsequently confirmed by the supergiant progenitor's disappearance \citep{VanDyk2013b}. Furthermore, detailed theoretical modeling of the progenitor by \citet{Benvenuto2013} supported the binary origin for the SN. \citet{Folatelli2014} claimed detection of a possible blue companion to the progenitor, although \citet{Maund2015} and \cite{Maund2019} cast some doubt on that possibility.

\begin{figure*}[htb]
\gridline{\fig{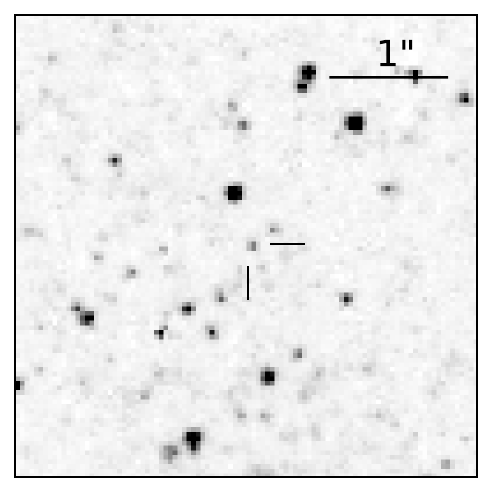}{0.25\textwidth}{(a) F555W}
          \fig{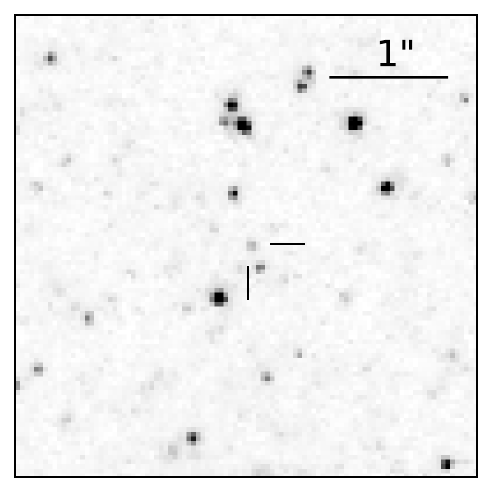}{0.25\textwidth}{(b) F814W}
          \fig{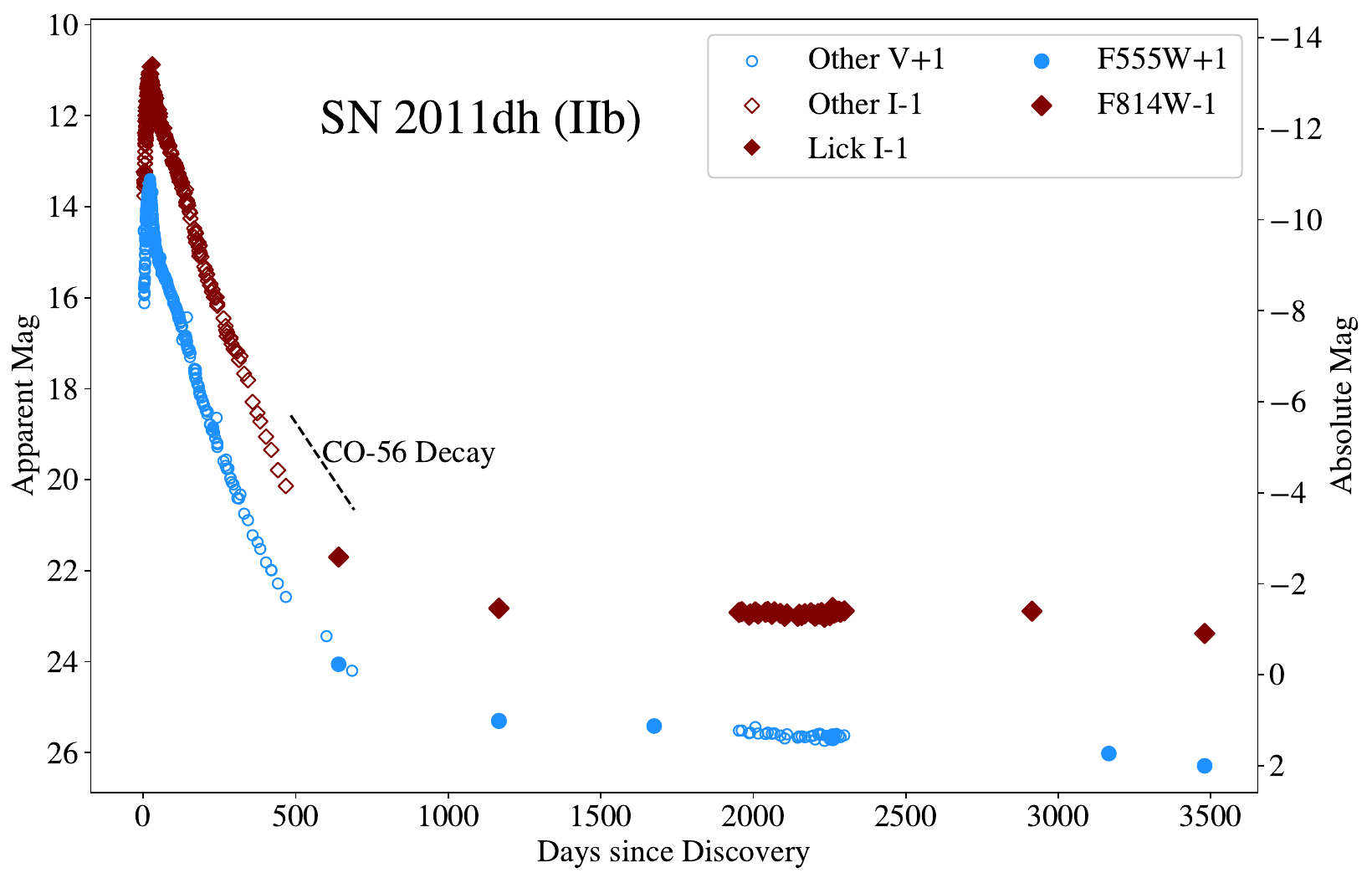}{0.4\textwidth}{(c) light curve}}
\caption{A portion of the WFC3 image mosaic containing SN 2011dh, from observations on 2020 December 10, in (a) F555W and (b) F814W. 
Also shown are the Lick \citep{Zheng2022} $V$ and $I$ (c) light curves, along with (``Other'') data in these bands from \citet{Arcavi2011}, \citet{Tsvetkov2012}, \citet{Sahu2013}, \citet{Marion2014}, and \citet{Ergon2015}, together with F555W and F814W data from \citet{Maund2019} and \citet{Maund2015}, {\sl HST\/} programs GO-15645 (PI D.~Sand) and GO-16024 (PI A.~Filippenko), and our Snapshot observations.}
\label{fig:11dh}
\end{figure*}

\cite{Maund2019}, analyzing a veritable treasure trove of {\sl HST\/} imaging serendipitously covering the SN site, argued that a light echo originating from dust with a preferred disk geometry could be responsible for the observed extended late-time emission. We pinpointed the location of the SN in our Snapshot images in F555W and F814W from 2020 December 10 (3480~d $\approx 9.5$~yr after discovery), using a number of these prior {\sl HST\/} images for comparison, and as can be seen, the emission is significantly fainter than from the analysis by \cite{Maund2019}; see Figure \ref{fig:11dh}. Since that study, previously-unpublished {\sl HST\/} observations also have been obtained of the SN site in 2019 on November 24 with the Advanced Camera for Surveys (ACS) Wide-Field Channel (WFC) in F814W by GO-15645 (PI D.~Sand; $m_{\rm F814W} = 23.89 \pm 0.02$~mag) and in 2020 January 29 with WFC3/UVIS in F555W by GO-16024 (PI A.~Filippenko; $m_{\rm F555W} = 25.02 \pm 0.05$~mag). (We note that data were also obtained by GO-16024 in F225W, not shown; however, the SN was not detected, to a limit of 24.0~mag; cf.~\citealt{Maund2015}.) Taken together, these late-time {\sl HST\/} data indicate a slow, steady fading of the SN emission --- but clearly the SN at very late times has not followed a radioactive decay-powered decline.

\subsection{SN 2012A}

\citet{Tomasella2013} extensively monitored the normal SN~II-P 2012A, beginning within a few days after discovery. The SN was also monitored in multiple bands and spectroscopically from Lick Observatory \citep{deJaeger2019}. \citet{Silverman2017} analyzed a late-time Keck Observatory spectrum. \citet{Prieto2012} detected a progenitor candidate for the SN in pre-explosion Gemini ground-based adaptive optics (AO) imaging in the $K$ band and characterized the star as a red supergiant (RSG). \citet{Tomasella2013} reanalyzed these data and estimated that the progenitor's initial mass was $\sim 10~M_{\odot}$. \citet{Utrobin2015} also estimated the progenitor initial mass at $13.1 \pm 0.7~M_{\odot}$ from hydrodynamical modeling of the clumpiness of the ejecta. 

\begin{figure*}[htb]
\gridline{\fig{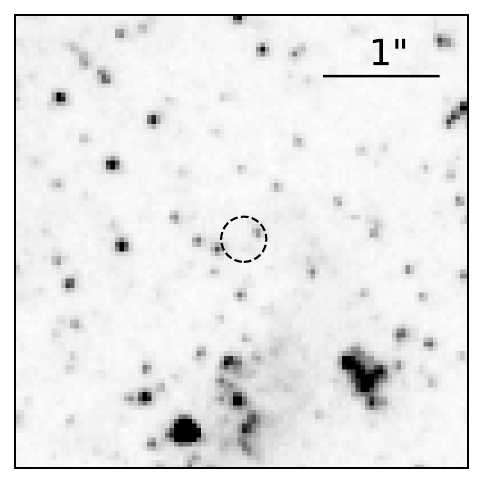}{0.25\linewidth}{(a) F606W}
          \fig{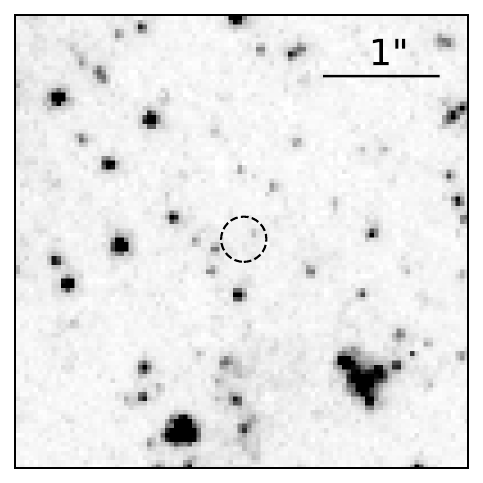}{0.25\linewidth}{(b) F814W}
          \fig{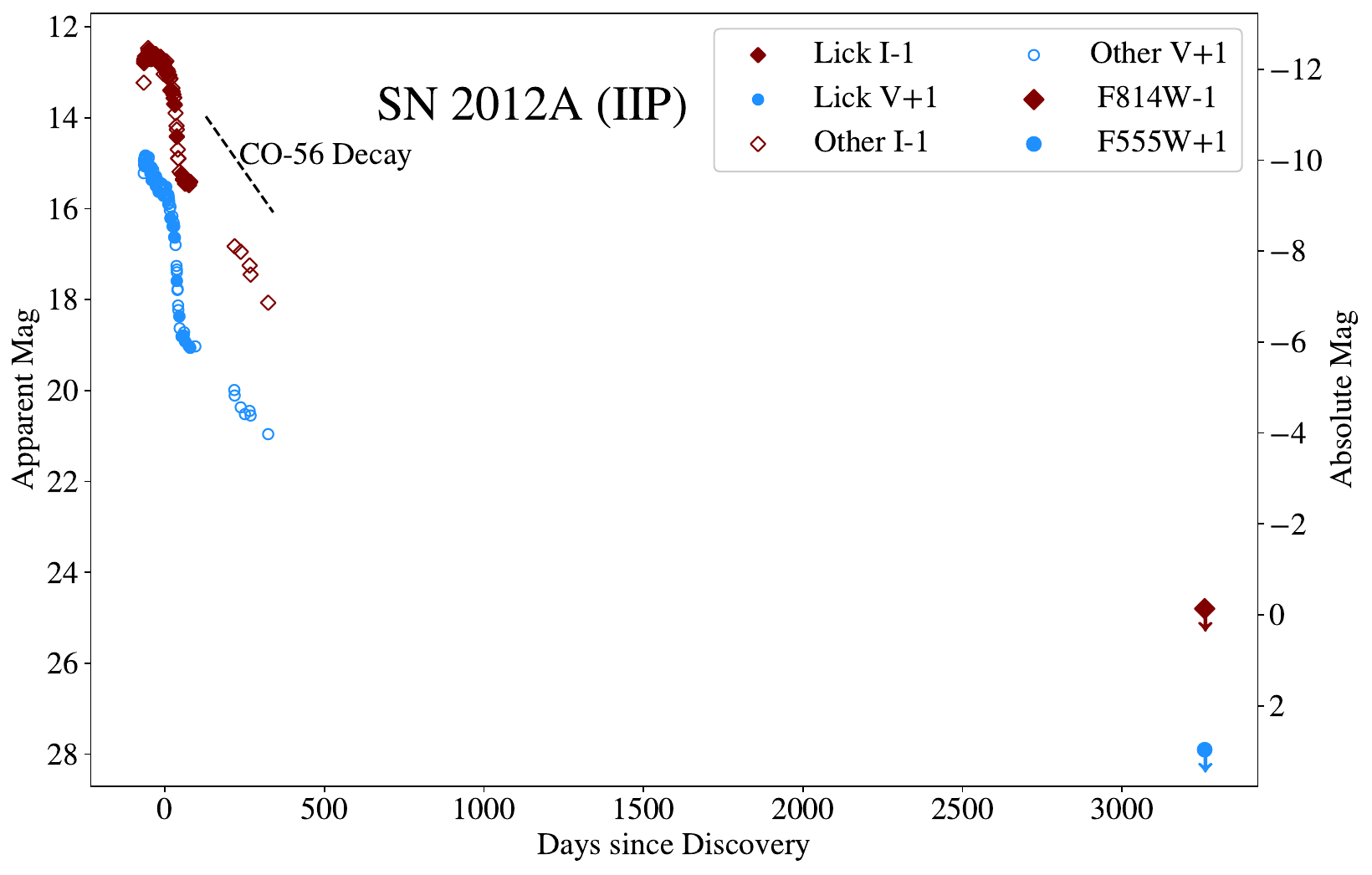}{0.4\linewidth}{(c) light curve}}
\caption{A portion of the WFC3 image mosaic containing SN 2012A, from observations on 2021 February 16, in (a) F606W and (b) F814W. 
The SN was not detected in either band; the site is indicated by the dashed circle. Also shown are the Lick \citep{deJaeger2019} $V$ and $I$ (c) light curves, along with (``Other'') data in these bands from \citet{Tomasella2013}, together with the detection upper limits from our Snapshot data.}
\label{fig:12A}
\end{figure*}

We obtained our Snapshot observations in F606W (``Wide'' $\sim V$) and F814W on 2021 February 16, 3329~d (9.1~yr) after discovery. We had purposely selected F606W in this case, rather than F555W, since the former is the preferred bandpass in which to acquire data, together with F814W, for use in tip-of-the-red-giant-branch (TRGB) distance estimates \citep{Anand2021}. Our intention was to obtain data that can better constrain the distance to SN 2012A, although that TRGB estimation is beyond the scope of this paper. (Our focus for the Snapshots was on the SN itself, which is within the main body of the host galaxy, and not the galaxy halo in which the TRGB would likely be more apparent.) We astrometrically aligned KAIT images from \citet{deJaeger2019} with our Snapshot images, in order to isolate the SN site, and concluded from that analysis that the SN was no longer detectable, to $26.9$ and $25.8$~mag in F606W and F814W, respectively; see Figure \ref{fig:12A}. \citet{VanDyk2023} performed a more precise alignment of the Snapshot F814W image with the AO data and concluded that the RSG progenitor star had vanished.

\subsection{SN 2012aw}

SN 2012aw in M95 is a nearby, well-studied normal SN~II-P. Multiwavelength follow-up observations have been conducted by a number of investigators, including \citet{Bose2013}, \citet{Dall'Ora2014}, \citet{Jerkstrand2014}, and \citet{deJaeger2019}). Not long after discovery both \citet{VanDyk2012b} and \citet{Fraser2012} identified a candidate progenitor RSG in pre-explosion {\sl HST\/} WFPC2 images. The progenitor was confirmed, as it had disappeared in late-time {\sl HST\/} imaging \citep{Fraser2016}. In the same late-time imaging, \citet{VanDyk2015} discovered a resolved light echo around the SN.  

\begin{figure*}[htb]
\gridline{\fig{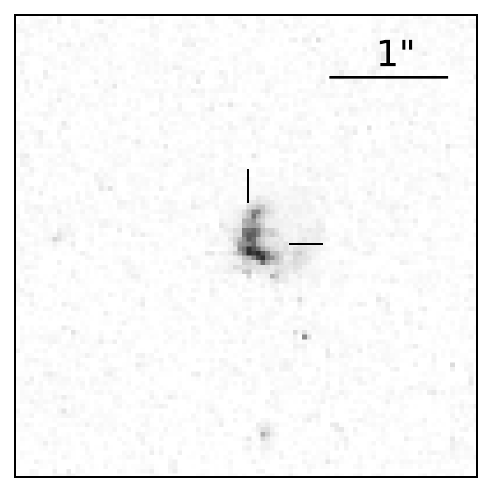}{0.25\textwidth}{(a) F555W}
          \fig{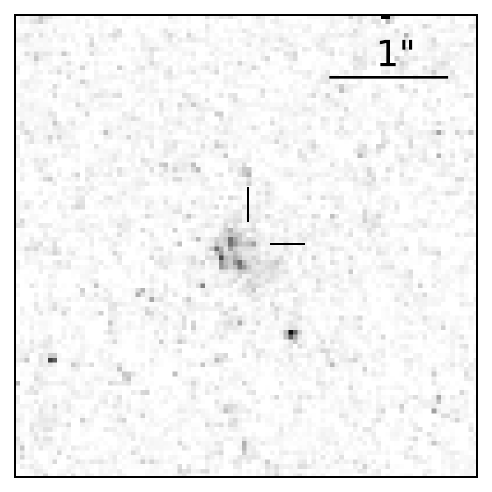}{0.25\textwidth}{(b) F814W}
          \fig{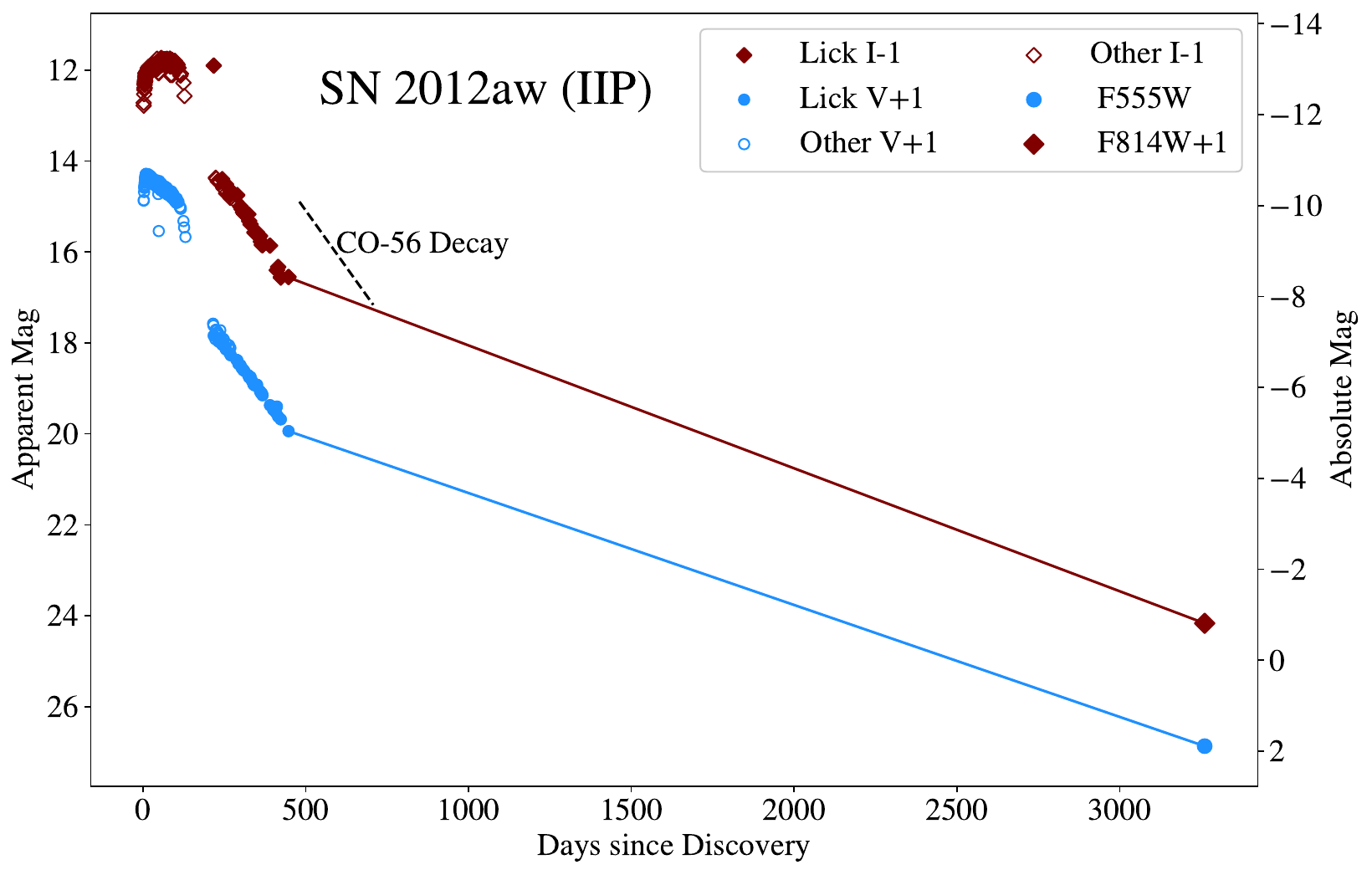}{0.4\textwidth}{(c) light curve}}
\caption{A portion of the WFC3 image mosaic containing SN 2012aw, from observations on 2021 February 17, in (a) F555W and (b) F814W. 
One can also clearly see the light echo surrounding the SN. Also shown are the Lick \citep{deJaeger2019} $V$ and $I$ (c) light curves, along with (``Other'') data in these bands from \citet{Spogli2020}, \citet{Bose2013}, and \citet{Dall'Ora2014}, together with the Snapshot detections.}
\label{fig:12aw}
\end{figure*}

We pinpointed the location of the SN in our Snapshot data in F555W and F814W from 2021 February 17, using {\sl HST\/} data obtained on 2016 October 24 by GO-14668 (PI A.~Filippenko), as well as the pre-explosion WFPC2 images in which the progenitor had been identified; see Figure \ref{fig:12aw}. What is most strikingly apparent is the continued presence of the light echo in both bands around the SN site. Whereas the SN was obscured by the echo at earlier times \citep{VanDyk2015}, the SN had become recoverable in both the {\sl HST\/} F814W and F555W images 3261~d (8.9~yr) after discovery.  

The echo itself is seen almost as a perfect ring, although asymmetric relative to the SN position, with the SN offset by $\sim 2.6$ pixels ($\sim 0{\farcs}103$ at the UVIS pixel scale) southeast from the ring center. The surface brightness of the echo is still highest to the east and southeast, as reported by \citet{VanDyk2015}. We estimate that the radius of the echo is $\sim 7.4$ pixels ($\sim 0{\farcs}293$). This is nearly double the radius, at $\sim 2492$~d later than when first discovered by \citet{VanDyk2015} in 2014.
A detailed analysis of the echo and its evolution is beyond the scope of this paper.

\subsection{SN 2013df}\label{sec:sn2013df}

SN 2013df is an SN IIb in NGC 4414 and was first studied in detail by \citet{VanDyk2014}, who also identified its yellow supergiant progenitor.
\citet{Morales2014}, \citet{Maeda2015}, and \citet{Szalai2016} performed additional optical and near-IR follow-up observations. It was established early that SN 2013df strongly resembled SN 1993J (Section~\ref{sec:sn1993J}), both in SN and progenitor properties. The radio and X-ray emission from the SN \citep{Kamble2016}, together with the late-time ($\sim 670$~d) optical spectral characteristics \citep{Maeda2015}, were indicative of circumstellar interaction.

\begin{figure*}[htb]
\gridline{\fig{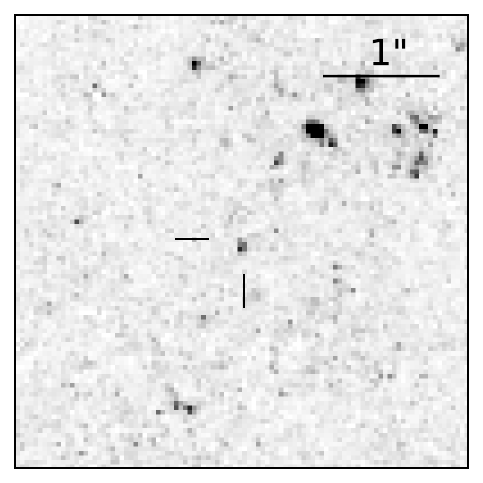}{0.25\textwidth}{(a) F336W}
          \fig{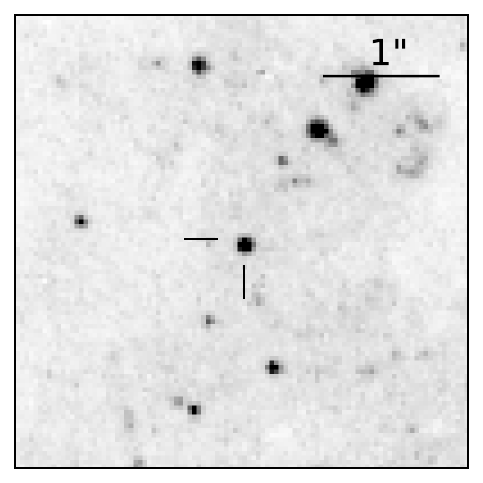}{0.25\textwidth}{(b) F555W}
          \fig{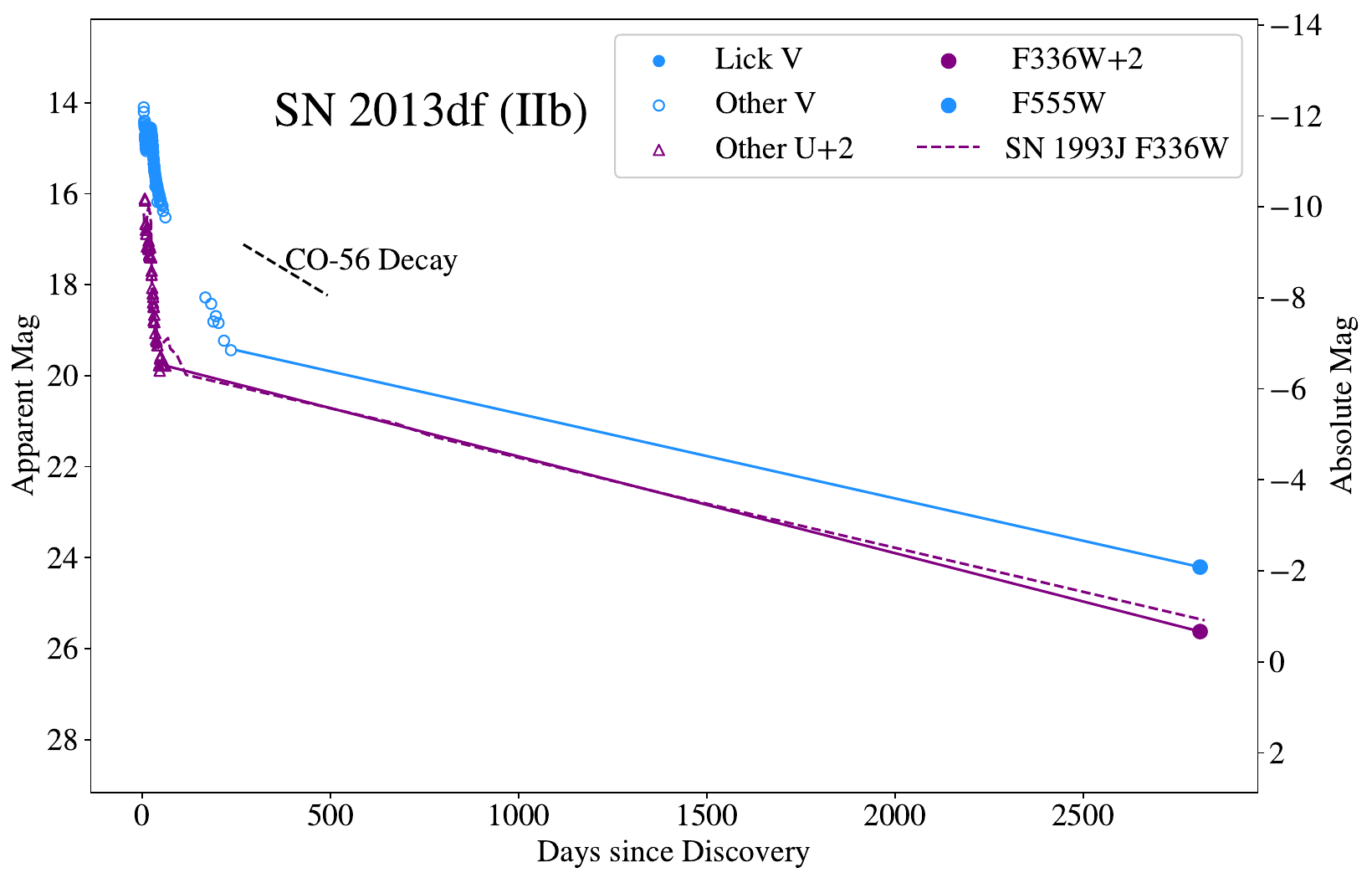}{0.4\textwidth}{(c) light curve}}
\caption{A portion of the WFC3 image mosaic containing SN 2013df, from observations on 2021 February 15, in (a) F336W and (b) F555W. 
Also shown is the Lick + RATIR \citep{VanDyk2014} $V$ (c) light curve, along with (``Other'') data in $U$ and $V$ from \citet{Morales2014} and \citet{Szalai2016}, together with the Snapshot detections. Furthermore, we have overlaid, by adjusting in time and in magnitude, the F336W light curve (dashed line) of SN 1993J (Section~\ref{sec:sn1993J}) on the F336W curve of SN 2013df.}
\label{fig:13df}
\end{figure*}

We pinpointed the exact location of SN 2013df by comparing directly with {\sl HST\/} observations from 2013 July 15 (GO-12888; PI S.~Van Dyk), when the SN had $m_{\rm F555W}=16.15 \pm 0.01$~mag. As one can see from Figure \ref{fig:13df}, the SN was clearly detected in our Snapshot images from 2021 February 15, both in F336W and F555W, 2811~d (7.7~yr) after discovery. That the SN is still relatively bright in F336W indicates that the interaction was still ongoing at the time of our observations. The brightness in F555W is likely dominated within the bandpass by continued luminous H$\alpha$ emission, as well as less prominent He~{\sc i}/Na~{\sc i} emission, as seen in the late-time spectra \citep{Maeda2015}. In the figure we have overlaid the F336W light curve of SN 1993J (see Section~\ref{sec:sn1993J}) on the light curve of SN 2013df in this same band. As one can see, the two agree amazingly well, which would imply that, based on the photometric evolution over more than 2800~d (7.7~yr), SN 2013df is essentially a twin of SN 1993J, as the early-time data, including the progenitor identification, tended to indicate as well.

\subsection{SN 2013ej}

\begin{figure*}[htb]
\gridline{\fig{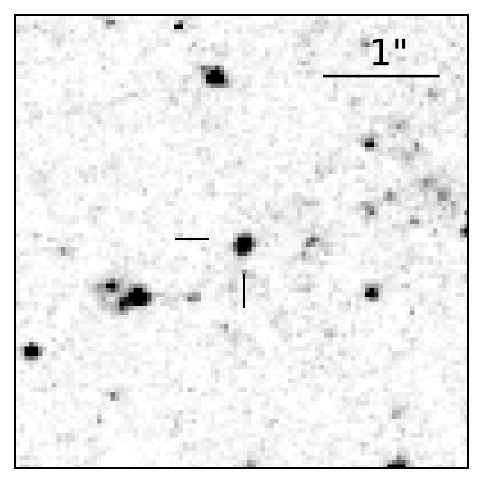}{0.25\textwidth}{(a) F555W}
          \fig{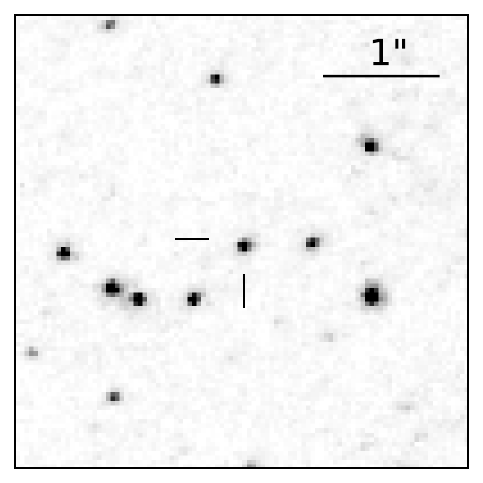}{0.25\textwidth}{(b) F814W}
          \fig{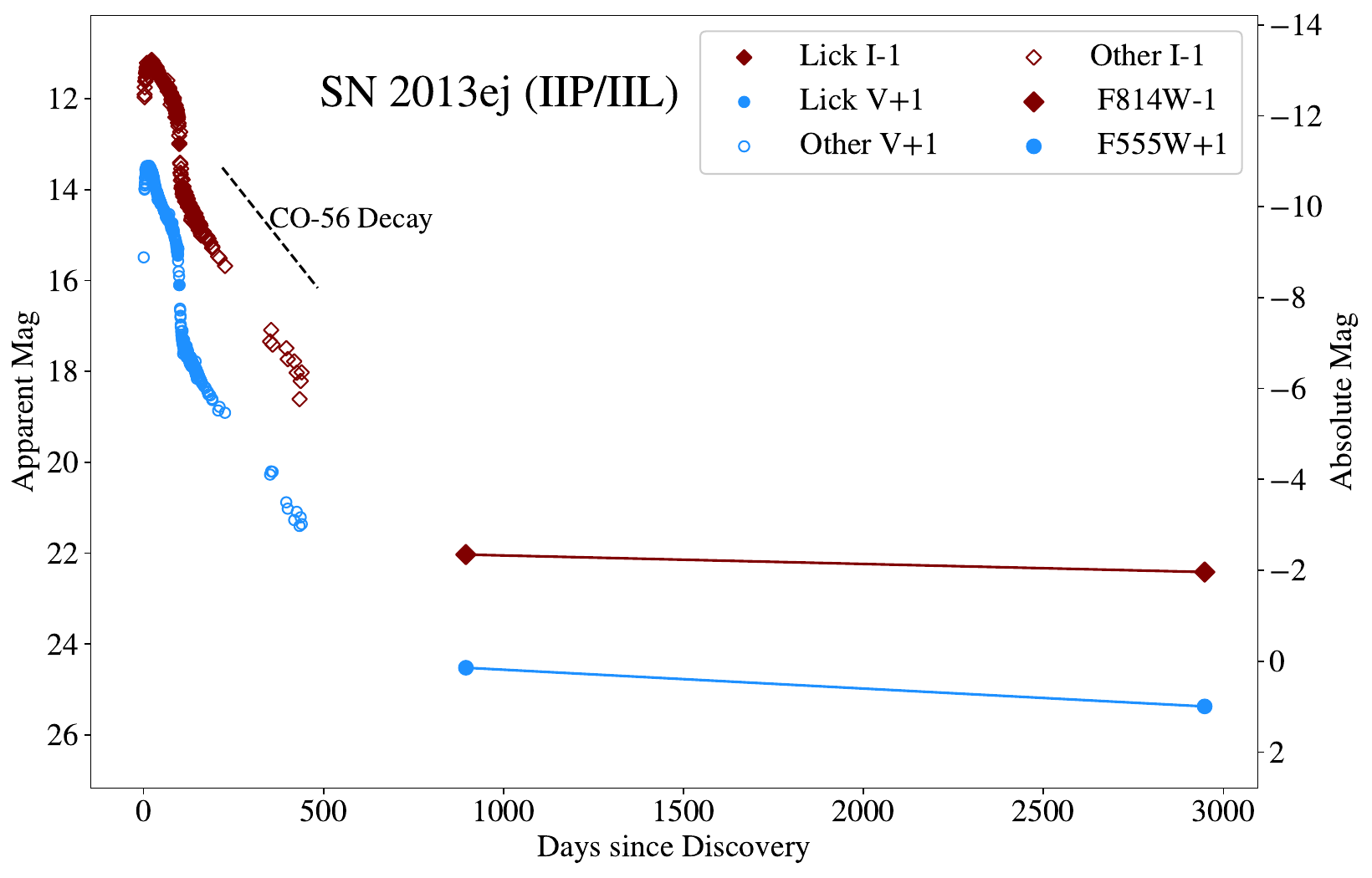}{0.4\textwidth}{(c) light curve}}
\caption{A portion of the WFC3 image mosaic containing SN 2013ej, from observations on 2021 August 19, in (a) F555W and (b) F814W. 
Also shown are the Lick \citep{deJaeger2019} $V$ and $I$ (c) light curves, along with (``Other'') data from \citet{Richmond2014}, \citet{Bose2015}, \citet{Huang2015}, \citet{Dhungana2016}, \citet{Yuan2016}, and \citet{Mauerhan2017}, together with the Snapshot detections.}
\label{fig:13ej}
\end{figure*}

SN 2013ej in M74 has been considered an atypical SN~II, possibly an SN~II-P/II-L hybrid \citep{Mauerhan2017}. A number of investigators have observed and analyzed the SN; see \citet{VanDyk2023} and references therein. Early-time monitoring was also undertaken by KAIT; see \citet{deJaeger2019}. Our {\sl HST\/} Snapshots were obtained in F555W and F814W on 2021 August 19, 2948~d (8.1~yr) after discovery. The location of the SN in our data was established based on \citet[][ their Figure 11]{Mauerhan2017}; see Figure \ref{fig:13ej}. The SN 2013ej light curves in both bands have flattened out significantly at late times, showing essentially no decline in brightness ($< 1$~mag decline in both F555W and F814W) over more than 2000~d (5.5~yr).  \citet{VanDyk2023}, based on the brightness of the SN in these Snapshot data, concluded that the progenitor identified by \citet{Fraser2014} had vanished (confirming an earlier inference made by \citealt{Mauerhan2017}).

Technically, we covered the field again in bands F438W ($\sim B$) and F625W when we observed AT 2019krl (see Section~\ref{sec:2019krl}); however, unfortunately the SN 2013ej site fell within the chip gap for both of those bands.

\subsection{SN 2014C}\label{sec:2014C}

SN 2014C in NGC 7331 is a fascinating event, having been classified soon after explosion as an SN~Ib (without H) and, after $\sim 1$~yr, exhibited distinct and strong signs of circumstellar interaction, similar to an SN~IIn, with strong H$\alpha$ emission \citep[e.g.,][]{Milisavljevic2015}. From  radio and X-ray monitoring \citet{Margutti2017} inferred that the progenitor star had ejected a massive ($\sim 1\ M_{\odot}$) H shell decades to centuries before explosion, and that possibly as many as $\sim 10$\% of all SN~Ib progenitors might experience a similar history. \citet{Brethauer2022} have since interpreted that the shell, with as much as $\sim 2\ M_{\odot}$, has a radius of $\sim 2 \times 10^{16}$ to $\sim 10^{17}$~cm. \citet{Milisavljevic2015} identified in pre-explosion F658N {\sl HST\/} imaging a luminous H$\alpha$ source at the SN's position, which they inferred was a stellar cluster that was home to the progenitor. \citet{Sun2020} performed a detailed analysis of this cluster and estimated an age for it of $\sim 20$~Myr, which they found consistent with a $\sim 11\ M_{\odot}$ star stripped partially of its envelope via mass transfer with a companion in a relatively wide binary system, followed by an eruptive ejection of the remaining H prior to explosion. 

\begin{figure*}[htb]
\gridline{\fig{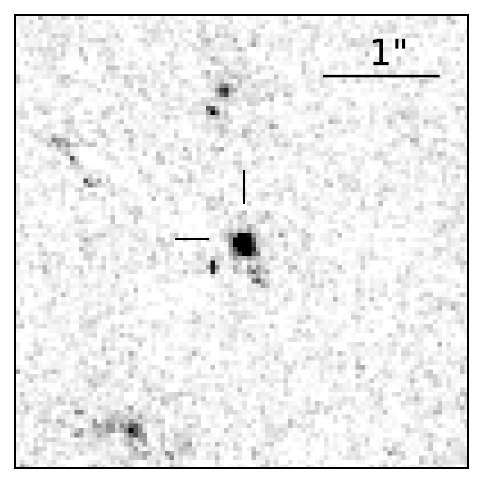}{0.25\textwidth}{(a) F336W}
          \fig{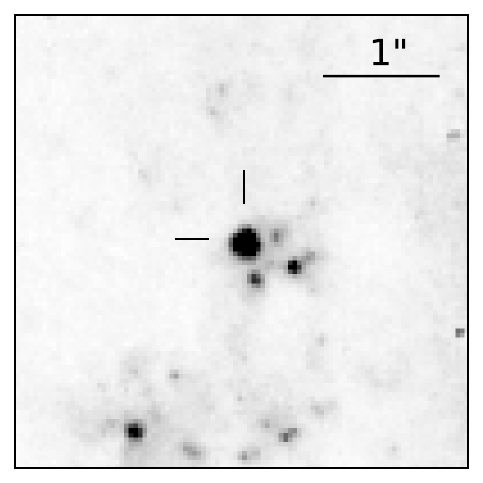}{0.25\textwidth}{(b) F625W}
          \fig{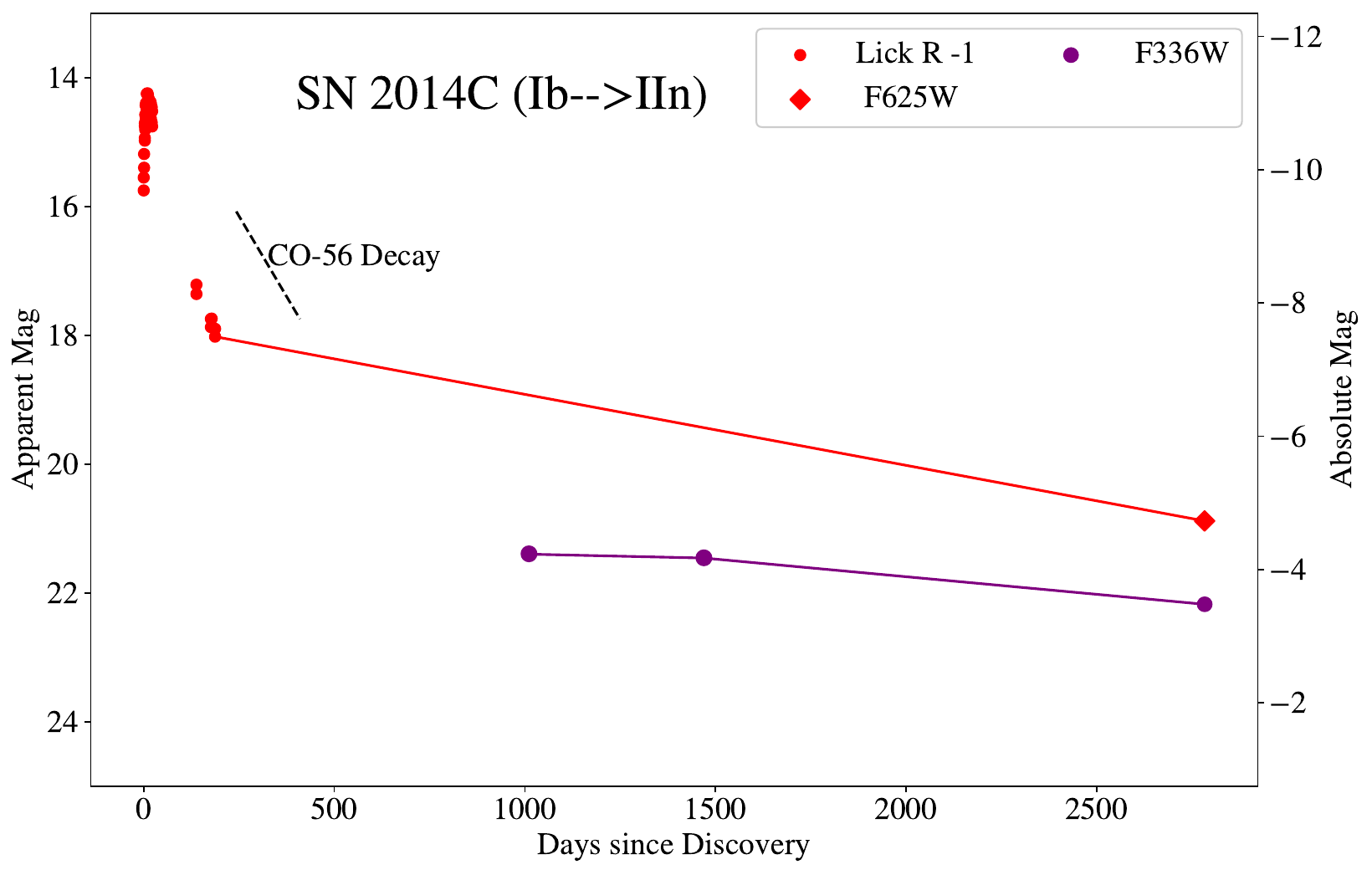}{0.4\textwidth}{(c) light curve}}
\caption{A portion of the WFC3 image mosaic containing SN 2014C, from observations on 2021 August 20, in (a) F336W and (b) F625W. 
Also shown is the Lick \citep{Zheng2022} $R$ (c) light curve, together with measurements from late-time F336W observations by \citet[][which were originally from our previous Snapshot programs GO-14668 and GO-15166, PI A.~Filippenko]{Sun2020}, as well as our Snapshot detections.}
\label{fig:14C}
\end{figure*}

We located the SN in our Snapshot images obtained in F336W and F625W on 2021 August 20 (2786~d $\approx 7.6$~yr after discovery), using {\sl HST\/} data obtained in 2016 October for program GO-14668 (PI A.~Filippenko); see Figure \ref{fig:14C}. We intentionally selected the F336W band, to sample any late-time UV emission from the SN, and F625W, to sample H$\alpha$ emission, both being indicators of ongoing interaction. \citet{Zheng2022} undertook early-time optical monitoring with KAIT, during the ``Ib'' phase of the SN, prior to the onset of strong interaction, and we combine that photometry here with our Snapshots and other available late-time {\sl HST\/} data. What can be seen in the figure is the contribution to the light curves from the circumstellar interaction and that the UV emission, in particular, has declined somewhat in strength since day $\sim 1500$. This behavior would be consistent with the decline in the observed X-ray luminosity after day $\sim 1000$ \citep{Brethauer2022}.

\bibpunct[;]{(}{)}{;}{a}{}{;}

\subsection{SN 2015cp}

SN 2015cp (also known as PS15dpq), in a host at $z \approx 0.04$, was originally classified as a ``91T-like'' overluminous SN~Ia at $\sim 40$~d post-peak, but was shown to be experiencing circumstellar interaction at 681~d after explosion, based on a near-UV (NUV) {\sl HST\/} detection (\citealt{Graham2019} and references therein). \citet{Harris2018}, based on radio and X-ray follow-up observations of this ``SN Ia-CSM,'' constrained the total circumstellar mass at $< 0.5\ M_{\odot}$. \citet{Graham2019} estimated constraints on the inner radius of the CSM of $R_{\rm CSM} > 10^{16}$ and $< 10^{17}$~cm. There was little early-time optical follow-up photometry of the SN, beyond a minimal iPTF light curve in the $g$ and $R$ bands.

\begin{figure*}[htb]
\gridline{\fig{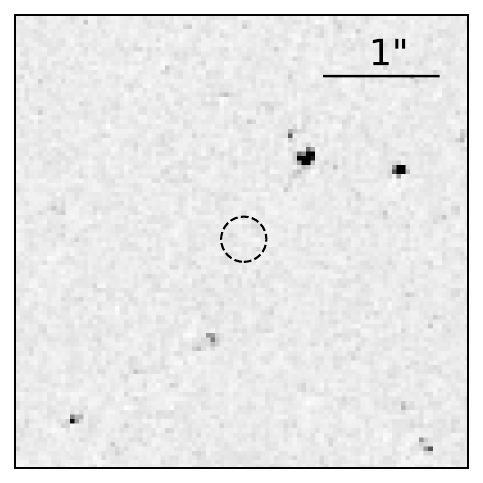}{0.25\textwidth}{(a) F275W}
          \fig{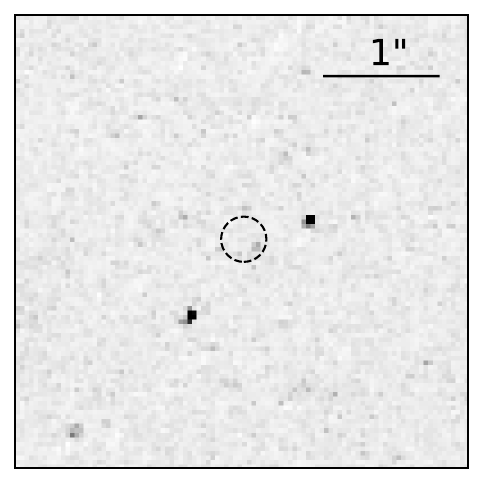}{0.25\textwidth}{(b) F625W}
          \fig{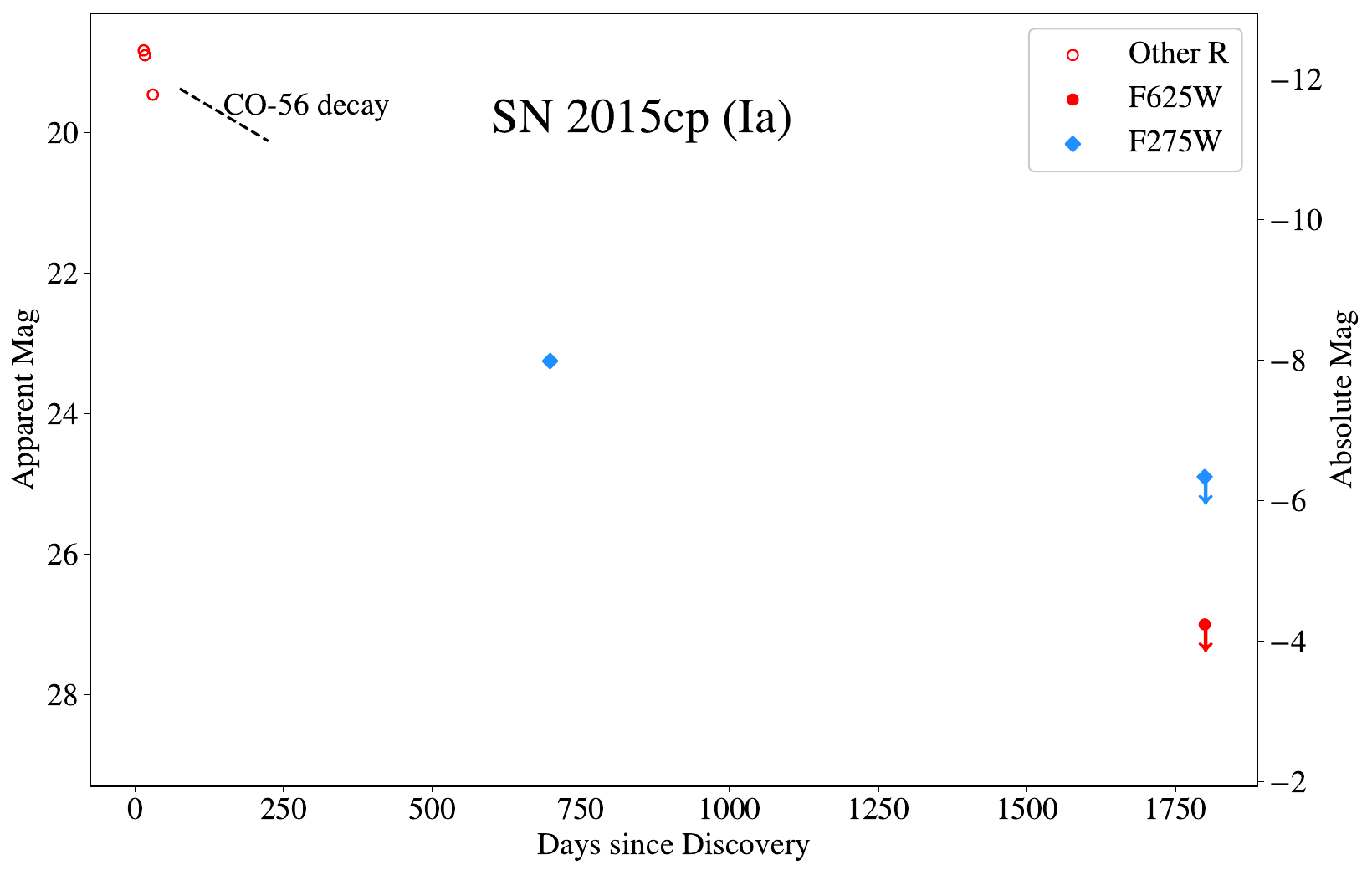}{0.4\textwidth}{(c) light curve}}
\caption{A portion of the WFC3 image mosaic containing SN 2015cp, from observations on 2020 November 30, in (a) F275W and (b) F625W. 
The SN was not detected in either band; the site is indicated by the dashed circle. Also shown is the iPTF $R$ (c) light curve (adjusted from AB mag to Vegamag) and our rereduction of the {\sl HST\/} F275W detection on 2017 September 12 from \citet{Graham2019}, together with our Snapshot upper limits.}
\label{fig:15cp}
\end{figure*}

Our Snapshot observations were obtained on 2020 November 30 (1798~d $\approx 4.9$~yr after discovery) in F275W (NUV) and F625W. We first rereduced with {\tt Astrodrizzle} \citep{STSCI2012} and {\tt Dolphot} the {\sl HST\/} F275W image  (858~s) from \citet[][GO-14779, PI M.~Graham]{Graham2019} and confirmed the detection of the SN, at $m_{\rm F275W}=23.25 \pm 0.08$~mag (in this case, Vega mag, whereas \citealt{Graham2019} present the brightness in AB mag). We used the resulting mosaic to isolate the location of SN 2015cp in our Snapshot mosaics. As can be seen in Figure \ref{fig:15cp} the SN was not detected to a limit of 25.1 and 27.2~mag in F275W and F625W, respectively. The upper limit to detection in F275W implies that the SN shock may have ceased interacting with the CSM some time between 681~d and 1798~d, when our Snapshots were executed (at least to the detection depth of our F275W data). If we assume that the SN shock expanded through the CSM at $\sim 2000$~km~s$^{-1}$ \citep{Graham2019}, then we can place a limit on the outer radius of the CSM at $\lesssim 3.1 \times 10^{16}$~cm based on our nondetection. 

\subsection{SN 2016G}

\begin{figure*}[htb]
\gridline{\fig{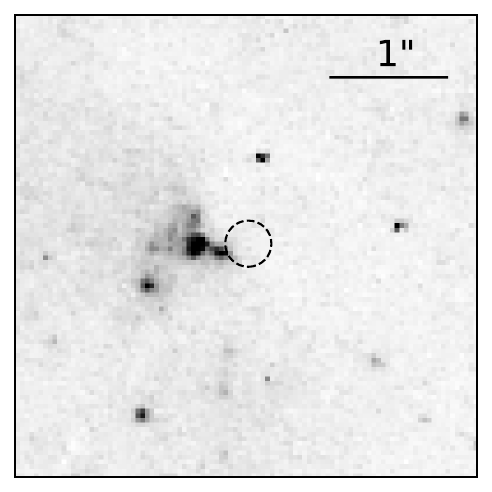}{0.25\textwidth}{(a) F555W}
          \fig{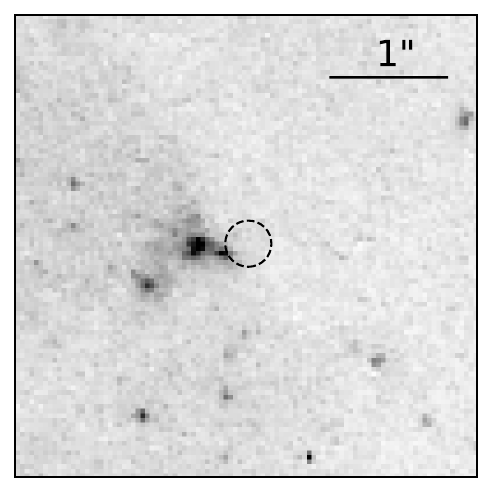}{0.25\textwidth}{(b) F814W}
          \fig{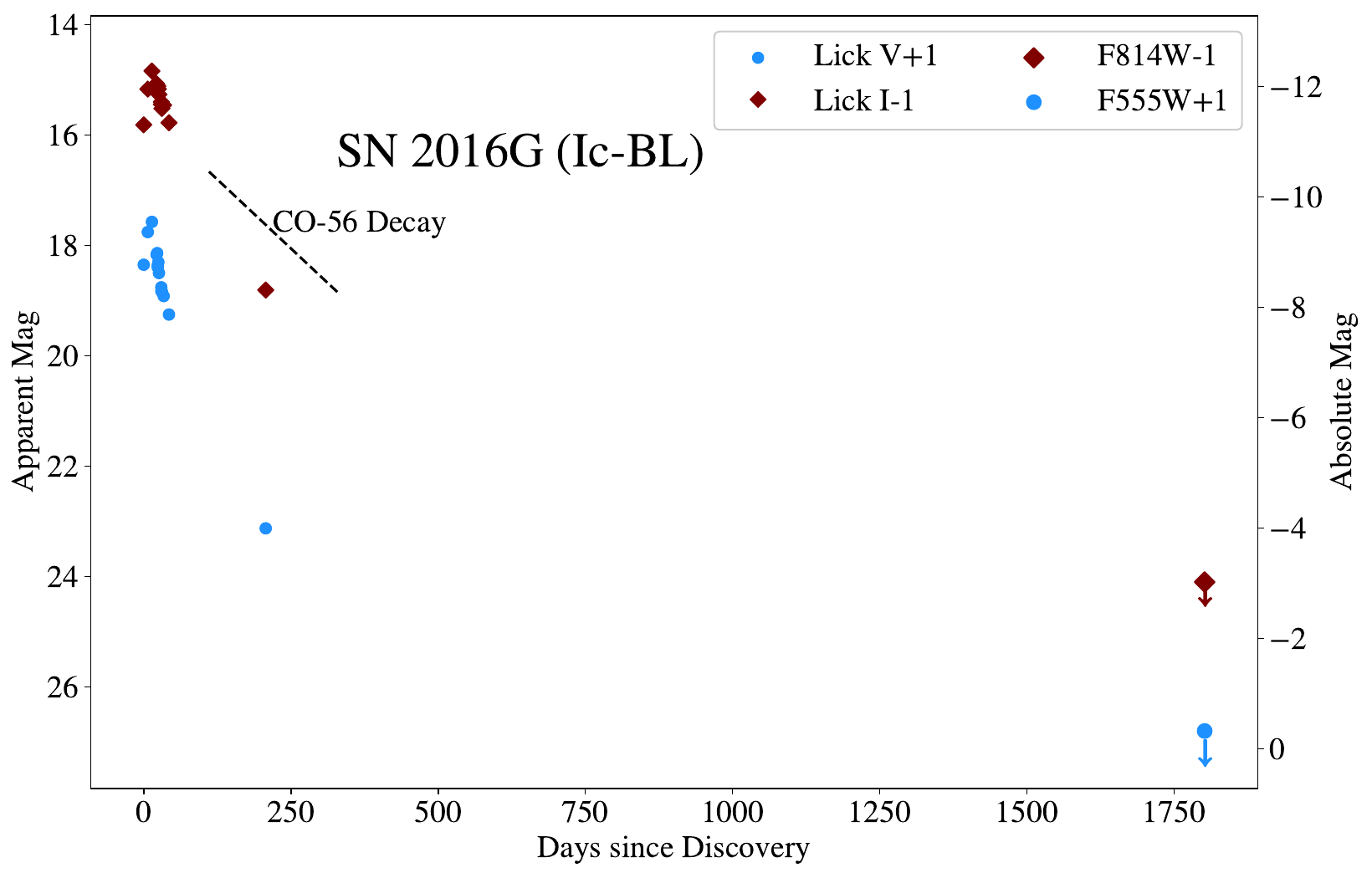}{0.4\textwidth}{(c) light curve}}
\caption{A portion of the WFC3 image mosaic containing SN 2016G, from observations on 2020 December 20, in (a) F555W and (b) F814W. 
As the SN was not detected in either band, its location is denoted by the dashed circle. Also shown are the Lick \citep{Zheng2022} $V$ and $I$ (c) light curves, together with the Snapshot upper limits.}
\label{fig:16G}
\end{figure*}

SN 2016G in NGC 1171 was classified as a broad-lined SN~Ic (Ic-BL) by \cite{Zhang2016}. \citet{Zheng2022} conducted early-time, multiband optical monitoring of the SN with KAIT. Our Snapshot observations were obtained on 2020 December 20, 1807~d (5.1~yr) after discovery, in F555W and F814W. We attempted to locate the SN in the {\sl HST\/} data astrometrically aligning to the ground-based KAIT images from 2016 February. The SN was not detected in either {\sl HST\/} band; see Figure \ref{fig:16G}. It appears from our data that the SN may have been located in a dust lane within the host galaxy, possibly complicating our recovery of the SN at late times; the $B-V$ color curve for SN 2016G in \citet{Zheng2022} implies that the SN may have experienced significant internal reddening (also consistent with the lack of detection of the SN in early times in the UV with {\sl Swift}; \citealt{Campana2016}).

\subsection{SN 2016adj}

SN 2016adj in NGC 5128 (Centaurus A) is certainly one of the nearest SNe (at 3.42~Mpc) of any type in recent years. The object was classified as a core-collapse SN, potentially with a stripped-envelope progenitor, with a C-rich SN~Ic classification inevitably proposed \citep{Stritzinger2023}. It became readily apparent at early stages that SN 2016adj was heavily reddened ($A_V \approx 2$--4~mag) by internal host dust \citep{Stritzinger2016}. In time it also became obvious that there was, at first, one \citep{Sugerman2016} and then several prominent light echoes apparent around the SN \citep{Stritzinger2022}.

\begin{figure*}[htb]
\gridline{\fig{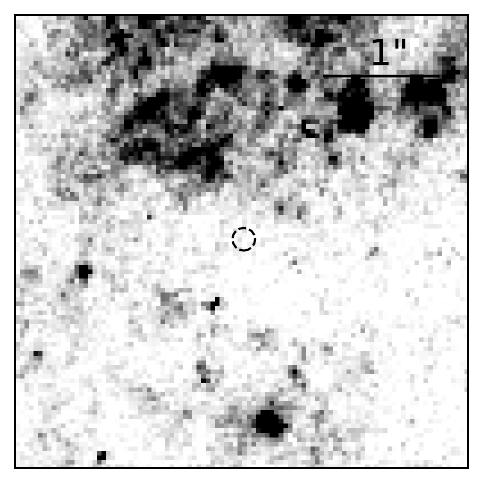}{0.25\textwidth}{(a) F438W}
          \fig{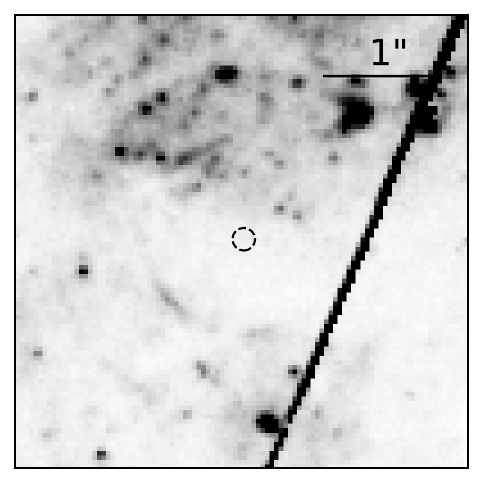}{0.25\textwidth}{(b) F555W}
          \fig{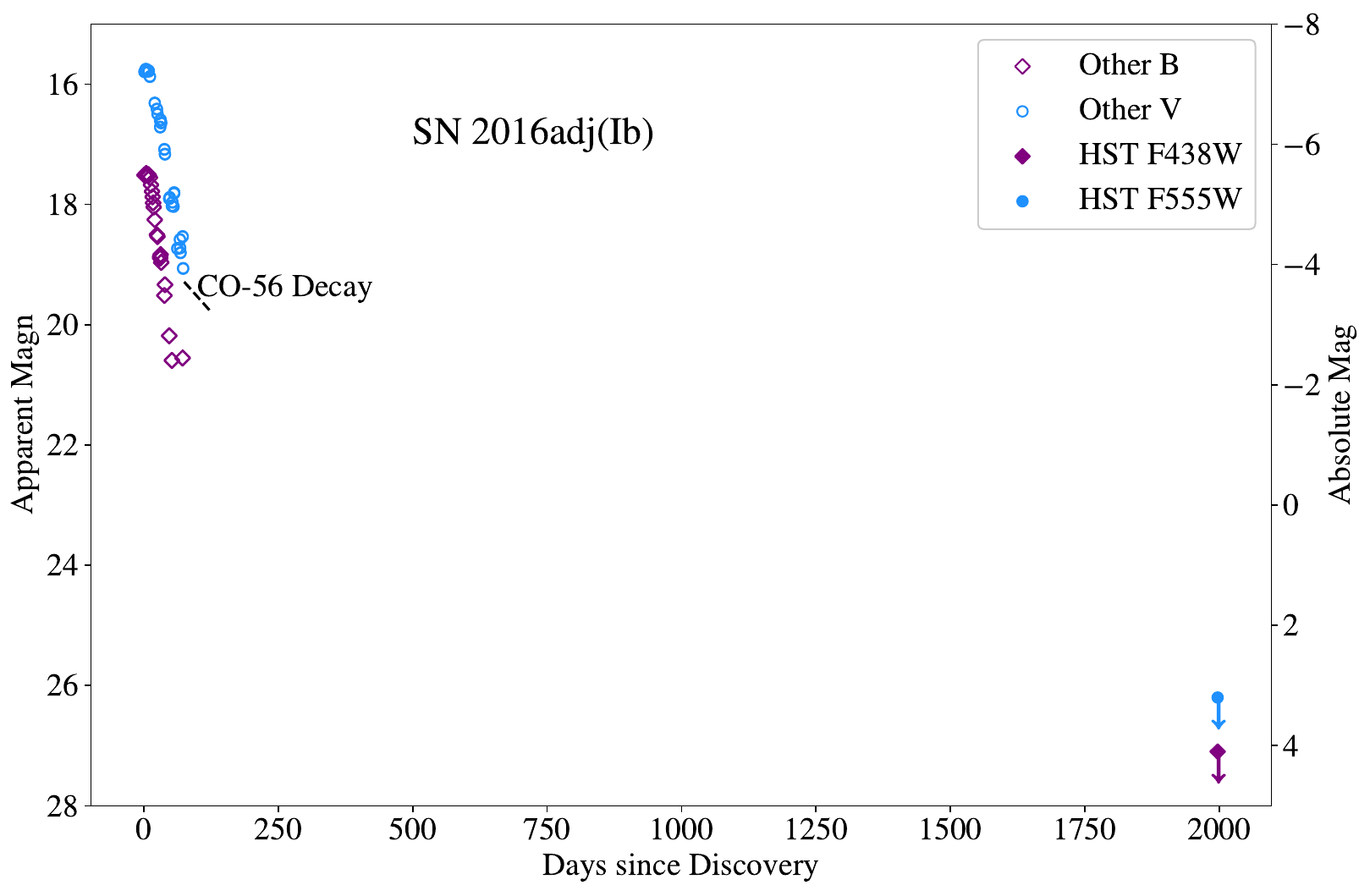}{0.4\textwidth}{(c) light curve}}
\caption{A portion of the WFC3 image mosaic containing SN 2016adj, from observations on 2021 July 28, in (a) F438W and (b) F555W. 
The SN was not detected in either band; the site is indicated by the dashed circle. What is most obvious in both bands are the light echoes around the SN site; see also \citet{Stritzinger2022}. Also shown are (``Other'') $B$ and $V$ (c) light-curve data  \citep{Stritzinger2023}, together  with our upper limits. We see a diffraction spike going straight through the F555W image, but it does not affect the SN site. }
\label{fig:16adj}
\end{figure*}

We observed the SN as part of our Snapshot program on 2021 July 28, 1998~d (5.5~yr) after discovery, in F438W and F555W. The former band was chosen intentionally to capture the blue light from the echoes, whereas the latter was used to establish the echo colors. We found the precise position of the SN in our {\sl HST\/} data comparing with images obtained on 2016 February 22 for GO-14115 (PI S.~Van Dyk) in F438W and F814W; see Figure \ref{fig:16adj}. The SN was not detected in the current Snapshots, at 27.4 and 26.5~mag in F438W and F555W, respectively. The echoes, however, are quite prominent (see also \citealt{Stritzinger2022}, who used our Snapshot data as well in their work). A detailed analysis of the light echoes is beyond the scope of this paper.

\subsection{SN 2016bkv}\label{sec:2016bkv}

SN 2016bkv is an exceptional example of a low-luminosity SN~II-P, with an extraordinarily long plateau phase ($\gtrsim 140$~d) and very low expansion velocities, in addition to a strong initial bump in the light curve, as well as ``flash-ionization'' features, all signs of short-lived, early-time circumstellar interaction \citep{Nakaoka2018,Hosseinzadeh2018}. \citet{Nakaoka2018} concluded that the progenitor mass-loss rate within a few years of explosion was quite high, $\sim 1.7 \times 10^{-2}\ M_{\odot}$~yr$^{-1}$ (although see \citealt{Deckers2021}), possibly indicating that the star had experienced a violent outburst. \citet{Hosseinzadeh2018} further suggested that SN 2016bkv is an example of an electron-capture (EC) SN. Through radiative-transfer modeling of the spectra, \citet{Deckers2021} inferred an odd surface composition for the progenitor, implying that it was more likely a binary rather than a single star, with the primary either accreting unprocessed material from its companion or undergoing a merger before explosion.

\begin{figure*}[htb]
\gridline{\fig{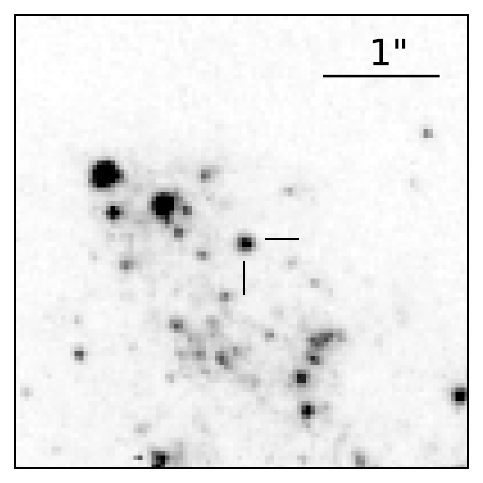}{0.25\textwidth}{(a) F555W}
          \fig{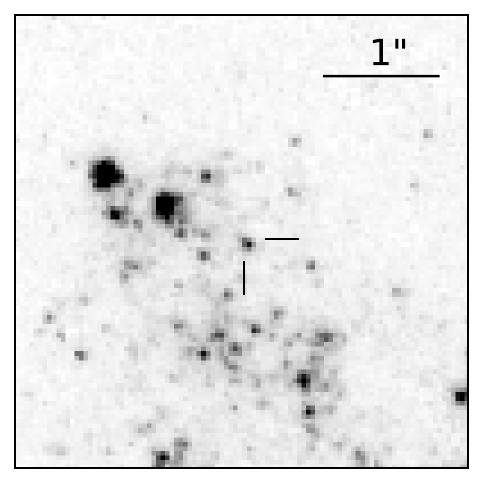}{0.25\textwidth}{(b) F814W}
          \fig{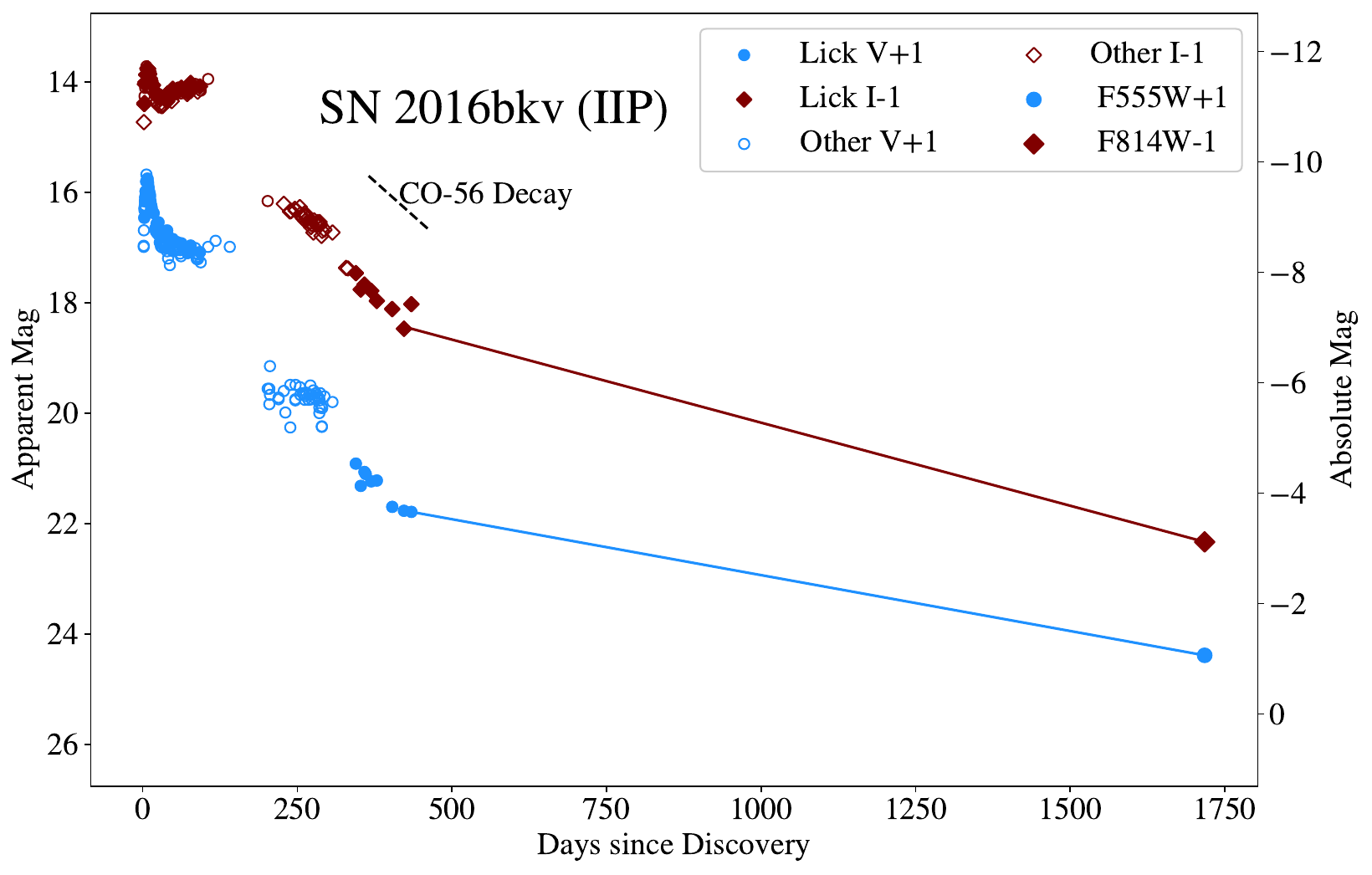}{0.4\textwidth}{(c) light curve}}
\caption{A portion of the WFC3 image mosaic containing SN 2016bkv, from observations on 2020 December 13, in (a) F555W and (b) F814W. 
Also shown are previously unpublished Lick $V$ and $I$ (c) light curves, along with (``Other'') data including upper limits from \citet{Nakaoka2018} and \citet{Hosseinzadeh2018}, together with the Snapshot detections.}
\label{fig:16bkv}
\end{figure*}

Our Snapshot data were obtained on 2020 December 13, 1721~d (4.7~yr) after discovery, in F555W and F814W. The location of the SN was pinpointed by referring to early-time {\sl HST\/} F555W data from 2016 April 14 (GO-14115, PI S.~Van Dyk), when the SN was at $m_{\rm F555W} = 16.06 \pm 0.01$~mag. The SN was still clearly detected in our Snapshots in both bands; see Figure \ref{fig:16bkv}. 

\subsection{AT 2016blu}

AT 2016blu, also known as NGC 4559-OT1, PSN J12355230+2755559, and Gaia16ada, was actually discovered by the Lick Observatory Supernova Search earlier, in 2012 \citep{Kandrashoff2012}, and classified as a luminous blue variable (LBV) or SN impostor (see also \citealt{Sheehan2014}). The object is highly variable and has been ``rediscovered'' a number of times over the years thereafter (e.g., \citealt{Vinokurov2021}) --- hence, the multiple identifiers for the same object. Not long after discovery, \citet{VanDyk2012a} identified a possible precursor in {\sl HST\/} images from 2005 and, based on preliminary photometry, estimated that the star had $M_V=-9.4$~mag with intrinsic colors $B-V=0.10$ and $V-I=0.36$~mag, consistent with an early-F spectral type. \citet{Aghankaloo2022b} recently conducted an analysis of the recurring outbursts from the transient.  
They found a periodicity to the outbursts, and proposed that AT~2016blu is probably an LBV in an eccentric interacting binary very much like SN~2000ch.

Our {\sl HST\/} program obtained observations on 2021 February 17 (3320~d since discovery) in F606W and F814W. The F606W band was used rather than F555W, expressly in order to probe the TRGB for estimation of the distance to the host galaxy (NGC 4559; however, \citealt{McQuinn2017} had already performed a TRGB analysis, with different {\sl HST\/} data, and found a distance of 8.91~Mpc).

\begin{figure*}[htb]
\gridline{\fig{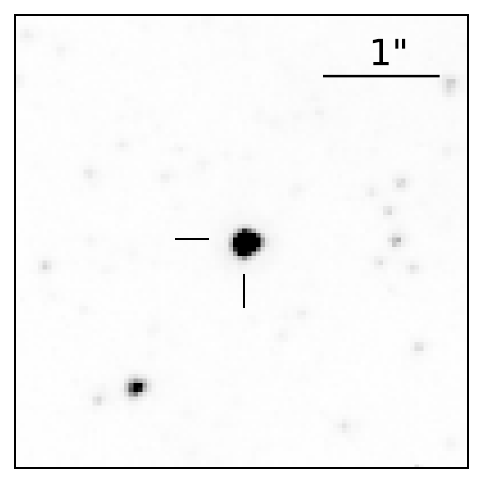}{0.25\textwidth}{(a) F606W}
          \fig{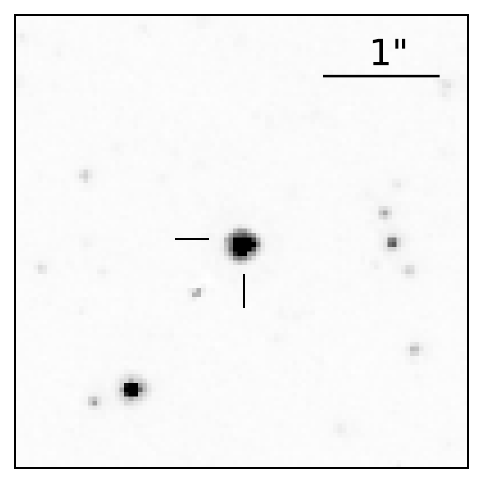}{0.25\textwidth}{(b) F814W}
          \fig{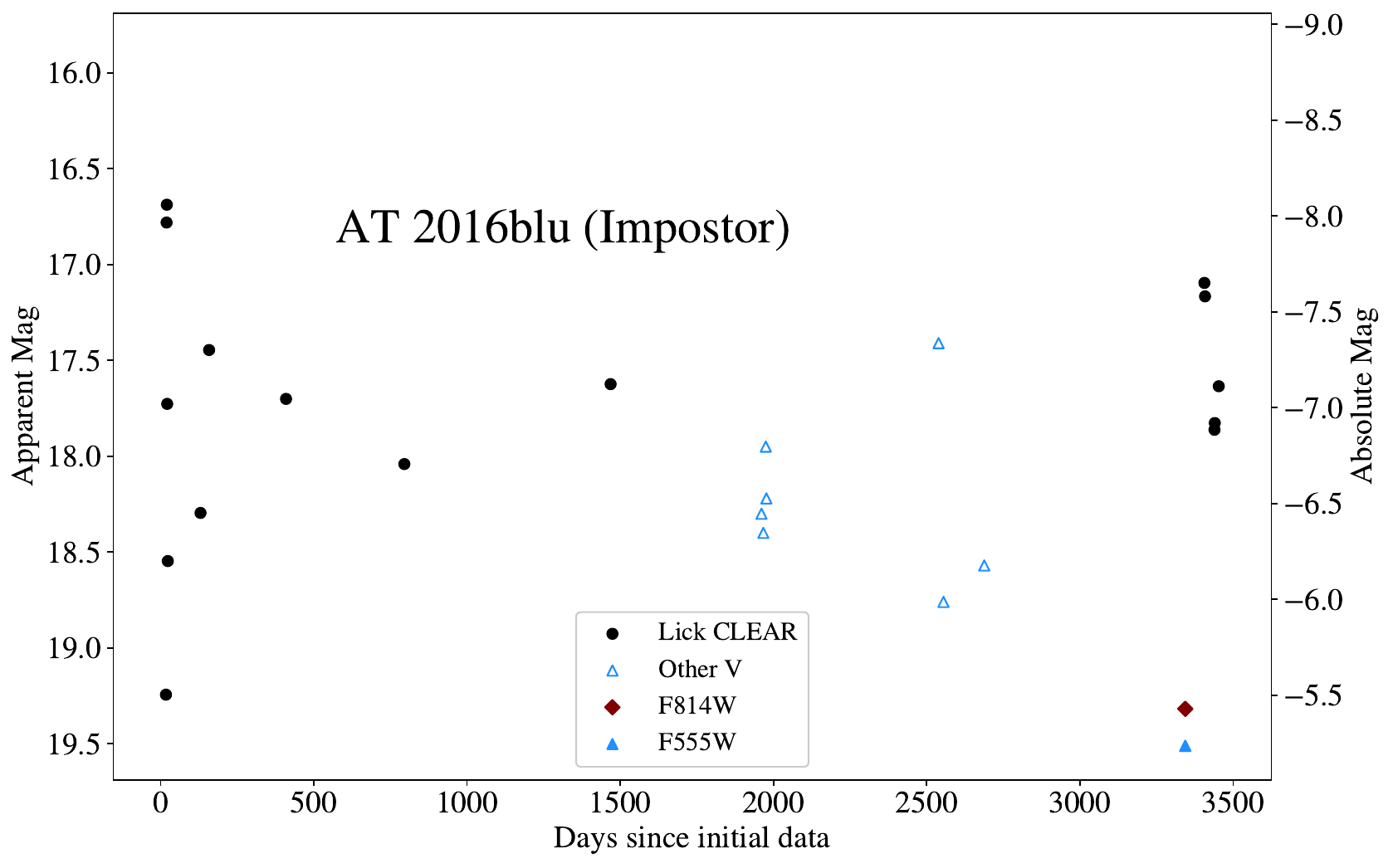}{0.4\textwidth}{(c) light curve}}
\caption{A portion of the WFC3 image mosaic containing AT 2016blu, from observations on 2021 February 17, in (a) F606W and (b) F814W. 
Also shown is the Lick ``clear'' (unfiltered) and $V$ \citep{Aghankaloo2022b} (c) light curves, together with the Snapshot detections in the two {\sl HST\/} bands.}
\label{fig:ngc4559}
\end{figure*}

We pinpointed the AT's location by astrometrically aligning with KAIT ground-based data from 2016 April, as well as precursor {\sl HST\/} images from 2005 March (GO-10214, PI R.~Soria). AT 2016blu is still strongly detected in our Snapshot images in both bands (see Figure \ref{fig:ngc4559}).

\subsection{SN 2016coi}

SN 2016coi (ASASSN-16fp) in UGC 11868 is an SN~Ic-BL, or possibly a transitional event between a normal and broad-lined SN~Ic, with evidence for some residual He --- extensive multiwavelength follow-up observations were performed by \citet{Yamanaka2017}, \citet{Kumar2018}, \citet{Prentice2018},  \citet{Terreran2019}, and \citet{Tsvetkov2020}. Early-time photometry was obtained with KAIT as well \citep{Zheng2022}. 

\begin{figure*}[htb]
\gridline{\fig{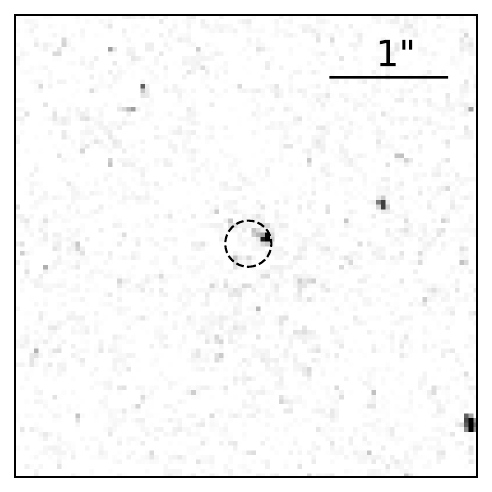}{0.25\textwidth}{(a) F336W}
          \fig{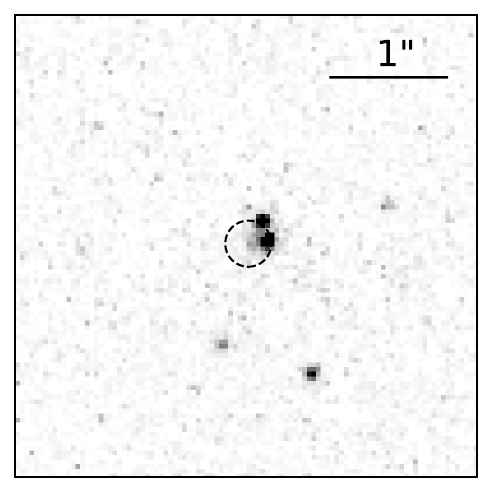}{0.25\textwidth}{(b) F814W}
          \fig{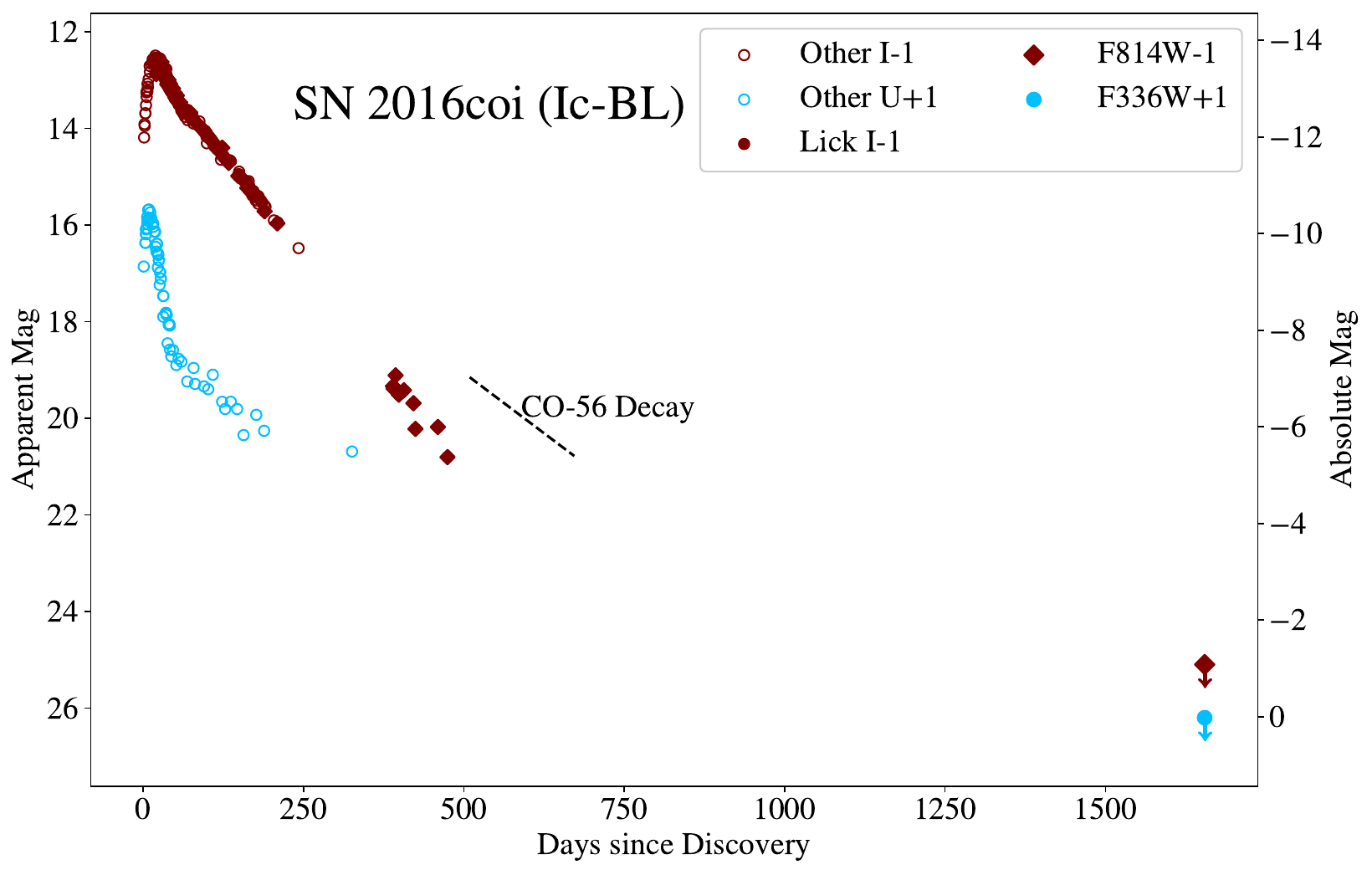}{0.4\textwidth}{(c) light curve}}
\caption{A portion of the WFC3 image mosaic containing SN 2016coi, from observations on 2020 December 6, in (a) F336W and (b) F814W. 
The SN was not detected in either band; the site is indicated by the dashed circle. Also shown is the Lick \citep{Zheng2022} $I$ (c) light curve, along with (``Other'') data in $I$, as well as $U$, from \citet{Kumar2018}, \citet{Terreran2019},   \citet{Prentice2018}, and \citet{Tsvetkov2020}, together with the Snapshot upper limits.}
\label{fig:16coi}
\end{figure*}

Our Snapshot observations of the SN were executed on 2020 December 6, 1655~d (4.5~yr) after discovery, in F336W and F814W. In order to pinpoint the SN location in the Snapshots, we compared with {\sl HST\/} data from 2016 October 4 for our previous Snapshot program GO-14668 (PI A.~Filippenko), when the younger SN was at $m_{\rm F555W}= 16.79 \pm 0.01$ and $m_{\rm F814W}= 16.18 \pm 0.01$~mag. The SN was not detected in our late-time Snapshots, to limits of 26.1 and 26.0 mag in F336W and F814W, respectively; see Figure \ref{fig:16coi}.

\subsection{SN 2016coj}

SN 2016coj in NGC 4125 was discovered on 2016 May 28 with KAIT and was subsequently classified as a normal SN~Ia \citep{Zheng2017}. \citet{Richmond2017} and \citet{Stahl2019} have presented further optical multiband follow-up photometry.

The Snapshot observations occurred on 2020 December 9, 1656~d (4.5~yr) after discovery, in F555W and F814W. The location of the SN was pinpointed using {\sl HST\/} data from 2017 December 25 as part of our previous Snapshot program GO-15166 (PI A.~Filippenko), when SN 2016coj was at $m_{\rm F555W} = 24.20 \pm 0.04$ and $m_{\rm F814W} = 23.46 \pm 0.07$~mag. The SN was not detected in the current Snapshots to limits of 26.8 and 25.5~mag in F555W and F814W, respectively.

\begin{figure*}[htb]
\gridline{\fig{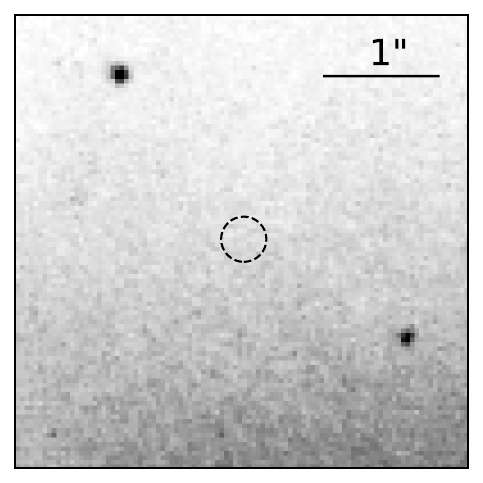}{0.25\textwidth}{(a) F555W}
          \fig{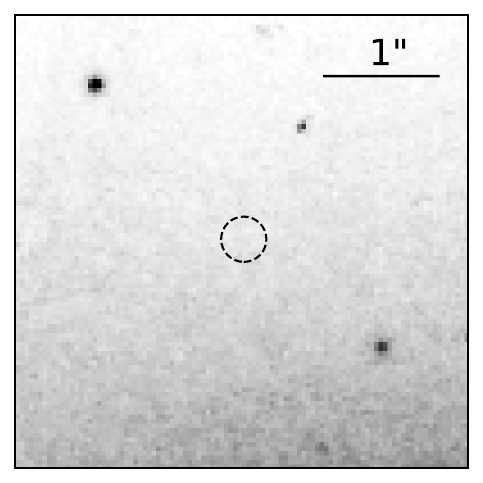}{0.25\textwidth}{(b) F814W}
          \fig{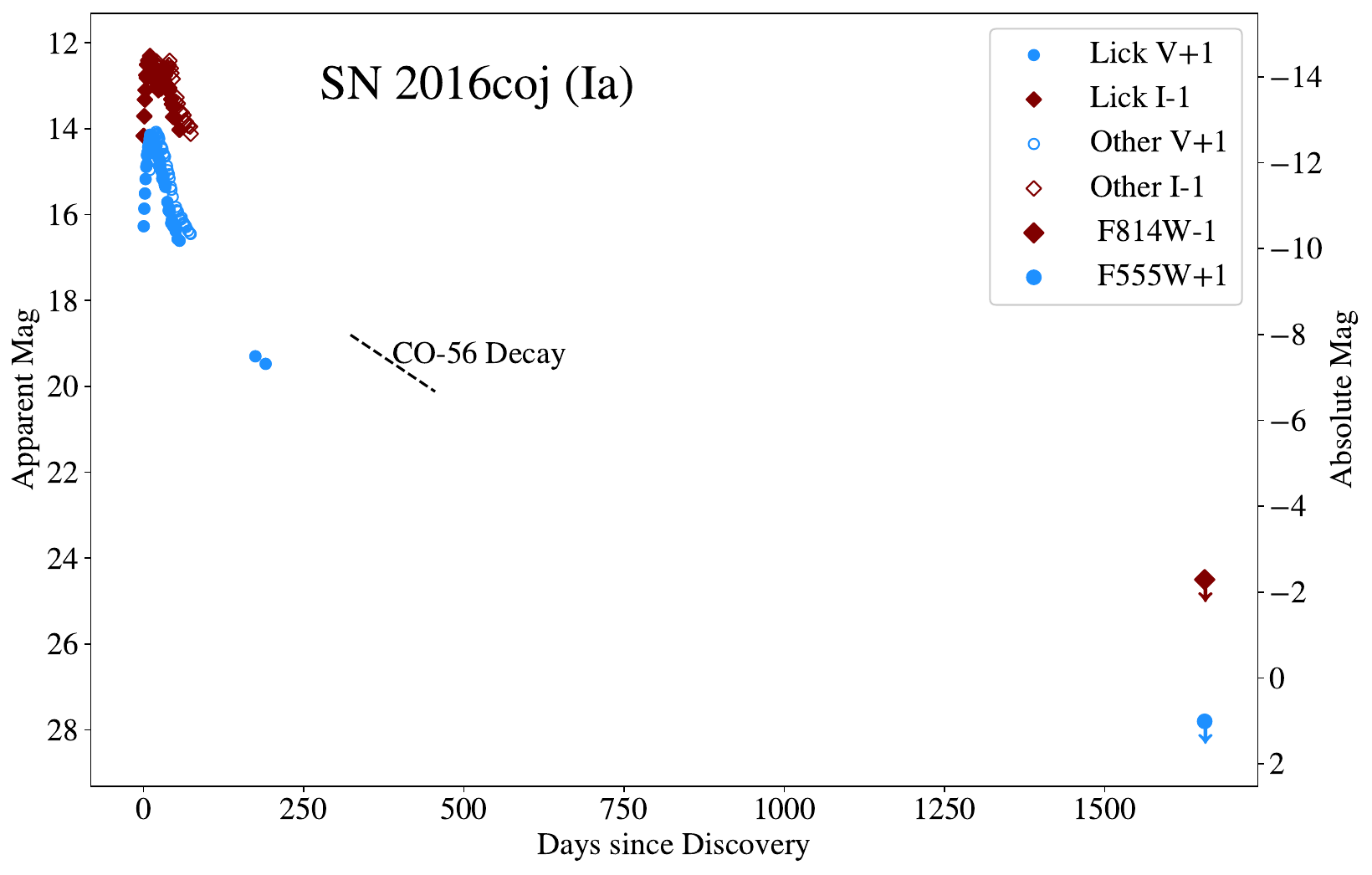}{0.4\textwidth}{(c) light curve}}
\caption{A portion of the WFC3 image mosaic containing SN 2016coj, from observations on 2020 December 9, in (a) F555W and (b) F814W. 
The SN was not detected in either band; the site is indicated by the dashed circle. Also shown are the Lick \citep{Stahl2019} $V$ and $I$ (c) light curves, along with (``Other'') data from \citet{Richmond2017}, together with the Snapshot upper limits.}
\label{fig:16coj}
\end{figure*}

\subsection{SN 2016gkg}

SN 2016gkg in NGC 613 is an SN~IIb with properties that are intermediate between those of SN 1993J (Section~\ref{sec:sn1993J} and SN 2011dh (Section~\ref{sec:sn2011dh}; \citealt{Tartaglia2017a}). Both \citet{Arcavi2017} and \citet{Piro2017} modeled the first cooling peak of the SN to infer properties of the progenitor. \citet{Tartaglia2017a} and \citet{Kilpatrick2017} independently identified a progenitor candidate in pre-explosion {\sl HST\/} images. \citet{Bersten2018} also identified and characterized the progenitor, as well as the SN itself (which included KAIT photometry, enhanced with further data by \citealt{Zheng2022}).

\begin{figure*}[htb]
\gridline{\fig{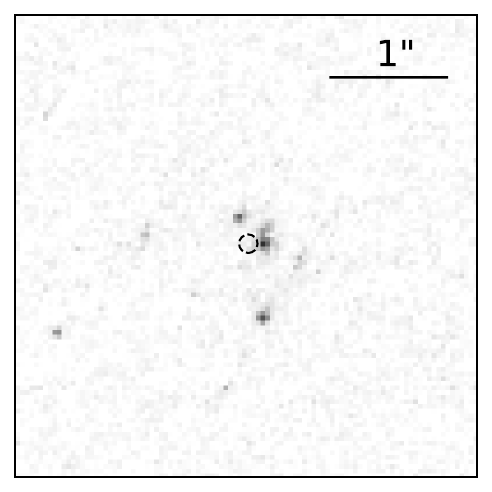}{0.25\textwidth}{(a) F438W}
          \fig{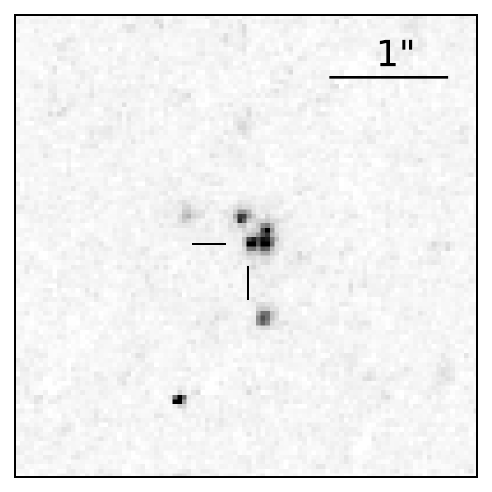}{0.25\textwidth}{(b) F606W}
          \fig{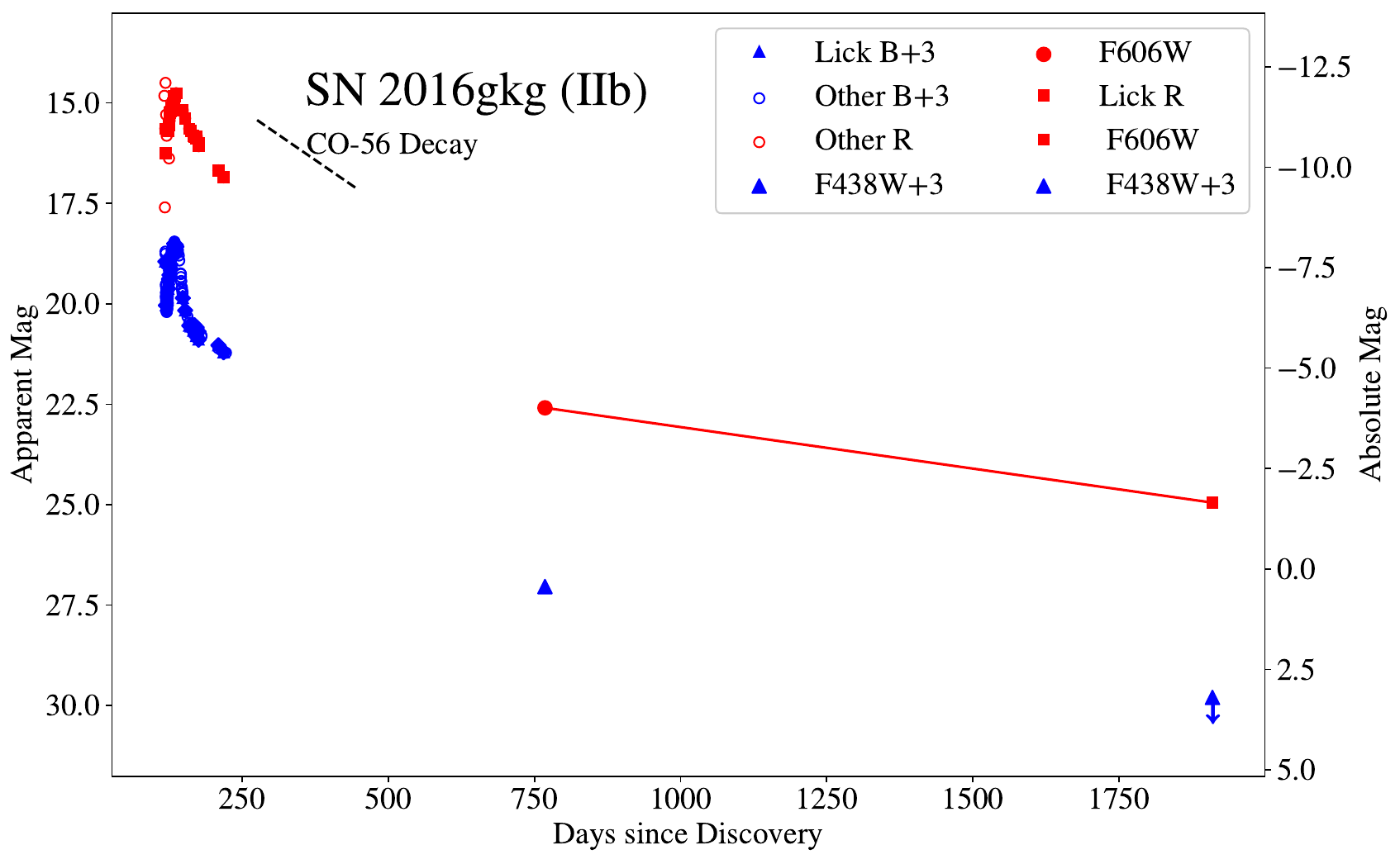}{0.4\textwidth}{(c) light curve}}
\caption{A portion of the WFC3 image mosaic containing SN 2016gkg, from observations on 2021 August 19, in (a) F438W and (b) F606W. 
The SN is detected in F606W, as indicated by tick marks, but not detected in F438W; the site in that band is indicated by the dashed circle. Also shown are the Lick \citep{Bersten2018,Zheng2022} $V$ and $I$ (c) light curves, along with (``Other'') data from \citet{Tartaglia2017a} and additional previous {\sl HST\/} and $B$ data from \citet{Kilpatrick2022}, together with the Snapshot detections.}
\label{fig:16gkg}
\end{figure*}

The {\sl HST\/} Snapshots were obtained on 2021 August 19, 1795~d (4.9~yr) after discovery, in F438W and F606W. The SN location was found using {\sl HST\/} data taken 2016 October 10 for GO-14116 (PI S.~Van Dyk), when the SN was young and bright, at $m_{\rm F555W}=15.11 \pm 0.01$~mag. \citet{Kilpatrick2022} revisited the SN and, using our Snapshots found that $m_{\rm F606W}=25.10 \pm 0.07$ and $m_{\rm F438W}=26.61 \pm 0.27$~mag. Our results differ from these, with $m_{\rm F606W}=24.95 \pm 0.04$ and $m_{\rm F438W}>26.1$~mag. We can potentially ascribe the discrepancy in F606W as due to differences in assumed {\tt Dolphot} input parameters; however, as can be seen in Figure \ref{fig:16gkg}, the SN is clearly not visually detected at F438W. Based on our results, \citet{VanDyk2023} concluded that the progenitor candidate had vanished. The behavior of the late-time light curve in F606W implies that CSM interaction may be a source of additional power.

\subsection{AT 2016jbu}

AT 2016jbu (Gaia16cfr) in NGC 2442 has been considered since early after discovery to be an SN impostor, although it has been argued that it should actually be considered a pre-explosion LBV \citep{Kilpatrick2018b} or simply an interacting, SN 2009ip-like transient \citep{Brennan2022a}. Both \citet{Kilpatrick2018b} and \citet{Brennan2022b} independently identified in pre-outburst {\sl HST\/} images and subsequently characterized the precursor: a massive ($\sim 22$--30~$M_{\odot}$) yellow supergiant enshrouded by a dusty circumstellar shell.

\begin{figure*}[htb]
\gridline{\fig{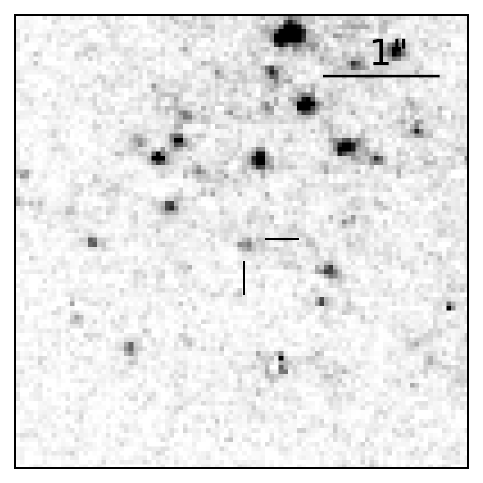}{0.25\textwidth}{(a) F555W}
          \fig{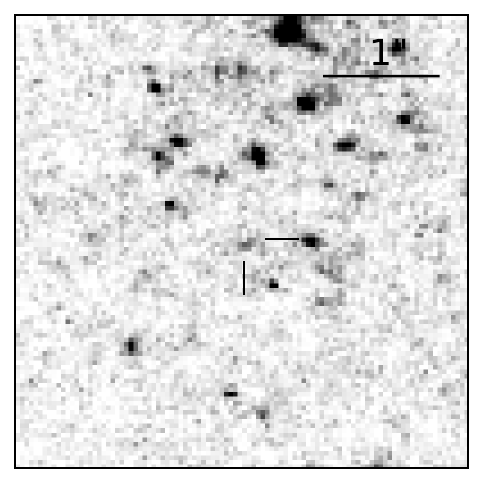}{0.25\textwidth}{(b) F814W}
          \fig{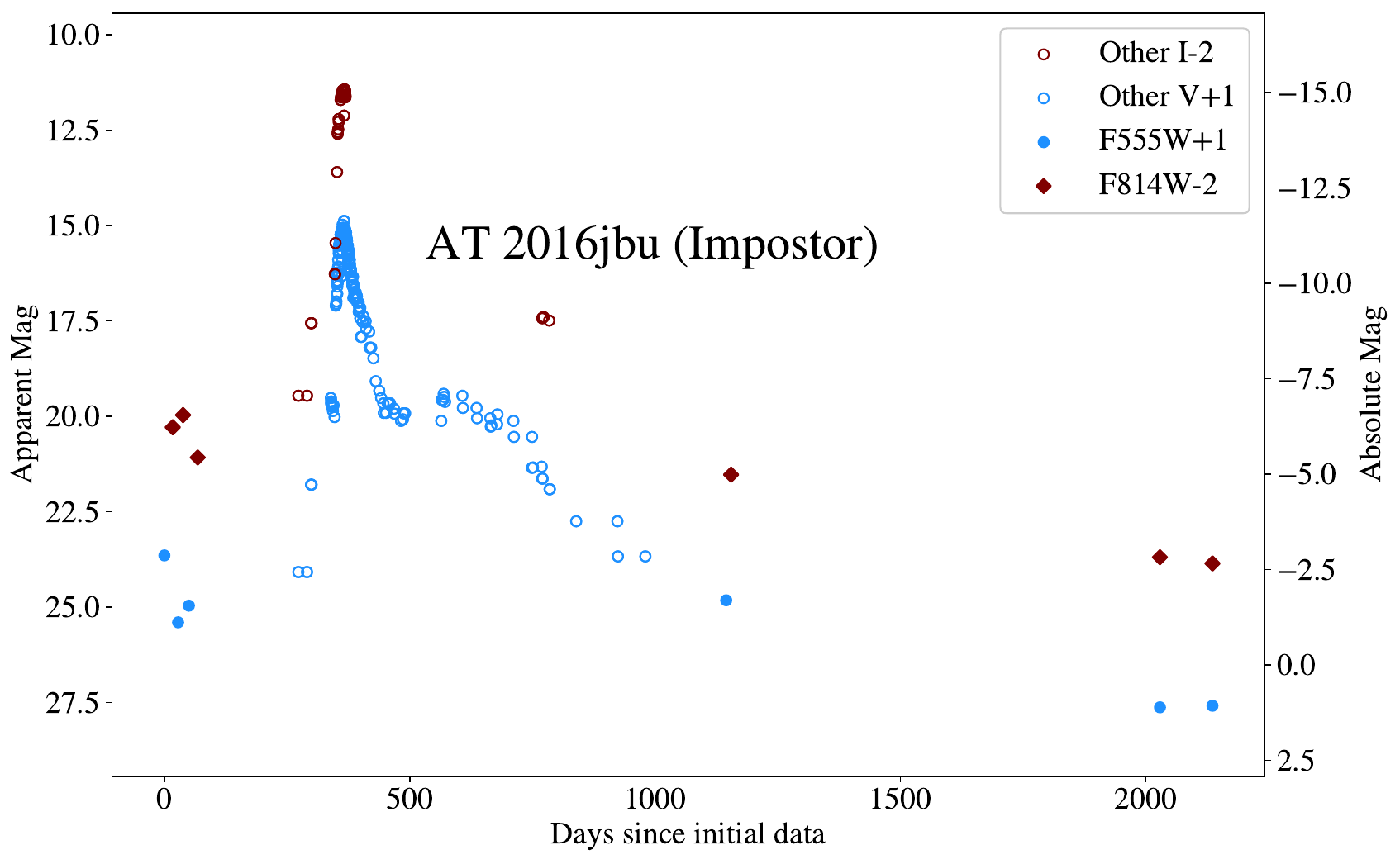}{0.4\textwidth}{(c) light curve}}
\caption{A portion of the WFC3 image mosaic containing AT 2016jbu, from observations on 2021 August 21, in (a) F555W and (b) F814W. 
Also shown are the combined $V$ (c) and $I$(c) light curves from \citet{Kilpatrick2018b} and \citet{Brennan2022c}, together with {\sl HST\/} data from \citet{Brennan2022b}, \citet{Brennan2022a}, and this paper (the second-to-latest data points).}
\label{fig:16jbu}
\end{figure*}

The object was detected in our Snapshots in F555W and F814W on 2021 August 21, 1725~d (4.7~yr) after discovery. We isolated the location of the object using {\sl HST\/} data obtained on 2019 March 21 for our previous Snapshot program GO-15166 (PI A.~Filippenko), together with finder charts from the ensemble of {\sl HST\/} observations presented by \citet{Brennan2022b}; see Figure \ref{fig:16jbu}. \citet{Brennan2022c} also obtained multiband {\sl HST\/} observations of AT 2016jbu on 2021 December 6, only 107~d later, and discovered that the observed brightness of the transient was less than the precursor levels. Those authors further found that it was difficult to explain this dimming in terms of increasing dust obscuration and therefore concluded that the precursor had likely vanished --- thus, AT 2016jbu may have actually been a terminal explosion. Our Snapshot data also confirm the dimming and precursor disappearance.

\subsection{SN 2017cfd}

SN 2017cfd in IC 511 was discovered on 2017 March 16 with KAIT and classified as a normal SN~Ia \citep{Han2020}. \citet{Stahl2019} presented further early-time multiband optical monitoring with KAIT.

\begin{figure*}[htb]
\gridline{\fig{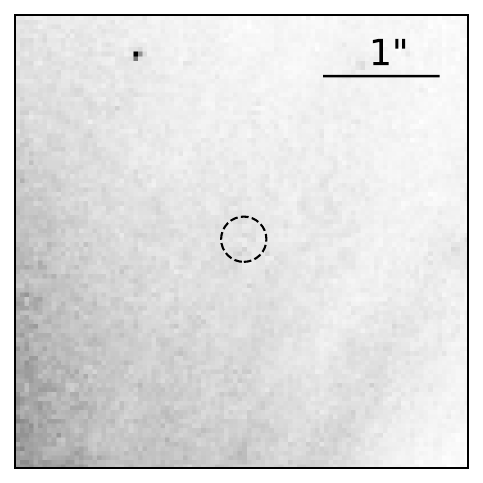}{0.25\textwidth}{(a) F555W}
          \fig{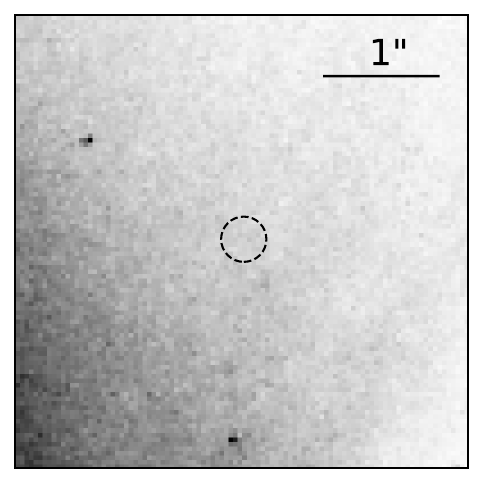}{0.25\textwidth}{(b) F814W}
          \fig{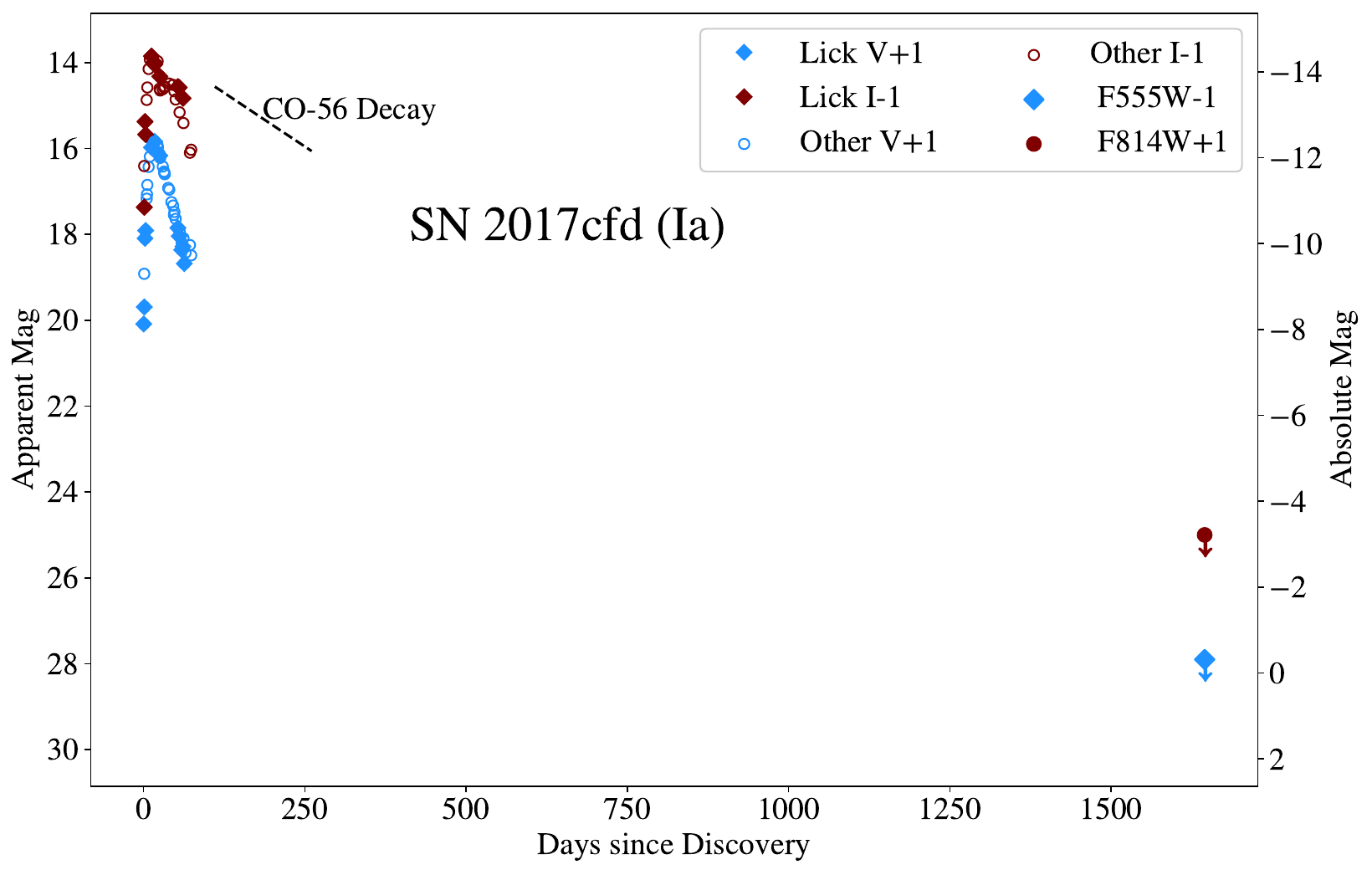}{0.4\textwidth}{(c) light curve}}
\caption{A portion of the WFC3 image mosaic containing SN 2017cfd, from observations on 2021 February 17, in (a) F555W and (b) F814W.  
The SN was not detected in either band; the site is indicated by the dashed circle. Also shown are the Lick \citep{Stahl2019} $V$ and $I$ (c) light curves, along with additional (``Other''; in this case, non-Lick) data from \citet{Han2020}, together with the Snapshot upper limits.}
\label{fig:2017cfd}
\end{figure*}

Our Snapshots were obtained on 2021 February 17, 1654~d (4.5~yr) after discovery, in F555W and F814W. We astrometrically aligned ground-based KAIT images from 2017 April with the Snapshot image mosaics, in order to locate the SN position; see Figure \ref{fig:2017cfd}. The SN was not detected in either of the {\sl HST\/} bands, to limits of 26.9 and 26.0~mag in F555W and F814W, respectively. 


\subsection{SN 2017eaw}

SN 2017eaw is the tenth SN discovered in the nearby galaxy NGC 6946 and was likely a normal SN~II-P of intermediate luminosity \citep{VanDyk2019}. \citet{Buta2019}, \citet{VanDyk2019}, \citet{Szalai2019b}, and \citet{Buta2019} all undertook extensive optical and near-IR follow-up campaigns. \citet{VanDyk2019} pointed out the early ``bump'' in the light curves, which \citet{Morozova2020} interpreted as CSM set up by a pre-SN outburst $\sim 50$--350~d induced by a nuclear-burning episode $\sim 150$--450~d prior to the SN. Furthermore, \citet{Weil2020} found evidence spectroscopically of continued circumstellar interaction at late times. \citet{Kilpatrick2018c}, \citet{VanDyk2019}, and \citet{Rui2019} all independently identified in pre-explosion {\sl HST\/} images and characterized the progenitor candidate for the SN; \citet{VanDyk2019} modeled the star as a dusty, luminous $M_{\rm ZAMS} \approx 15\ M_{\odot}$ RSG.  \citet{Bostroem2023} suggested that SN~2017eaw may arise from a binary progenitor, based on analysis of the surrounding stellar population.

\begin{figure*}[htb]
\gridline{\fig{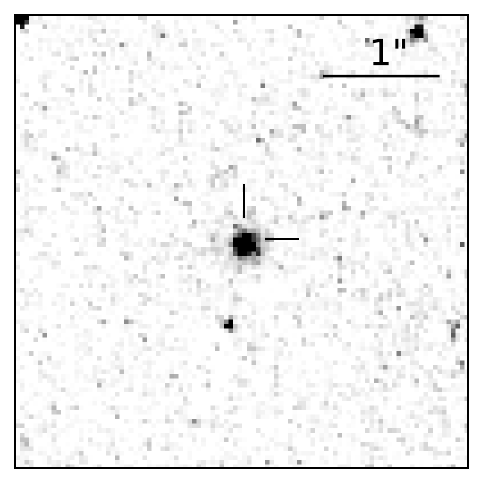}{0.25\textwidth}{(a) F555W}
          \fig{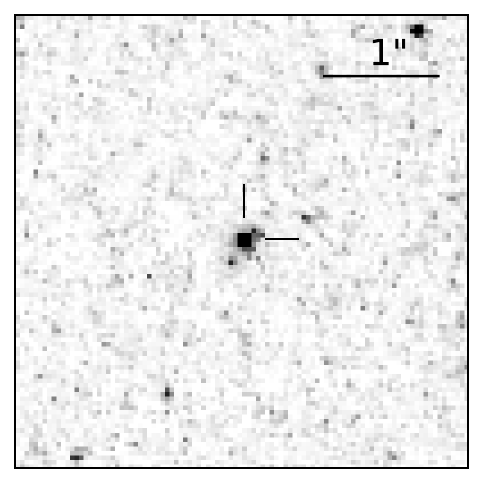}{0.25\textwidth}{(b) F814W}
          \fig{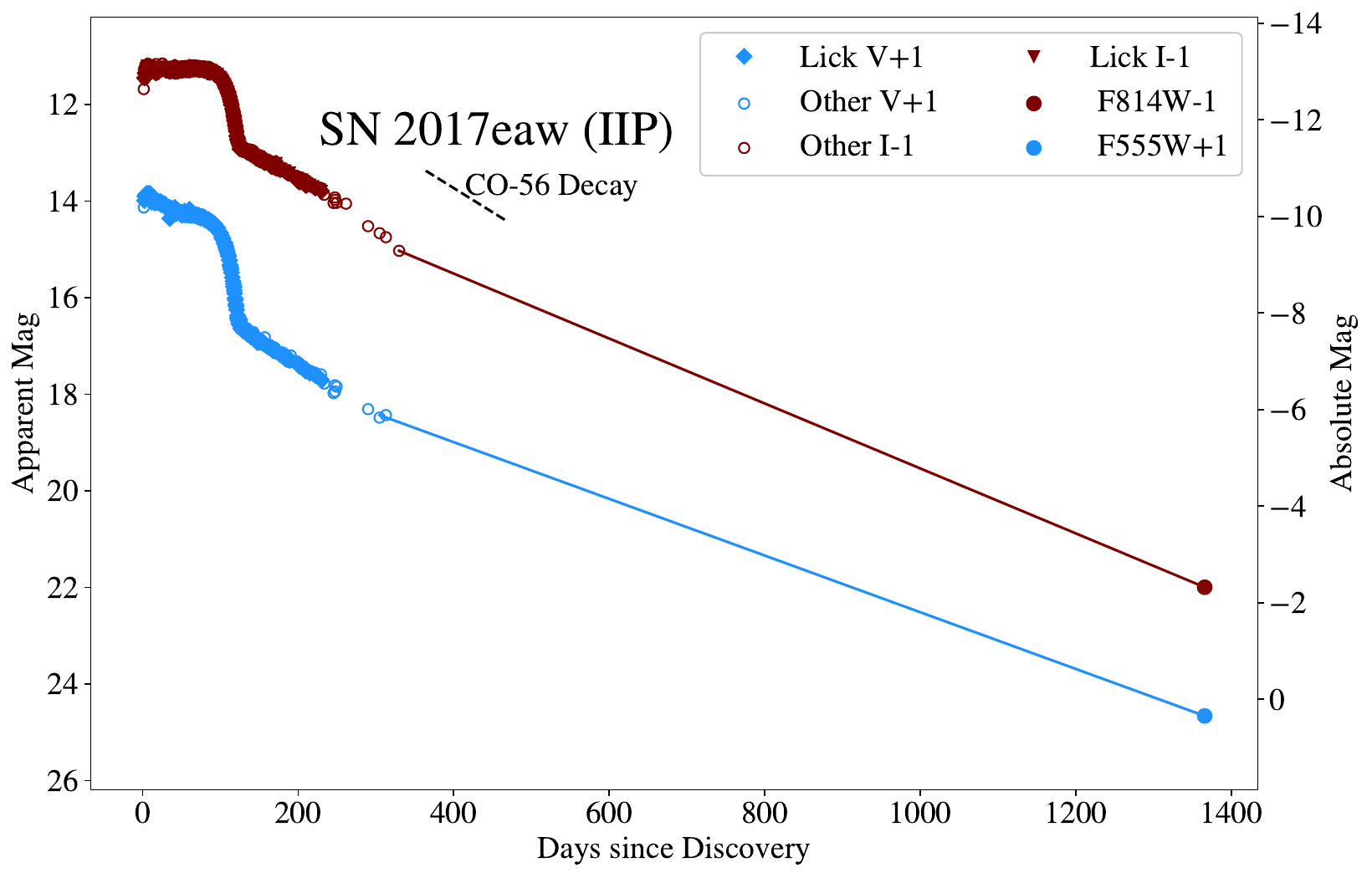}{0.4\textwidth}{(c) light curve}}
\caption{A portion of the WFC3 image mosaic containing SN 2017eaw, from observations on 2020 November 11, in (a) F555W and (b) F814W. 
Also shown are the Lick \citep{VanDyk2019} $V$ and $I$ (c) light curves, along with (``Other'') data from \citet{Tsvetkov2018}, \citet{Szalai2019b}, and \citet{Buta2019}, together with the Snapshot detections.}
\label{fig:17eaw}
\end{figure*}

Snapshots for our program were obtained on 2020 November 11, 1273~d (3.5~yr) after discovery, in F555W and F814W. The location of the SN was confirmed using {\sl HST\/} images taken 2018 January 5 for GO-15166 (PI A.~Fillipenko), when the SN was at $m_{\rm F814W}=18.62 \pm 0.01$~mag and $m_{\rm F555W}=19.83 \pm 0.01$~mag. The SN was still quite bright in the current images; see Figure \ref{fig:17eaw}. Based on the change in brightness of the SN in both bands in our Snapshots, relative to pre-SN observations, \citet{VanDyk2023} concluded that the RSG candidate was indeed the progenitor, and also confirmed the late-time CSM interaction, manifested as a UV excess.

\subsection{SN 2017gax}\label{sec:2017gax}

\begin{figure*}[htb]
\gridline{\fig{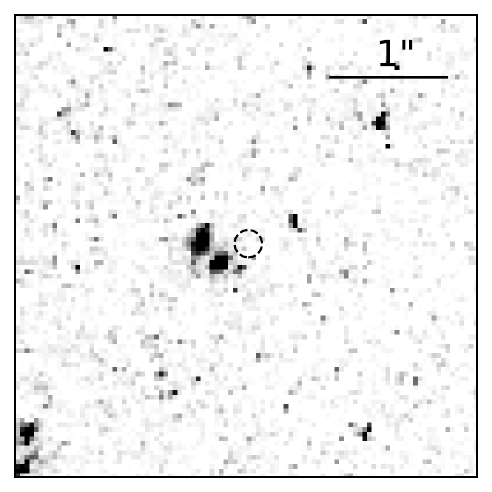}{0.25\textwidth}{(a) F336W}
          \fig{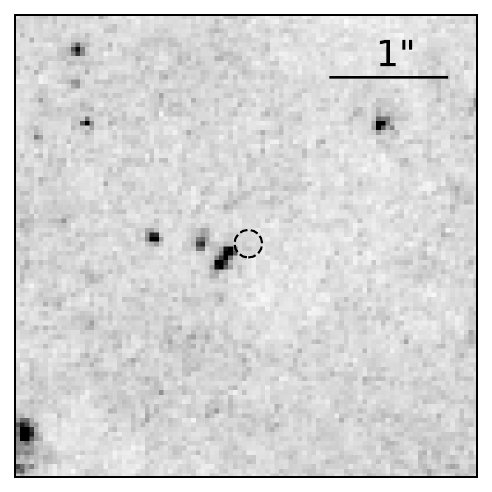}{0.25\textwidth}{(b) F814W}}
\caption{A portion of the WFC3 image mosaic containing SN 2017gax, from observations on 2020 November 27, in (a) F336W and (b) F814W. 
Nothing is detected in either band at the SN location, which is denoted by the dashed circle. No photometry has been published for this SN, aside from a single $V$ measurement.} 
\label{fig:17gax}
\end{figure*}

SN 2017gax (DLT17ch) was discovered in NGC 1672 by \citet{Tartaglia2017b} on 2017 August 14. The SN was spectroscopically classified, within a day of discovery, as an SN~Ic by \citet{Jha2017}. The SN location was established in our F336W and F814W Snapshots from 2020 November 27, 1202~d (3.5~yr) after discovery, using {\sl HST\/} data from 2017 October 19 (GO-14645, PI S.~Van Dyk), when the SN was at $m_{\rm F555W} = 15.94 \pm 0.01$~mag. The SN was not detected in the Snapshot data in either band; see Figure \ref{fig:17gax}. Unfortunately, no published photometry exists, beyond the report by \citet{Maguire2017} of the SN at $V = 16.1 \pm 0.1$~mag on 2017 November 9; thus, we are unable to show a light curve for this SN.

\subsection{SN2017gkk}\label{sec:2017gkk}

\begin{figure*}[htb]
\gridline{\fig{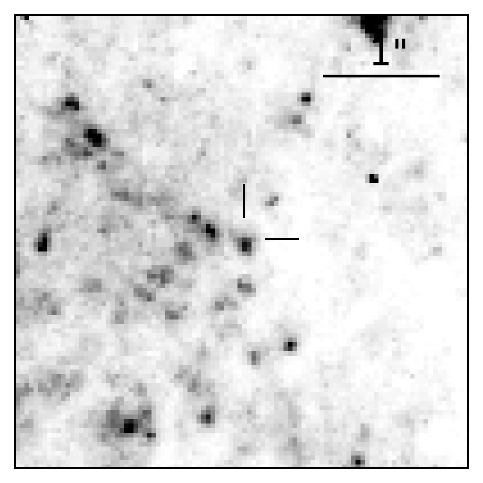}{0.25\textwidth}{(a) F555W}
          \fig{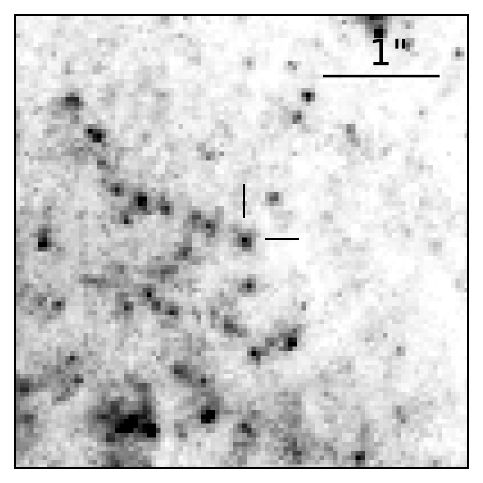}{0.25\textwidth}{(b) F814W}
          \fig{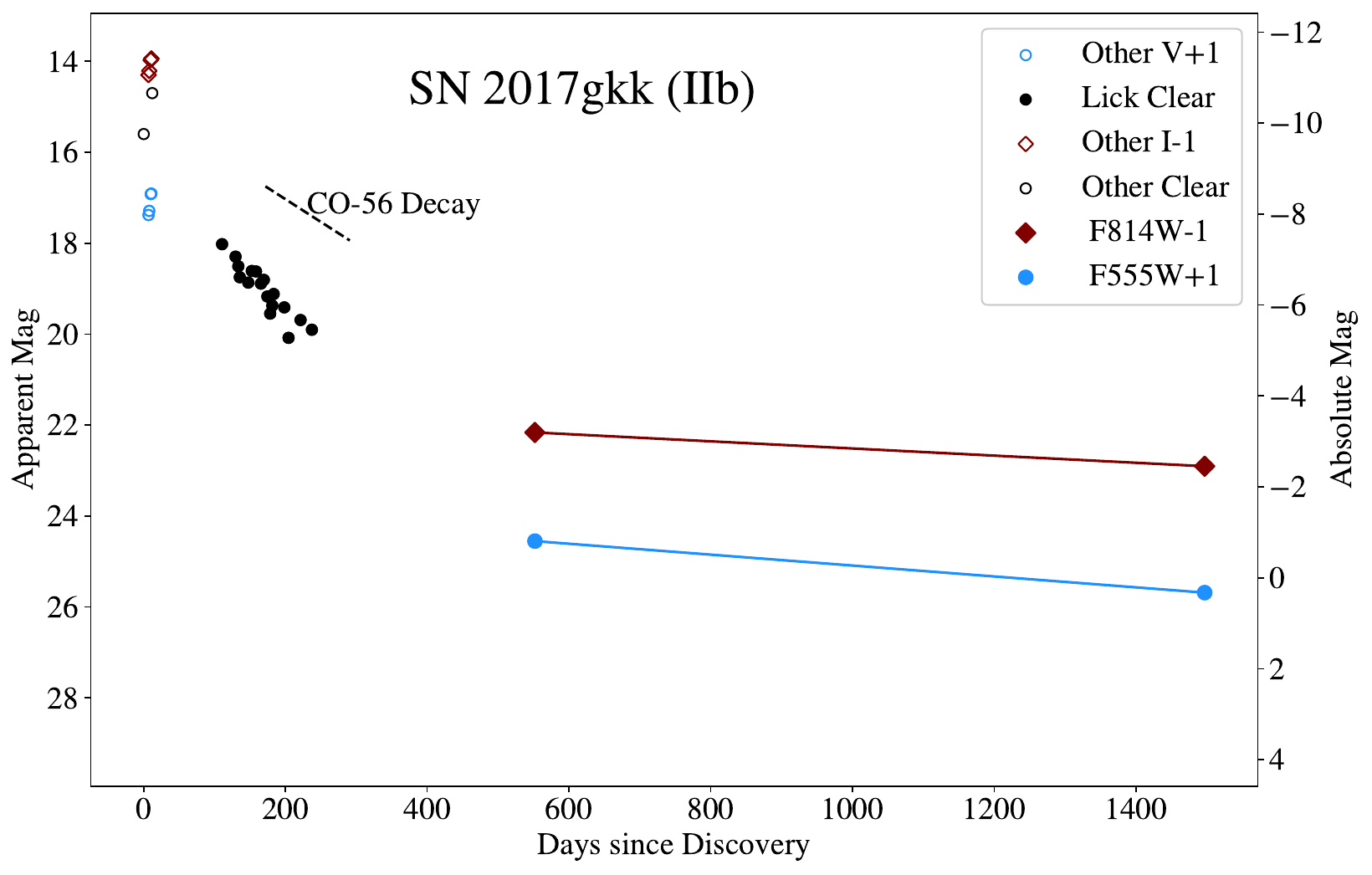}{0.4\textwidth}{(c) light curve}}
\caption{A portion of the WFC3 image mosaic containing SN 2017gkk, from observations on 2021 September 24, in (a) F555W and (b) F814W. 
Also shown is a light curve containing previously unpublished Lick ``clear'' (unfiltered) (c) points, together with data taken near discovery by \citet{Balanutsa2017}, \citet{Itagaki2017}, and  \citet{Vinko2017}, along with our Snapshot detections.}
\label{fig:17gkk}
\end{figure*}

SN 2017gkk was discovered in NGC 2748 on 2017 August 19 at 15.6 mag by \citet{Balanutsa2017}, and then later rediscovered at 14.7 mag (both unfiltered) by \citet{Itagaki2017}. The classification spectrum obtained just days after discovery by \citet{Onori2017} showed it was an SN~IIb. The SN was detected in our Snapshot images on 2021 September 24, 1485~d (4.1~yr) after discovery, in both F555W and F814W; see Figure \ref{fig:17gkk}. The location of the SN was established using data obtained on 2019 February 22 for our previous Snapshot program GO-15166 (PI A.~Filippenko), when the SN was at $m_{\rm F555W}=23.55 \pm 0.02$ and $m_{\rm F814W}=23.10 \pm 0.04$~mag. Limited early-time unfiltered (``clear''; $\sim R$) photometry was obtained with KAIT. The light-curve behavior at late times in both Snapshot bands may imply that CSM interaction is contributing to the SN luminosity.

\subsection{SN 2017ixv}\label{sec:2017ixv}

\begin{figure*}[htb]
\gridline{\fig{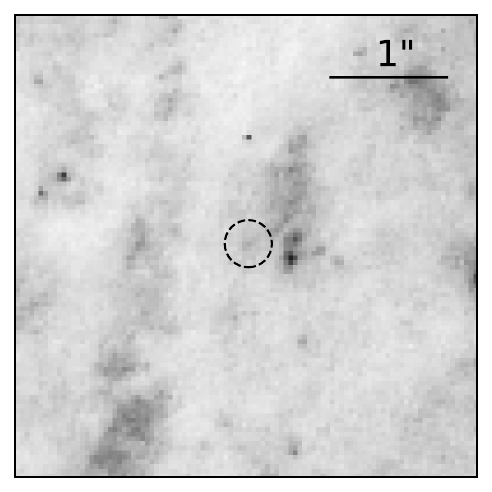}{0.25\textwidth}{(a) F555W}
          \fig{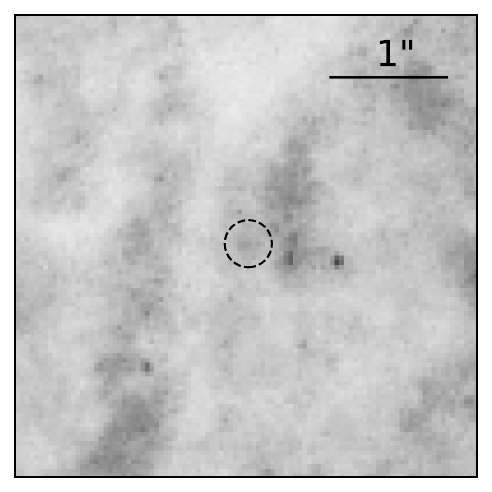}{0.25\textwidth}{(b) F814W}}
\caption{A portion of the WFC3 image mosaic containing SN 2017ixv, from observations on 2021 January 11, in (a) F555W and (b) F814W. 
The SN was not detected in either band; the site is indicated by the dashed circle (the radius of which, in this case, represents the uncertainty in the SN position). No photometry has been published for this SN.}
\label{fig:17ixv}
\end{figure*}

SN 2017ixv was discovered in NGC 6796 on 2017 December 17 by \citet{Cortini2017}. It was classified shortly thereafter as an SN~Ic-BL by \citet{Leadbeater2017}. Unfortunately, we are not aware of any published follow-up photometry. SN 2017ixv was not detectable in our F555W and F814W Snapshots from 2021 January 11, 1122~d (3.1~yr) after discovery. We note that the SN site is in an edge-on spiral galaxy, and therefore the exact location is difficult to confirm without any earlier imaging data, owing to the crowded environment. In order to locate the SN, we used the absolute position \citep{Cortini2017}, assuming a $0{\farcs}2$ uncertainty. We further added this in quadrature with a quoted uncertainty of $0{\farcs}03$ in the {\sl Gaia}-based {\sl HST\/} astrometric grid, and the total uncertainty is reflected in the radius of the dashed circle in Figure \ref{fig:17ixv}. Based on this position, it appears that the SN may have been in or near a patch of nebulosity in the host galaxy.

\subsection{SN 2018gj}

SN 2018gj was discovered in NGC 6217 by \citet{Wiggins2018} on 2018 January 1. It was classified as an SN~IIb (and possible II-P) by \citet{Bertrand2018} and as an SN~II by \citet{Kilpatrick2018a}. \citet{Teja2023} conducted extensive photometric and spectroscopic monitoring of the SN. Previously-unpublished early-time photometric monitoring also exists from KAIT. Our Snapshots were obtained on 2021 January 27, 1109~d (3.0~yr) after discovery. The exact location of the SN was isolated in our {\sl HST\/} images using data obtained on 2019 May 16 for GO-15151 (PI S.~Van Dyk), when the SN was at $m_{\rm F625W}=22.43 \pm 0.01$ and $m_{\rm F814W}=21.90 \pm 0.01$~mag.

\begin{figure*}[htb]
\gridline{\fig{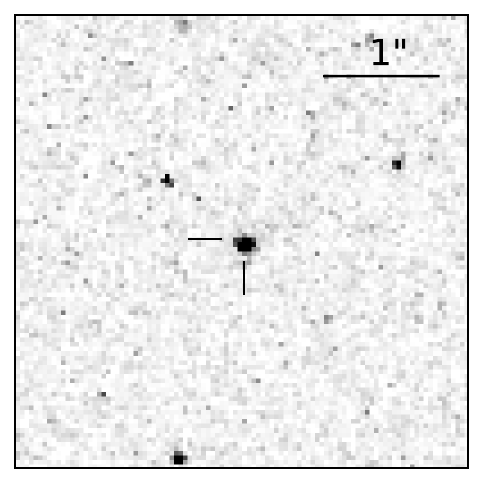}{0.25\textwidth}{(a) F555W}
          \fig{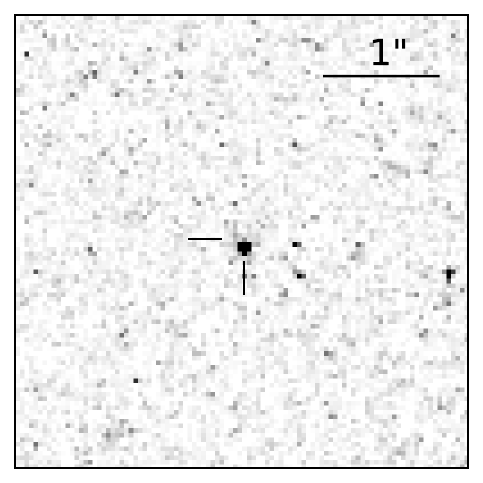}{0.25\textwidth}{(b) F814W}
          \fig{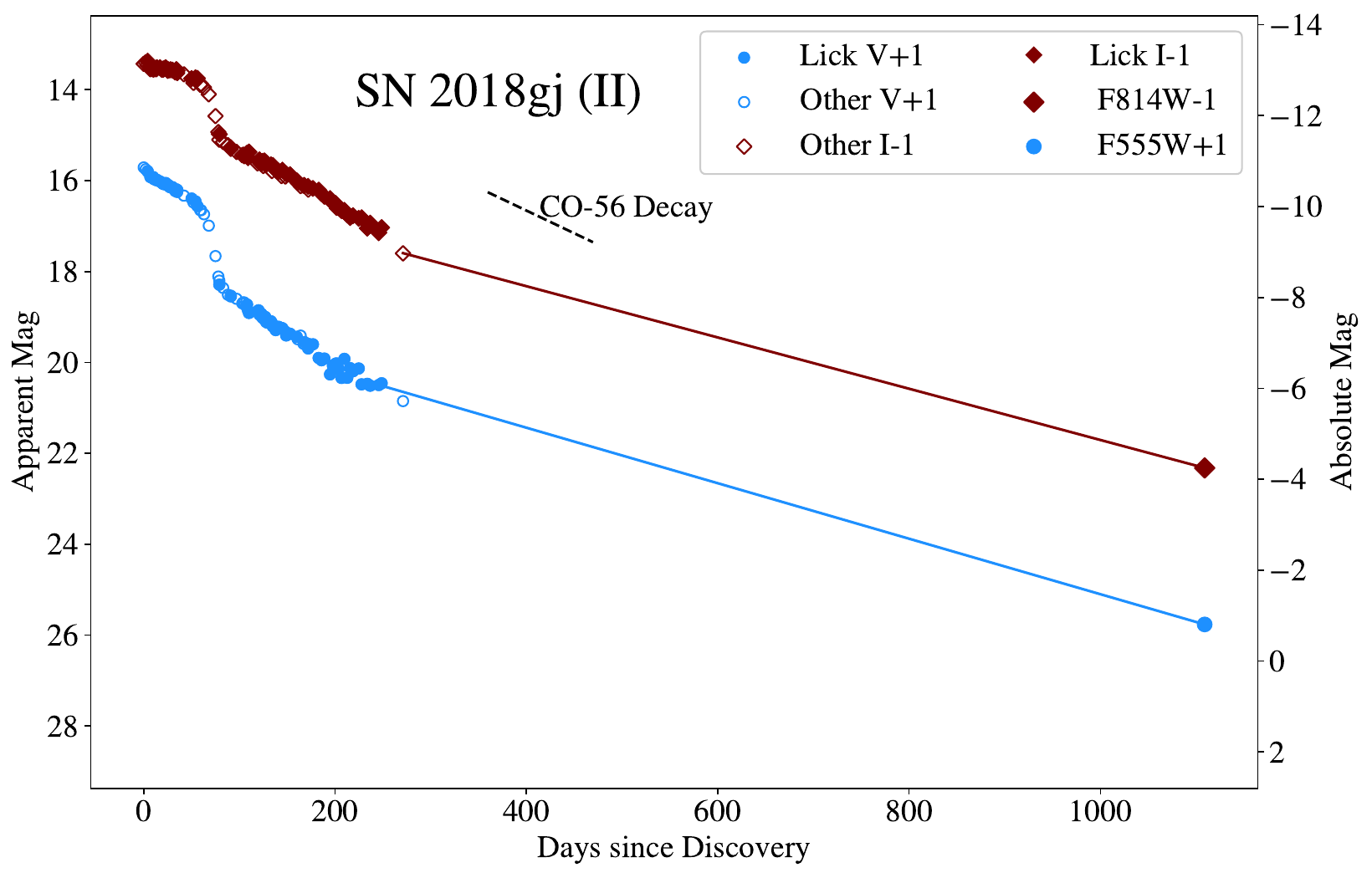}{0.4\textwidth}{(c) light curve}}
 \caption{A portion of the WFC3 image mosaic containing SN 2018gj, from observations on 2021 January 27, in (a) F555W and (b) F814W. 
 Also shown are unpublished Lick $V$ and $I$ (c) light curves, along with (``Other'') data from \citet{Teja2023}, together with the Snapshot detections.}
\label{fig:18gj}
\end{figure*}


\subsection{SN 2018zd}

SN 2018zd in NGC 2146 was monitored and analyzed independently by \citet{Zhang2020} and \citet{Hiramatsu2021}. Both studies considered the event to be a low-luminosity SN~II; however, the latter considered this to be the best example so far for an electon-capture SN, whereas the former found it to have properties more consistent with a normal SN~II-P. Both studies found it likely that the progenitor candidate, identified in pre-explosion {\sl HST\/} images, could have been a super-asymptotic-giant-branch star.

Our F606W and F814W Snapshots were obtained on 2021 February 7, 1074~d (2.9~yr) after discovery. (These two bands, and in particular F606W, were chosen for purposes of estimating a TRGB distance to the host galaxy, which is beyond the scope of this paper.) The location of the SN was isolated using {\sl HST\/} data obtained on 2019 May 19 for GO-15151 (PI S.~Van Dyk; \citealt{Hiramatsu2021}). The SN was not detected in either band. \citet{VanDyk2023} took advantage of this fact to conclude that the candidate was indeed the SN 2018zd progenitor.

\begin{figure*}[htb]
\gridline{\fig{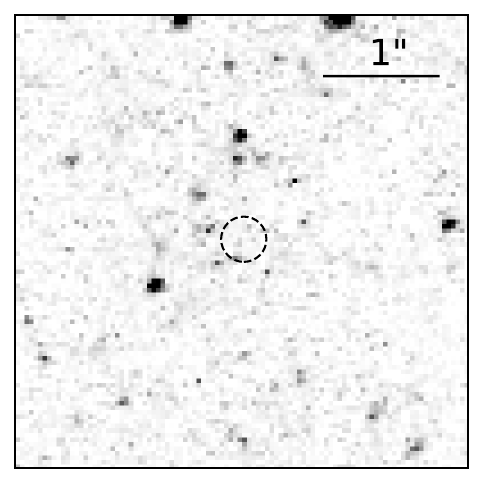}{0.25\textwidth}{(a) F606W}
          \fig{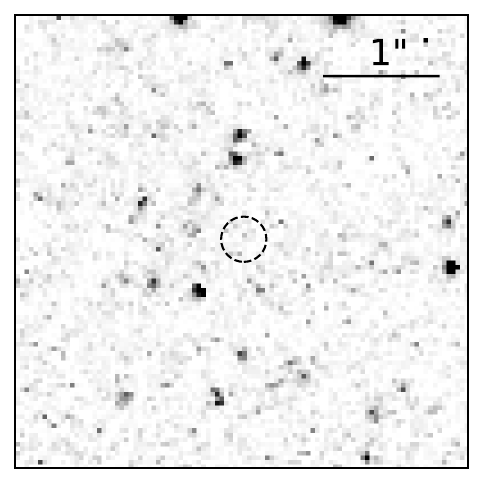}{0.25\textwidth}{(b) F814W}
          \fig{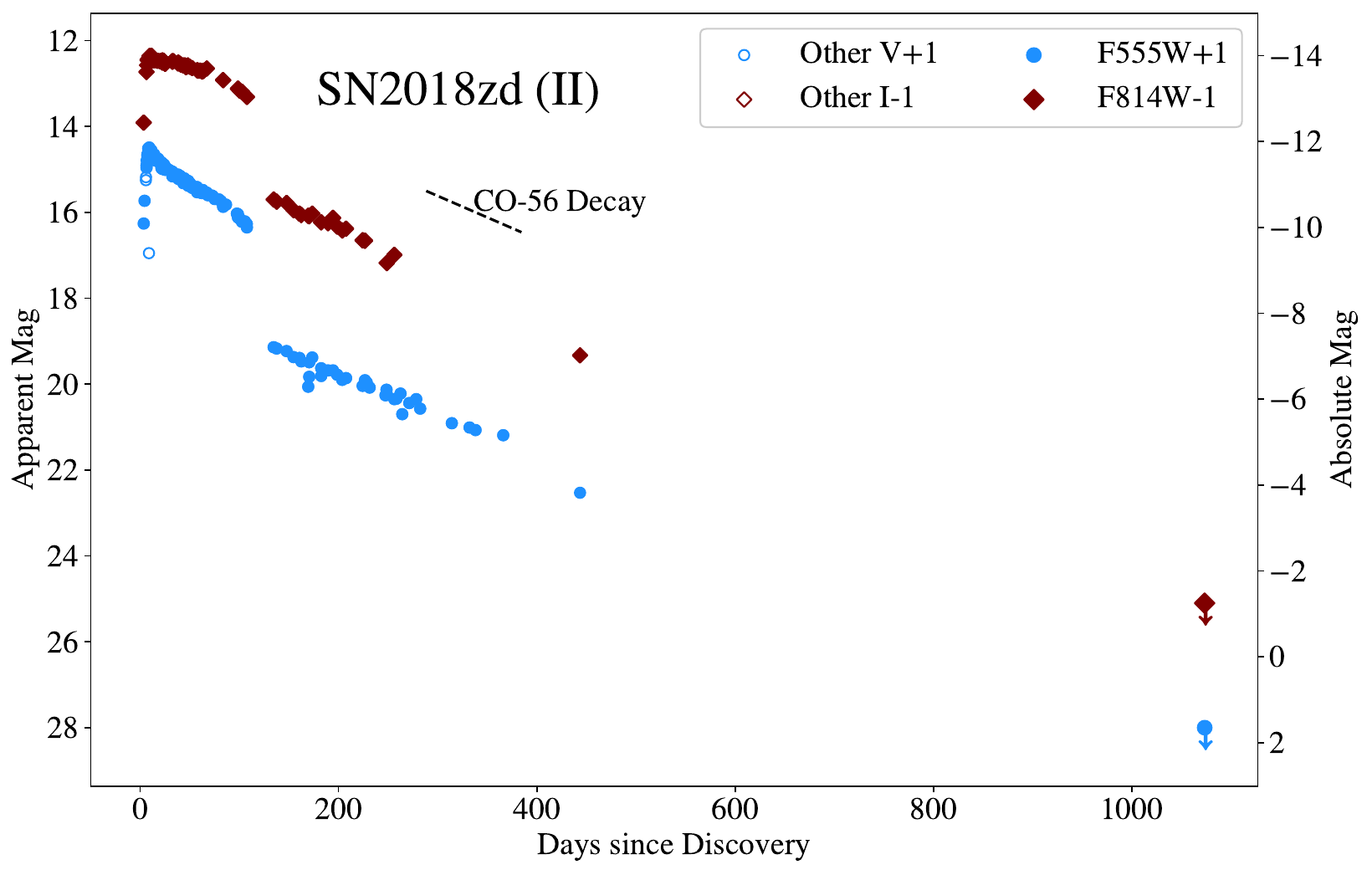}{0.4\textwidth}{(c) light curve}}
\caption{A portion of the WFC3 image mosaic containing SN 2018zd, from observations on 2021 February 7, in (a) F606W and (b) F814W. 
The SN was not detected in either band; the site is indicated by the dashed circle. Also shown are $V$ and $I$ (c) light curves based on (``Other'') data from \citet{Zhang2020}, \citet{Hiramatsu2021}, and \citet{Callis2021}, together with F555W and F814W data from  \citet{Hiramatsu2021} and the Snapshot upper limits.}
\label{fig:18zd}
\end{figure*}

\subsection{SN 2018aoq}

SN 2018aoq is a low-luminosity SN~II-P. Both \citet{ONeill2019} and \citet{Tsvetkov2019} undertook photometric monitoring campaigns of the SN, with the latter study including spectral coverage over the first $\sim 70$~d. Furthermore, based on the available multiband, pre-explosion {\sl HST\/} images, \citet{ONeill2019} estimated the initial mass of a RSG progenitor candidate at $\sim 10\ M_{\odot}$. 

Our Snapshot observations were executed in F555W and F814W on 2020 December 5, 984~d (2.7~yr) after discovery. We located the SN position in the Snapshots using {\sl HST\/} data from 2018 April 23 (GO-15151; PI S.~Van Dyk), when the SN was at $m_{\rm F555W}= 15.93 \pm 0.01$~mag. SN 2018aoq was not detected in either of our Snapshots. \citet{VanDyk2023} exploited that fact to conclude that the \citet{ONeill2019} candidate was indeed the progenitor.

\begin{figure*}[htb]
\gridline{\fig{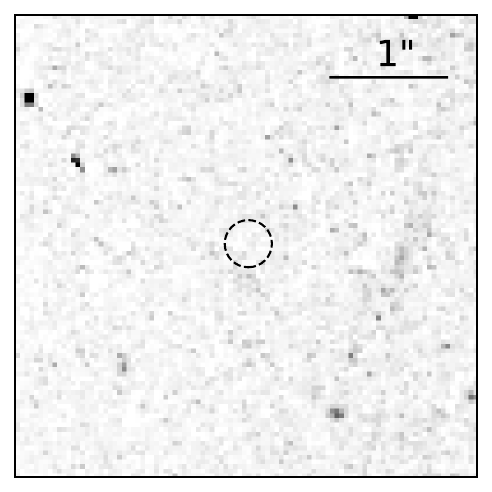}{0.25\textwidth}{(a) F555W}
          \fig{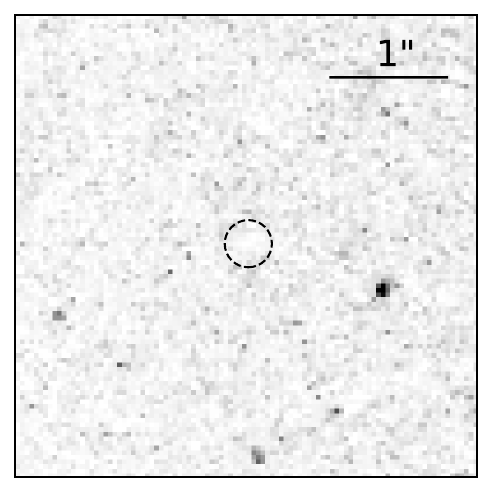}{0.25\textwidth}{(b) F814W}
          \fig{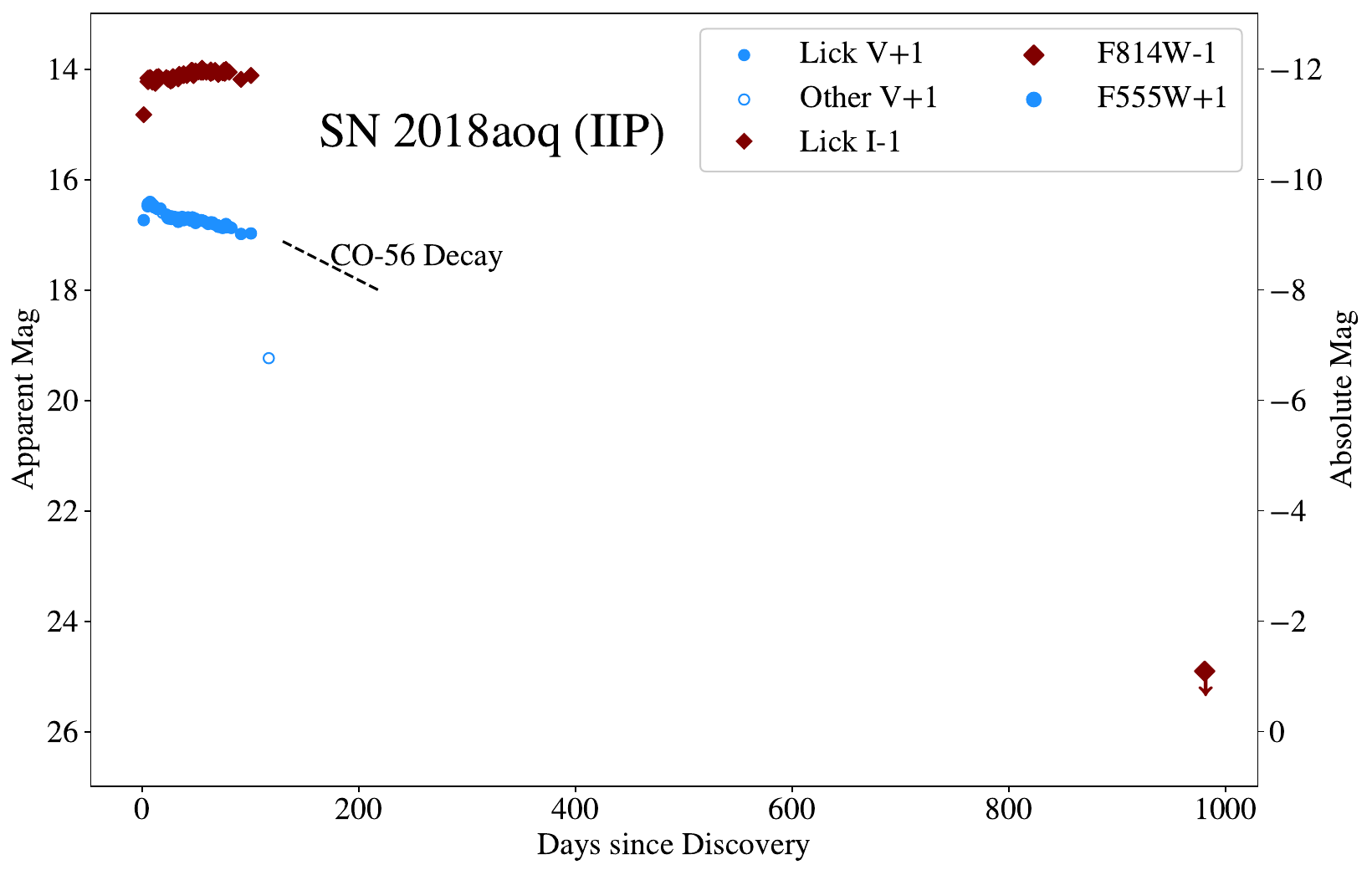}{0.4\textwidth}{(c) light curve}}
\caption{A portion of the WFC3 image mosaic containing SN 2018aoq, from observations on 2020 December 5, in (a) F555W and (b) F814W. 
The SN was not detected in either band; the site is indicated by the dashed circle. Also shown are previously unpublished Lick $V$ and $I$ (c) light curves, along with (``Other'') data from \citet{ONeill2019} and \citet{Tsvetkov2019}, together with the Snapshot upper limits.}
\label{fig:18aoq}
\end{figure*}

\subsection{AT 2018cow}

AT 2018cow, in CGCG 137$-$068 at $z = 0.0141$, is a particularly intriguing object. It very rapidly became exceedingly luminous ($\sim -22$~mag absolute) and blue, and is considered a prototypical ``fast blue optical transient,'' or FBOT. ``The Cow,'' as it has been dubbed, stimulated intense interest in the community, leading to several multiwavelength monitoring campaigns and theoretical analyses, including those of \citet{Perley2019}, \citet{Margutti2019}, and \citet{Xiang2021}. Despite all of the focused effort, the nature of AT 2018cow and its precursor is still not settled. For instance, \citet{Fox2019} surmised, based on similarities with various interacting SNe, that CSM interaction in a relatively H-depleted system could explain some its observed properties. \citet{Chen2023} concluded that a fading transient UV source persists, which may be from ejecta-CSM interaction or from a central engine, more specifically a precessing accretion disk. Additionally, \citet{Inkenhaag2023}  studied the late-time brightness of AT~2018cow and estimated  the potential black hole's mass using our Snapshot data.

\begin{figure*}[htb]
\gridline{\fig{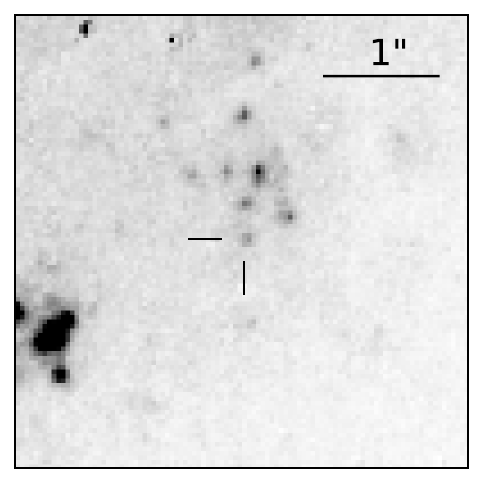}{0.25\textwidth}{(a) F555W}
          \fig{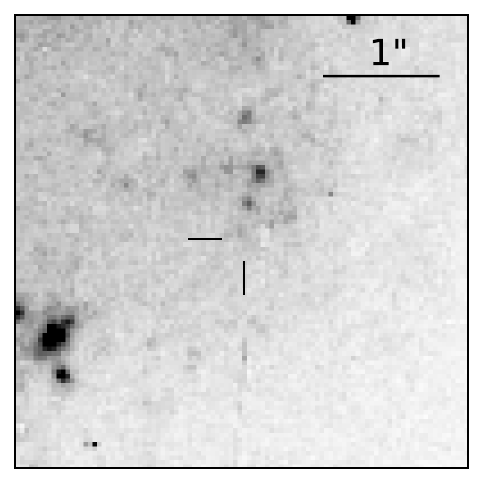}{0.25\textwidth}{(b) F814W}
          \fig{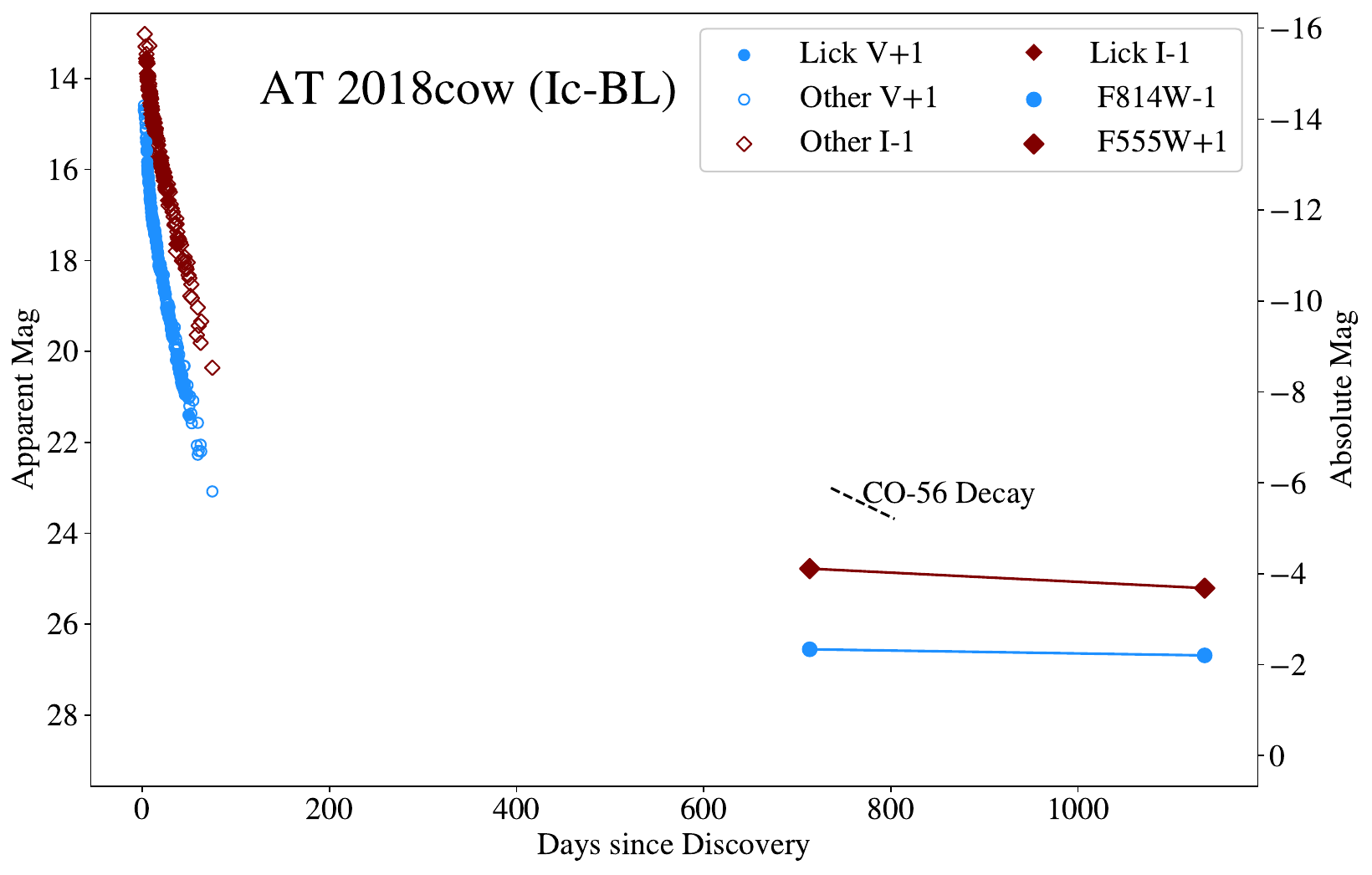}{0.4\textwidth}{(c) light curve}}
\caption{A portion of the WFC3 image mosaic containing AT 2018cow, from observations on 2021 July 25, in (a) F555W and (b) F814W. 
Also shown are Lick \citep{Zheng2022} $V$ and $I$ (c) light curves, along with (``Other'') data from \citet{Perley2019}, \citet{Margutti2019}, \citet{Xiang2021}, and \citet{Tsvetkov2022}, together with the Snapshot detections.}
\label{fig:18cow}
\end{figure*}

Our F555W and F814W Snapshots F555W and F814W were obtained on 2021 July 25, 1134~d (3.1~yr) after discovery. The location of the object was confirmed using prior {\sl HST\/} data taken on 2018 August 6 for GO-15600 (PI R.~Foley), as well as from finding charts in prior literature on the object, such as \citet{Perley2019} and \citet{Margutti2019}; see Figure \ref{fig:18cow}. \citet{Chen2023} used our Snapshots as part of a larger work, looking at a variety of late-time observations (from 50 to 1423~d post-discovery) to better understand the object. They concluded that the nature of a putative black hole at the center of the accretion disk is still up for debate, given the various intriguing properties of the late-time emission. (Both \citealt{Sun2022} and \citealt{Sun2023} also made use of our Snapshot data.) The final identity of the precursor object therefore remains unknown.

\subsection{SN 2018ivc}

From high-cadence follow-up spectroscopic and photometric observations since discovery, \citet{Bostroem2020} concluded that SN 2018ivc in NGC~1068 is an unusual SN~II. That study placed limits on the properties of the progenitor, based on available pre-explosion {\sl HST\/} of this famous Seyfert galaxy. \citet{Maeda2023a} considered SN 2018ivc as a possible variant of SN~II-L, with transitional characteristics between II-P and IIb. Interaction of the SN shock with pre-existing wind matter appears to be playing a strong role in the SN's emission \citep{Maeda2023a}, with origins in a progenitor experiencing an extreme form of binary evolution \citep{Maeda2023b}. The SN was detected in our {\sl HST\/} F555W and F814W Snapshots obtained on 2020 November 27, 739~d (2.0~yr) after discovery. The SN location was confirmed using {\sl HST\/} data obtained for GO-15151 (PI S.~Van Dyk), in which the SN was at $m_{\rm F555W}= 20.33 \pm 0.01$ and $m_{\rm F814W}= 19.03 \pm 0.01$~mag.

\begin{figure*}[htb]
\gridline{\fig{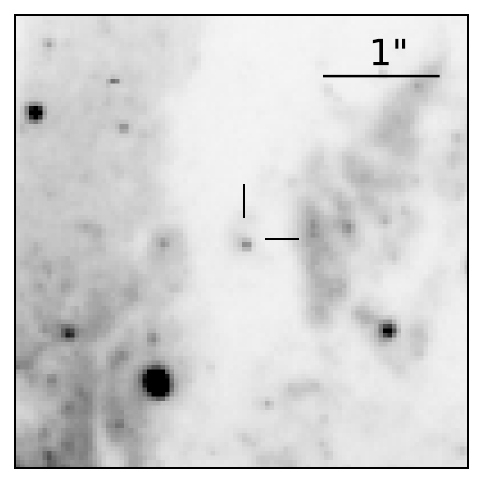}{0.25\textwidth}{(a) F555W}
          \fig{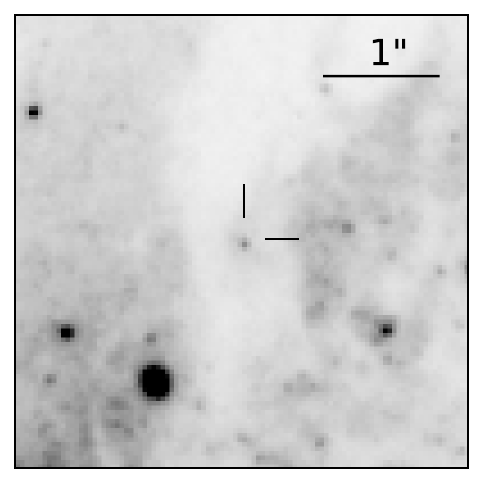}{0.25\textwidth}{(b) F814W}
          \fig{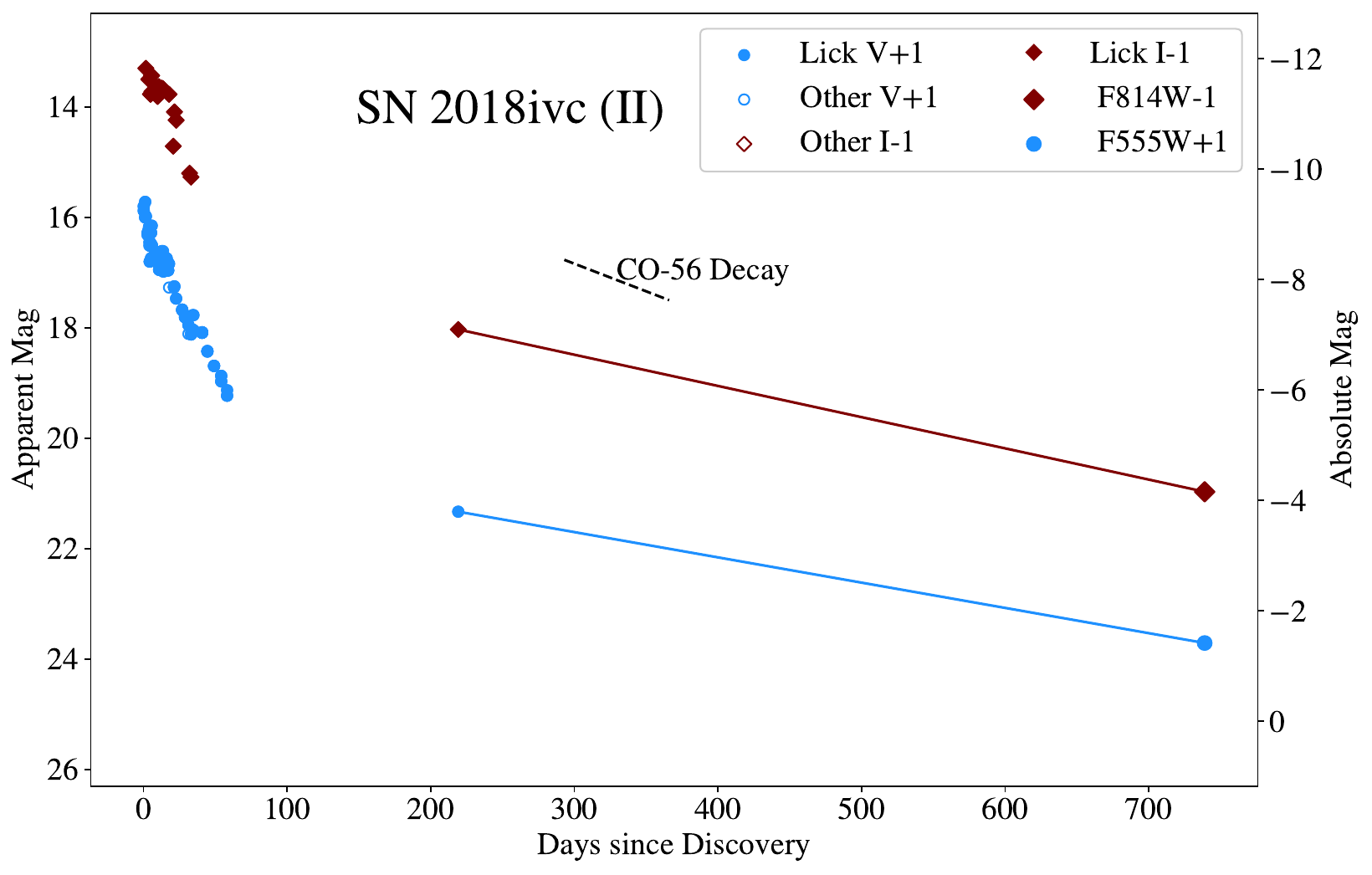}{0.4\textwidth}{(c) light curve}}
\caption{A portion of the WFC3 image mosaic containing SN2018ivc, from observations on 2020 November 27, in (a) F555W and (b) F814W. 
Also shown are previously unpublished Lick $V$ and $I$ (c) light curves, along with (``Other'') data from \citet[][ which include F555W and F814W data from GO-15151, PI S.~Van Dyk]{Bostroem2020}, together with the Snapshot detections.}
\label{fig:18ivc}
\end{figure*}

\subsection{SN 2019ehk}\label{sec:2019ehk}

\begin{figure*}[htb]
\gridline{\fig{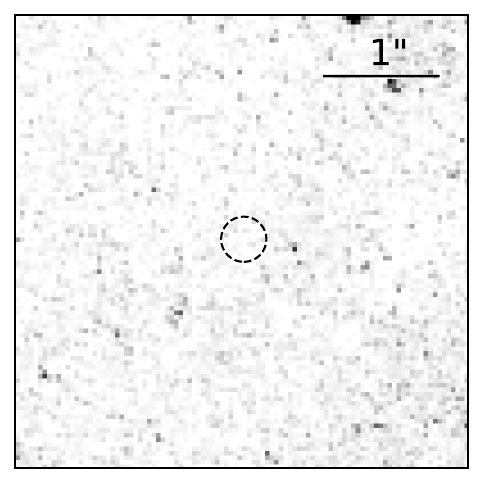}{0.25\textwidth}{(a) F438W}
          \fig{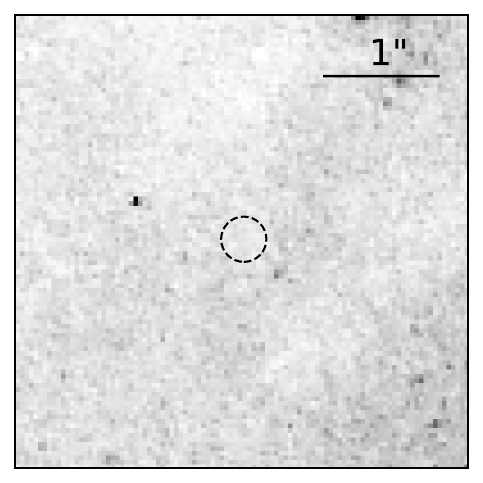}{0.25\textwidth}{(b) F625W}
          \fig{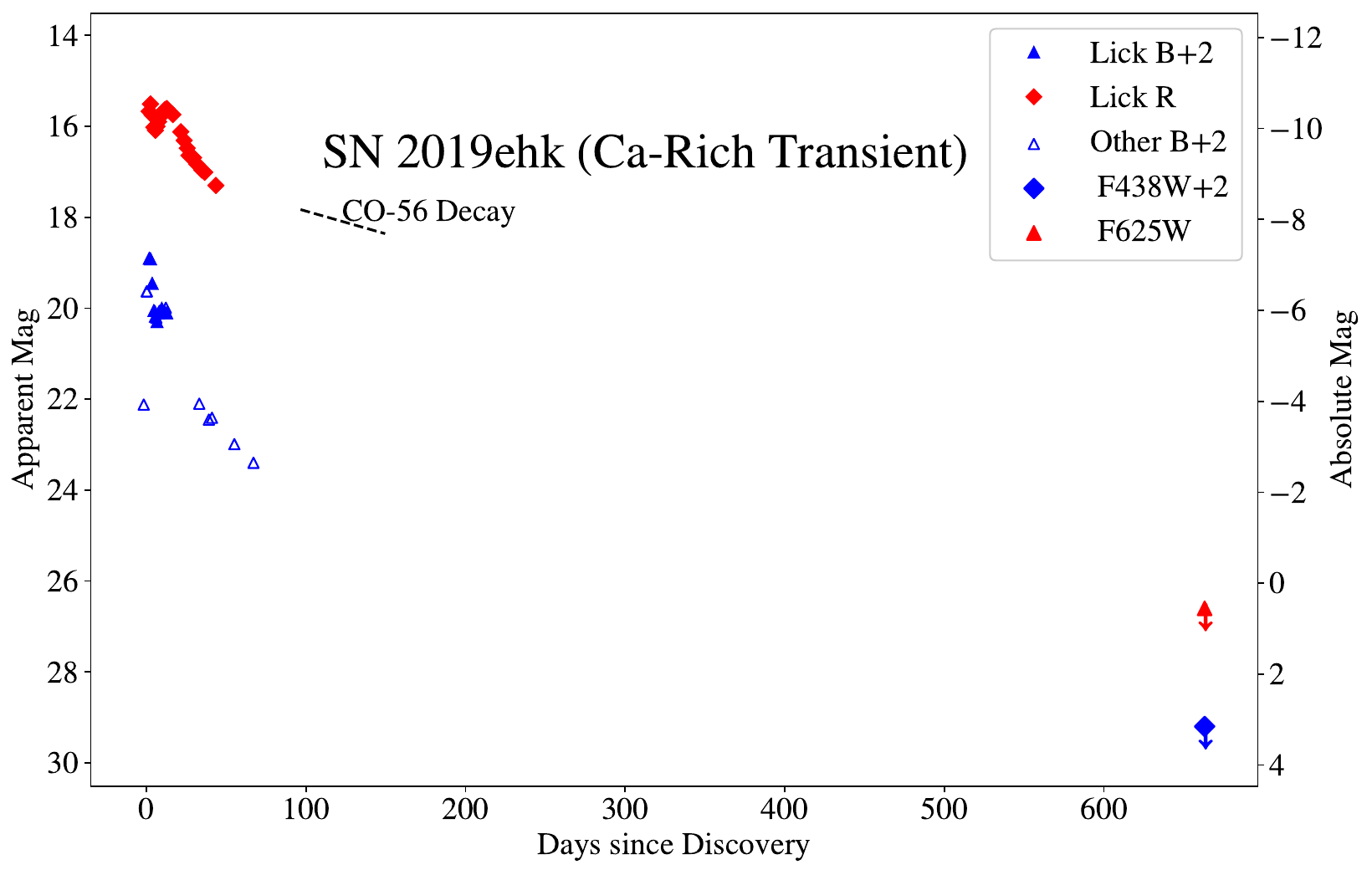}{0.4\textwidth}{(c) light curve}}
\caption{A portion of the WFC3 image mosaic containing SN 2019ehk, from observations on 2021 February 21, in (a) F438W and (b) F625W. 
The transient was not detected in either band; the site is indicated by the dashed circle. Also shown are previously unpublished Lick $V$ and $I$ (c) light curves, along with (``Other'') data from \citet{Nakaoka2021} and \citet{JacobsonGalan2021}, together with the Snapshot upper limits.}
\label{fig:19ehk}
\end{figure*}

SN 2019ehk in NGC 4321 (M100) has been characterized as a ``Ca-rich transient," possibly evolving from a Type Ib SN, based on photometric and spectroscopic monitoring of the event by \citet{JacobsonGalan2020,JacobsonGalan2021} and \citet{Nakaoka2021}. We located the SN 2019ehk site in our Snapshot data using {\sl HST\/} data from 2020 May for program GO-16075 (PI W.~Jacobson-Gal{\'a}n), in which the transient was at $m_{F555W} = 25.91 \pm 0.09$~mag. However, interestingly, it was not detected in our {\sl HST\/} F438W and F625W images obtained on 2021 February 21, 665~d (1.8~yr) after discovery; see Figure \ref{fig:19ehk}. The detection upper limits are 27.0 and 26.6~mag in F438W and F606W, respectively.
 
\subsection{AT 2019krl}\label{sec:2019krl}

\begin{figure*}[htb]
\gridline{\fig{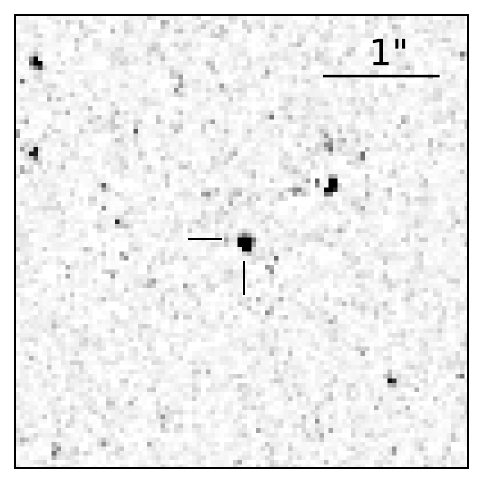}{0.25\textwidth}{(a) F438W}
          \fig{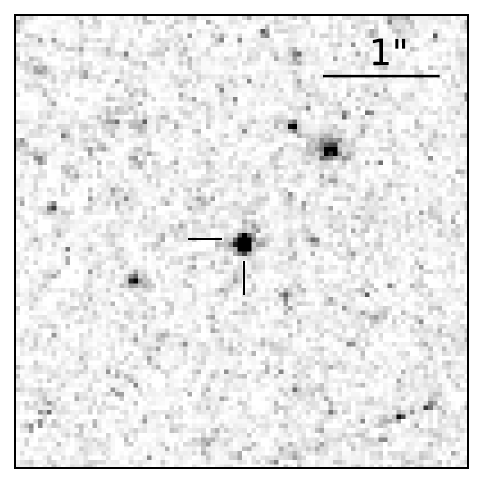}{0.25\textwidth}{(b) F625W}
          \fig{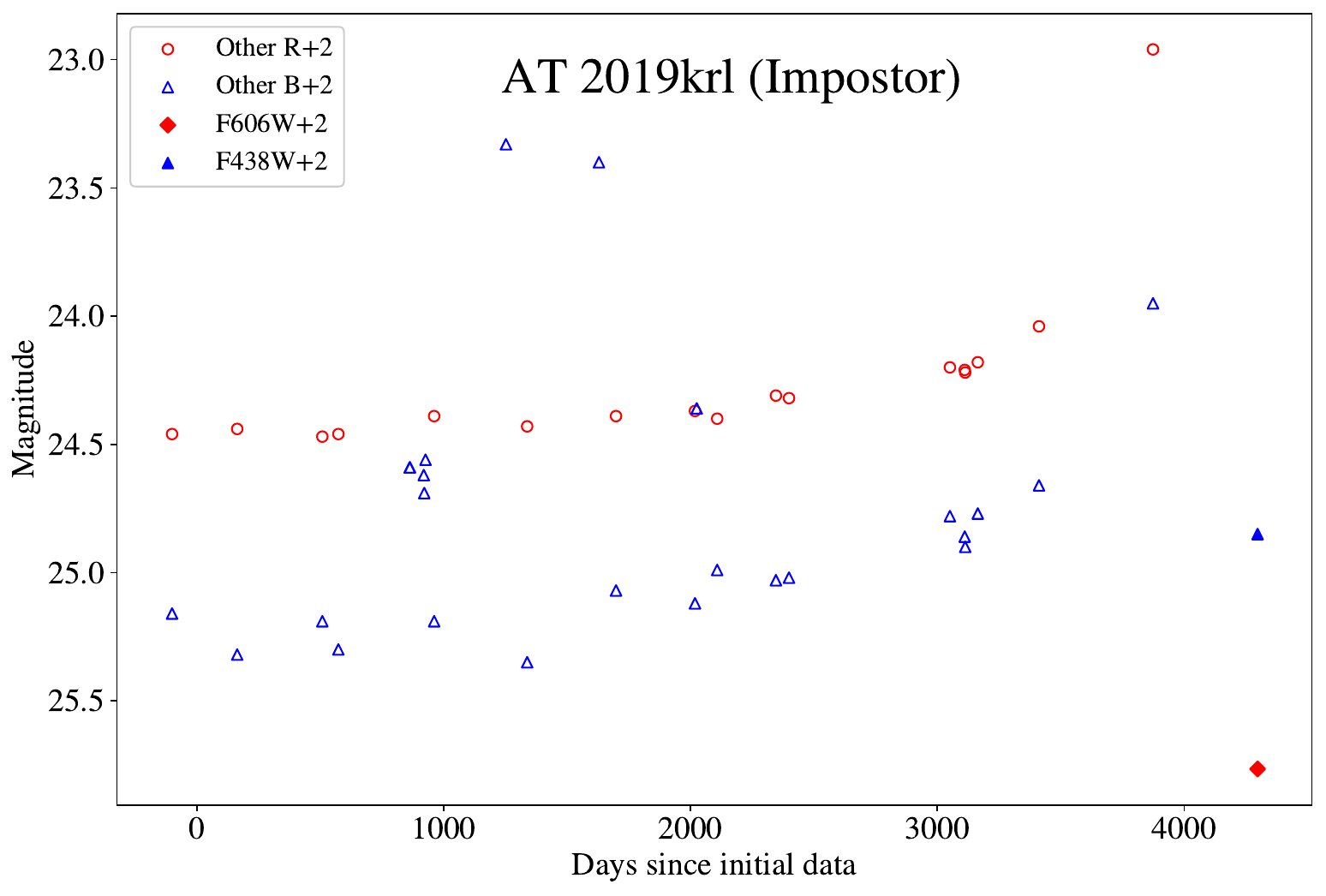}{0.4\textwidth}{(c) light curve}}
\caption{A portion of the WFC3 image mosaic containing AT 2019krl, from observations on 2021 February 15, in (a) F438W and (b) F625W. 
Also shown are $B$ and $R$ (``Other'') (c) light curves from \citet{Andrews2021}, together with the Snapshot detections.}
\label{fig:19krl}
\end{figure*}

\citet{Andrews2021} characterized the intermediate-luminosity transient AT 2019krl in NGC 628 (M74) as either a relatively unobscured blue supergiant or a more extinguished LBV in eruption. We located the position of the transient in our Snapshot images using 2019 {\sl HST\/} data from GO-15151 (PI S.~Van Dyk), in which the AT was at $m_{\rm F555W} = 21.91 \pm 0.02$~mag. The AT was clearly visible in the more recent F438W and F625W Snapshots from 2021 February 15; see Figure \ref{fig:19krl}.

\subsection{SN 2020dpw}\label{sec:2020dpw}

\begin{figure*}[htb]
\gridline{\fig{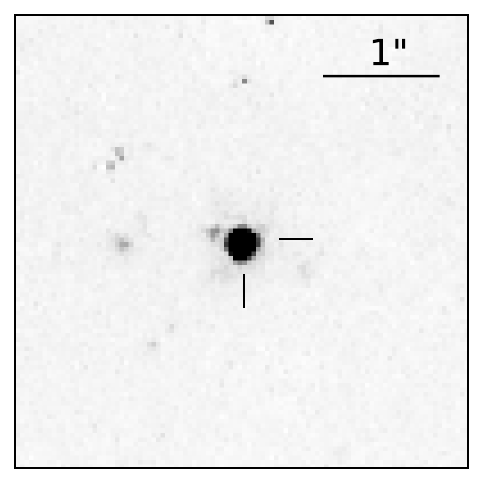}{0.25\textwidth}{(a) F555W}
          \fig{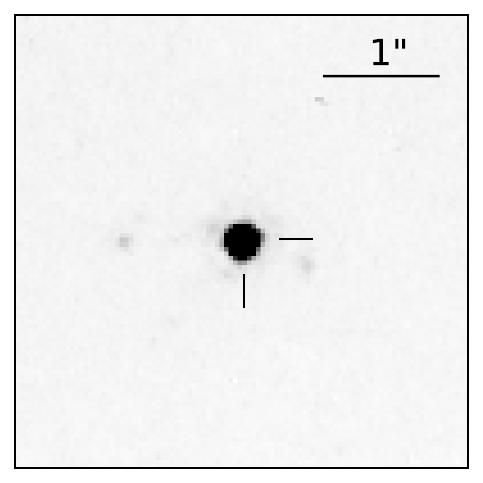}{0.25\textwidth}{(b) F814W}}
\caption{A portion of the WFC3 image mosaic containing SN 2020dpw, from observations on 2020 December 13, in (a) F555W and (b) F814W. 
No photometry has been published for this SN.}
\label{fig:20dpw}
\end{figure*}

\citet{Wiggins2020} discovered SN 2020dpw in NGC 6951 (based on the discovery position, the host must have been incorrectly reported as NGC 6952) and \citet{Kawabata2020} classified it as an SN~II-P. Unfortunately, we have no knowledge of existing early-time photometry, although the SN was easily detected in our {\sl HST\/} F555W and F814W Snapshots from 2020 December 13, near the reported discovery location 292~d (0.8~yr) after discovery; see Figure \ref{fig:20dpw}.

\subsection{SN 2020hvp}

\begin{figure*}[htb]
\gridline{\fig{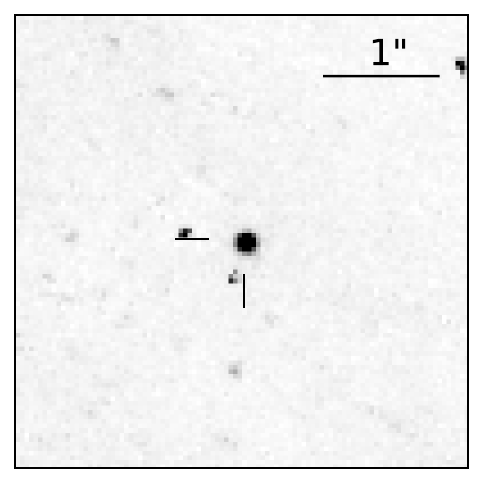}{0.25\textwidth}{(a) F555W}
          \fig{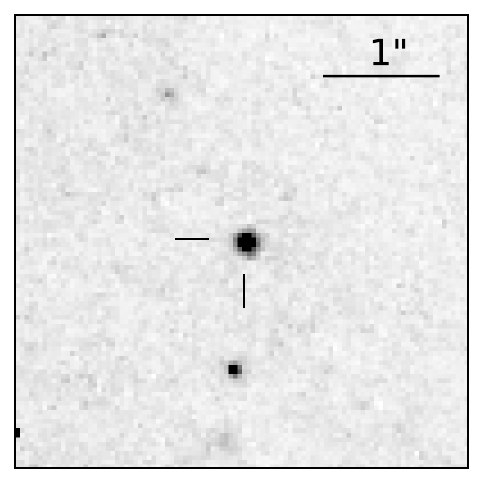}{0.25\textwidth}{(b) F814W}
          \fig{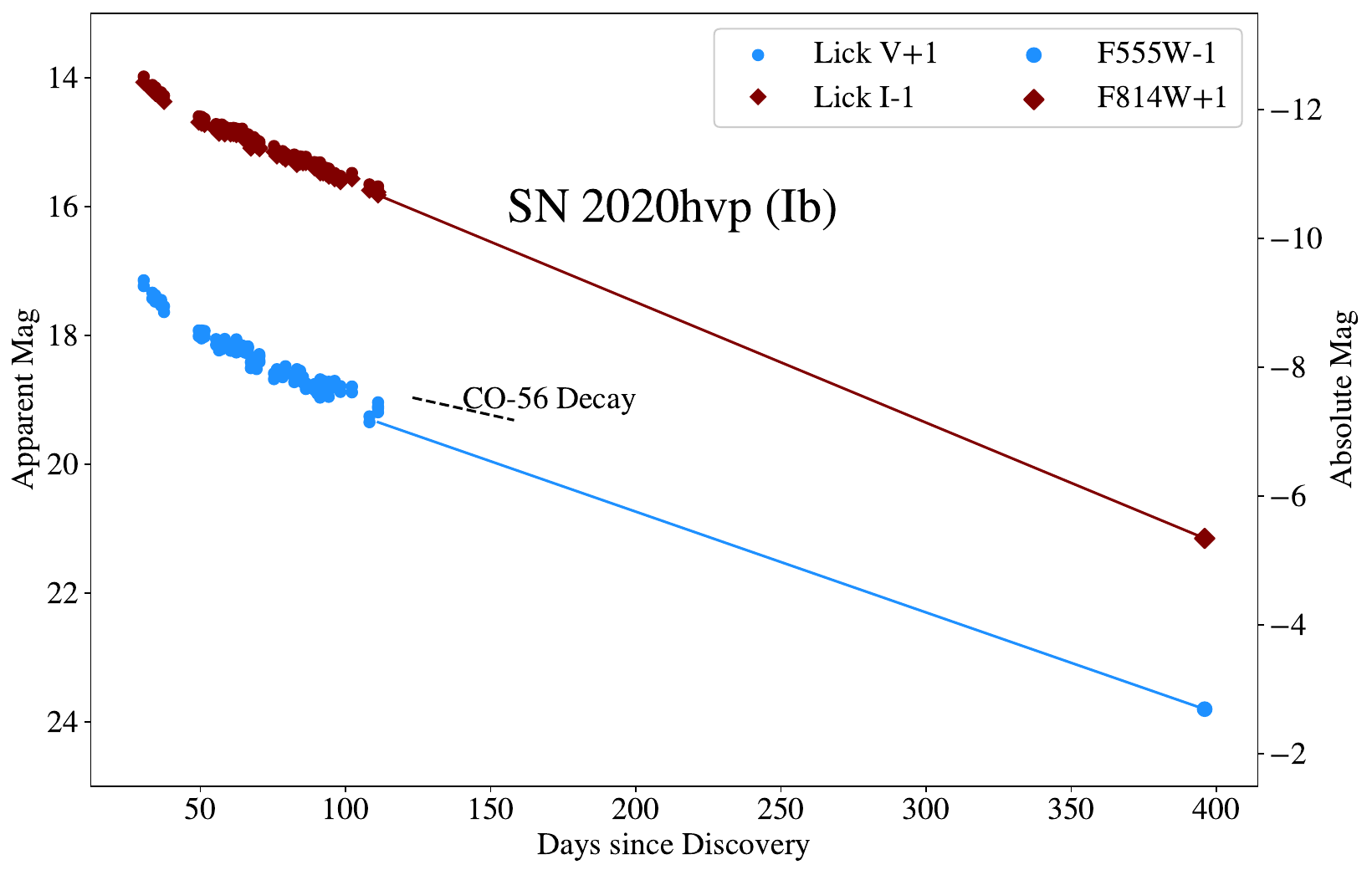}{0.4\textwidth}{(c) light curve}}
\caption{A portion of the WFC3 image mosaic containing SN 2020hvp, from observations on 2021 May 22, in (a) F555W and (b) F814W. 
Also shown are previously unpublished Lick $V$ and $I$ (c) light curves, together with the Snapshot detections.}
\label{fig:20hvp}
\end{figure*}

\citet{Tonry2020} discovered SN 2020hvp in NGC 6118 and \citet{Burke2020} subsequently classified it as an SN~Ib. The SN was easily detected near the reported discovery position in the Snapshot images obtained on 2021 May 22, 397~d (1.1~yr) after discovery; see Figure \ref{fig:20hvp}. In the figure we complement previously-unpublished early-time photometry obtained with KAIT with the {\sl HST\/} data.  

\subsection{SN 2020jfo}\label{sec:2020jfo}

\begin{figure*}[htb]
\gridline{\fig{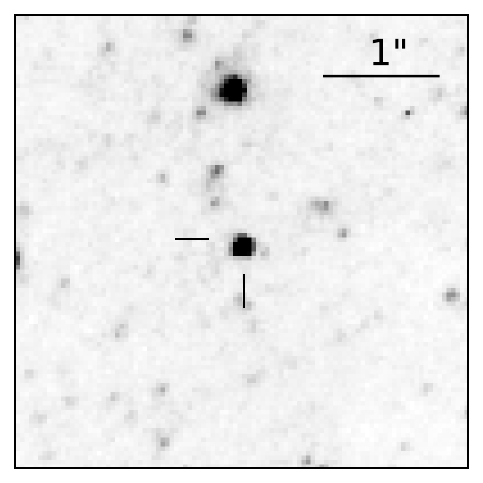}{0.25\textwidth}{(a) F555W}
          \fig{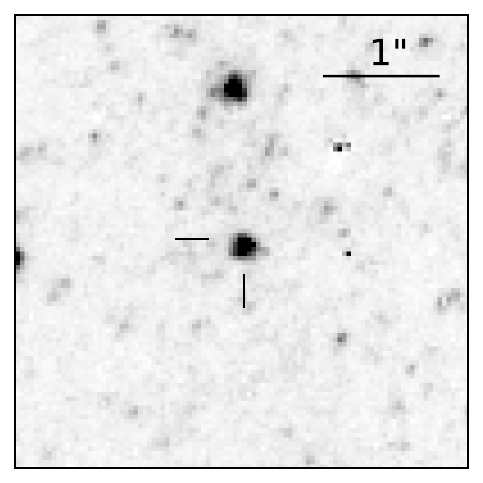}{0.25\textwidth}{(b) F814W}
          \fig{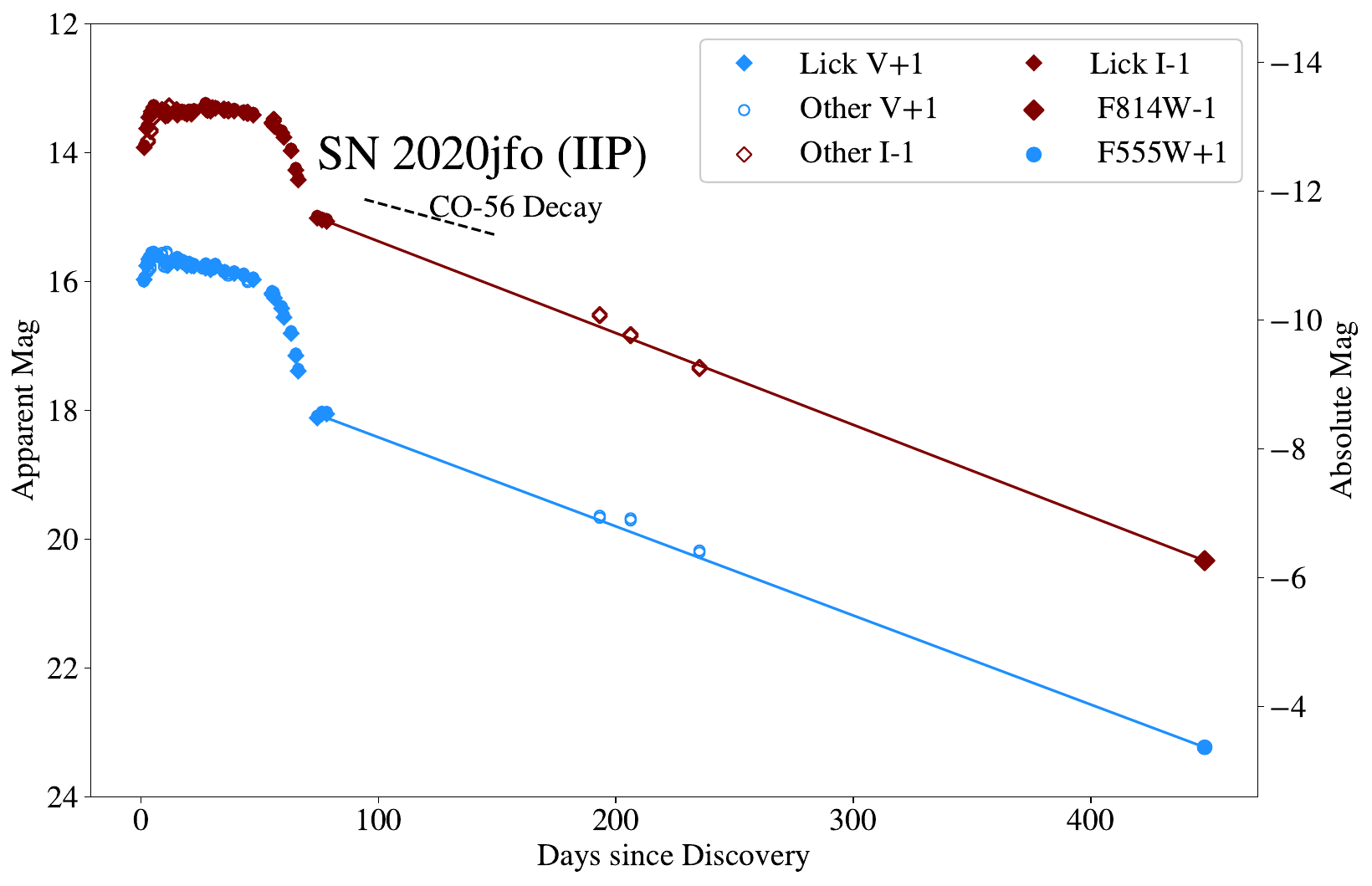}{0.4\textwidth}{(c) light curve}}
\caption{A portion of the WFC3 image mosaic containing SN 2020jfo, from observations on 2021 July 28, in (a) F555W and (b) F814W. 
Also shown are the Lick \citep{Ailawadhi2022} $V$ and $I$ (c) light curves, along with (``Other'') data from \citet{Sollerman2021} and \citet{Teja2022}, together with the Snapshot detections.}
\label{fig:20jfo}
\end{figure*}

The SN~II-P~2020jfo in NGC 4303 (M61) was monitored by \citet{Sollerman2021}, \citet{Teja2022}, and \citet{Ailawadhi2022}, in which early-time photometry obtained by KAIT was presented. The SN was clearly detected in both Snapshot bands F555W and F814W obtained on 2021 July 28, 449~d (1.2~yr) after discovery; see Figure \ref{fig:20jfo}. These {\sl HST\/} observations were first discussed by \citet{Sollerman2021}, as confirmation of the SN progenitor candidate identification. \citet{Kilpatrick2021} also identified the progenitor through pre-explosion images.

\section{Discussion and Conclusions}\label{conclusions}

We have conducted an analysis of images that we obtained from an {\sl HST\/} Snapshot survey during Cycle 28 of nearby SNe at late times. We were ultimately able to observe successfully the targeted sites of 31 SNe of various types and 4 SN impostors. The goal of the program was to reveal the possible origins of their late-time emission or lack thereof. Only 2 of the 31 SNe (SN 2020hvp and SN 2020jfo) listed in Table~\ref{table:phot} convincingly exhibited lingering emission most likely ascribed to radioactive decay of $^{56}$Co. For 12 of the remaining SNe (indicated by ``Yes?'' in Table~\ref{table:phot}) we could not determine the source of the late-time emission, since these events were no longer detectable, and upper limits to their luminosities were not sufficiently constraining. All three of the observed SNe~Ia fall in this category. For three of the SNe in the observed sample (SN 2017gax, SN 2017ixv, and SN 2020dpw), no early-time photometry was available, and the former two SNe were no longer detectable, so it was not possible to determine whether radioactive decay was powering the light at late times. A remaining 15 SNe were detected; however, it was clear from their extended light curves that the emission was in excess of what we would expect for radioactive decay. We can infer in these cases that the emission may arise, at least in part, from sustained CSM interaction or a light echo, or both. SN~2010jl had exhibited previous indications of CSM interaction, but was no longer detectable in our Snapshot data. It is also worth mentioning the possibility that the sustained late-time luminosity could at least partially be due to radioactive decay of elements with longer lifetimes.

We have also detected the known resolved light echoes around SN 2012aw and SN 2016adj, and we note that their geometries have evolved since they were first discovered \citep{VanDyk2015,Stritzinger2022}.

Of the four events that we consider to be SN impostors, all are still detectable in our Snapshots, implying that their eruptive behavior is persisting even at late times. Note that we have considered SN~2016jbu as an SN impostor, although \citet{Brennan2022c} concluded that the event may have actually been a terminal explosion and that the precursor has vanished.

Whereas we can likely infer from the few observed SNe~Ia that their late-time emission was consistent with radioactive decay, for a significant number of core-collapse SNe~II, CSM interaction may contribute to the luminosity even as 
late as $\sim 10,000$~d. This is not entirely surprising for the SNe~IIn in our sample, and also to some extent for the SNe~IIb, which have largely shown signs of CSM interaction at early times. However, for otherwise-normal SNe~II-P, such as SN 2016bkv and SN 2017eaw, the SN shock unexpectedly continues to interact at $\gtrsim 1000$~d with the pre-existing CSM lost by the progenitor prior to explosion. The presence of such interaction provides important information about the extent of the CSM and the duration and nature of the mass loss(which can be further constrained through information gathered from the SN spectrum), with further implications for the evolution of the massive progenitor. 

Snapshot surveys, such as ours, can efficiently provide a broad overview of the late-time properties of SNe and SN impostors and represent a reasonable use of valuable {\sl HST\/} observing time. Approximately 70\% of our originally proposed program was actually completed. The only wrinkle is that one has no control over which targets actually get executed, yet developing a relatively comprehensive sample is important in order to obtain a set of statistically significant results. Here we chose to target a large range of object types, to obtain knowledge of the late-time luminosity across a range of events from different astrophysical conditions. However, one could limit the sample to a large number of one particular SN type, nominally arising from a distinct progenitor population. Such is the case for {\sl HST\/} programs pointedly targeting samples of SNe~Ia (\citealt{Graham2019}; in Cycle 24) and SNe~II (PI C.~Kilpatrick in Cycle 30; PI W.~Jacobson-Gal\'an in Cycle 31). {\sl HST\/} Snapshot programs have been executed specifically to detect light echoes at late times around SNe (PI P.~Garnavich in Cycle 10, in this case those around SNe~Ia).

A number of investigators have already exploited our publicly-available Snapshot data, and we have cited those studies in this paper, including our own spin-off study on disappearing progenitors \citep{VanDyk2023}. We anticipate that other scientists will find this dataset valuable for their own use in the future, further proving that such surveys possess an archival legacy. To that end, we have examined our data for the possible detection of SNe other than the ones we had originally targeted, the sites of which are also serendipitously covered by our Snapshots. We provide a summary of those results in Section~\ref{appendix} in the Appendix.

\section{acknowledgments}
This research is based on observations, associated with programs GO-14668, GO-15166, GO-16179, and others, made with the NASA/ESA {\sl Hubble Space Telescope\/} and obtained from STScI, which is operated by the Association of Universities for Research in Astronomy, Inc., under NASA contract NAS 5–26555. {\sl HST\/} archival data were analyzed through program AR-14259. Support for these programs was provided by NASA through grants from STScI. This research has made use of NED, which is funded by NASA and operated by the California Institute of Technology. A.V.F.'s SN team at U.C. Berkeley also received generous support from the Miller Institute for Basic Research in Science (where A.V.F. was a Miller Senior Fellow), Gary and Cynthia Bengier (T.~deJ. was a Bengier Postdoctoral Fellow), the Christopher R.~Redlich Fund, Alan Eustace, Briggs and Kathleen Wood, and many other donors.

 KAIT and its ongoing operation were made possible by donations from Sun Microsystems, Inc., the Hewlett-Packard Company, AutoScope Corporation, Lick Observatory, the U.S. National Science Foundation (NSF), the University of California, the Sylvia \& Jim Katzman Foundation, and the TABASGO Foundation. Research at Lick Observatory is partially supported by a generous gift from Google, Inc.

 We thank (mostly U.C. Berkeley undergraduate students) Yukei Murakami, Kevin Tang, Benjamin Jeffers, Andrew Hoffman, Sanyum Channa, Sahana Kumar, Jeremy Wayland, Jeffrey Molloy, Julia Hestenes, James Sunseri, Goni Halevi, Costas Solar, Connor Jennings, Andrew Halle, Teagan Chapman, Shaunak Modak, Nick Choksi, Jackson Sipple, Heechan Yuk, Emily Ma, Edward Falcon, Nachiket Girish, Maxime de Kouchkovsky, Evelyn Liu, Derek Perera, Andrew Rikhter, Matt Chu, Kevin Hayakawa, Ivan Altunin, Haynes Stephens for their effort in taking Lick/Nickel data.

\facilities{HST(WFC3)}

\software{{\tt Drizzlepac}, {\tt Astrodrizzle} \citep{STSCI2012},  
          {\tt Dolphot} \citep{Dolphin2016}
          }

\appendix

\section{Serendipitous Events}\label{appendix}

A number of SNe or SN impostors could also have been caught serendipitously in our Snapshot data. Of those events that were in our Snapshot footprints, we provide a summary in the subsections below. We show only the images and dispense with measuring photometry and including it with earlier-time light curves, since these events were not originally targets of our Snapshot survey. We leave this for the interested reader.

Several SN sites were not in our Snapshot images despite being in targeted the host galaxies, as follows. For the SN 2011dh observation, the sites of neither the SN~Ic 1994I nor the SN~II-P 2005cs were covered; for both of the SN 2013ej and AT 2019krl observations, the sites of the SN~Ic 2002ap and SN~II-P 2003gd were not covered; for the SN 2016adj observation, the site of the SN~Ia 1986G was not covered; for the AT 2016jbu observation, the site of the SN~Ia 2015F was not covered; for the SN 2017eaw observation, of the other nine SNe in the host galaxy, none of the SNe within the last few decades are in the Snapshot pointing; the site of the intriguing interacting SN~Ib 2004dk \citep[e.g.,][]{Mauerhan2018}, which occurred in the same host as SN 2020hvp, was not in the Snapshot footprint; and, for the SN 2020jfo observation, the sites of the SN~II-P 2008in and SN~II-P 1999gn are not in the pointing, and the SN~Iax 2014dt site is too near the edge.

\subsection{SN 1999el}

\begin{figure*}[htb]
\gridline{\fig{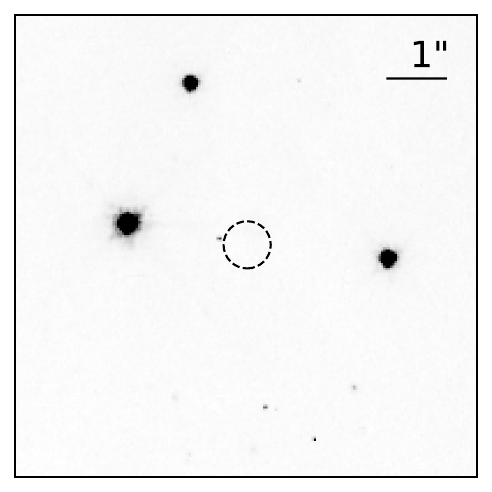}{0.25\textwidth}{(a) F555W}
          \fig{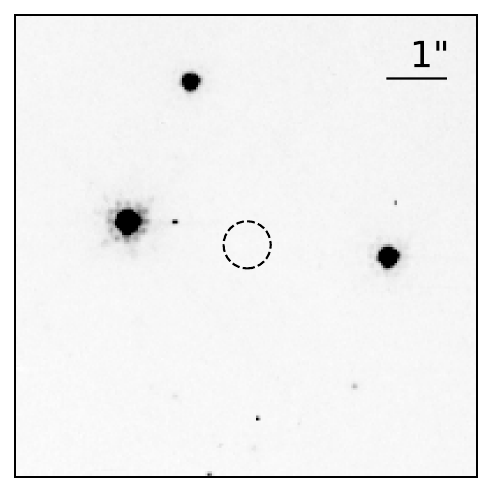}{0.25\textwidth}{(b) F814W}
          \fig{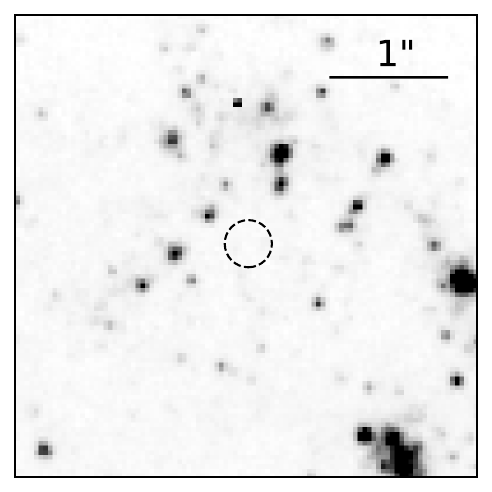}{0.25\textwidth}{(a) F555W}
          \fig{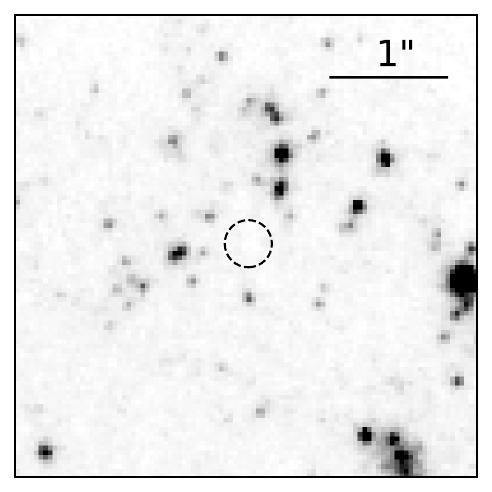}{0.25\textwidth}{(b) F814W}}
\caption{{\it Left two panels:} A portion of the WFC3 image mosaic containing SN 1999el, caught serendipitously in observations of SN 2020dpw (Section~\ref{sec:2020dpw}), in (a) F555W and (b) F814W. 
The SN is no longer detectable in either band; the site is indicated by the dashed circle.
{\it Right two panels:} A portion of the WFC3 image mosaic containing SN 1999gi \citep[e.g.,][]{Leonard2002}, caught serendipitously in observations of SN~2016bkv (Section~\ref{sec:2016bkv}), in (a) F555W and (b) F814W. 
The SN is no longer detectable; the site is indicated by the dashed circle.
}
\label{fig:1999el1999gi}
\end{figure*}

The site of the SN~IIn 1999el \citep{DiCarlo2002} was captured in our SN 2020dpw Snapshots (Section~\ref{sec:2020dpw}). However, it was not detected; see Figure \ref{fig:1999el1999gi}.

\subsection{SN 1999gi}

The site of the SN~II-P 1999gi \citep[e.g.,][]{Leonard2002} was in our Snapshots of SN~2016bkv (Section~\ref{sec:2016bkv}). However, the SN was no longer detectable at this late time. We pinpointed the location via comparison with WFPC2 images from one of our previous Snapshot programs (GO-8602; PI A.~Filippenko); see Figure \ref{fig:1999el1999gi}.


\subsection{SN 2000E}

\begin{figure*}[htb]
\gridline{\fig{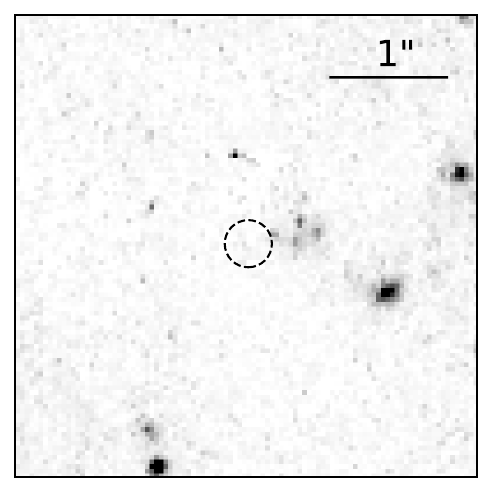}{0.25\textwidth}{(a) F555W}
          \fig{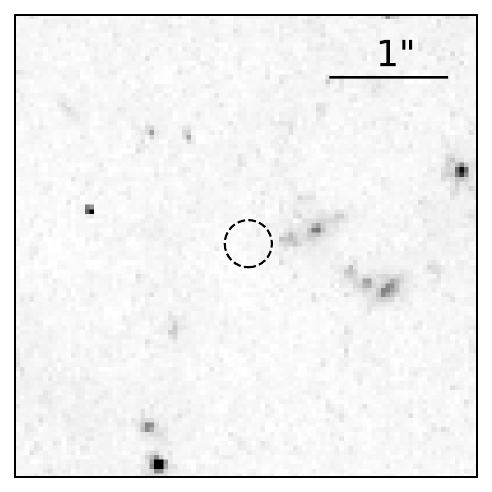}{0.25\textwidth}{(b) F814W}
          \fig{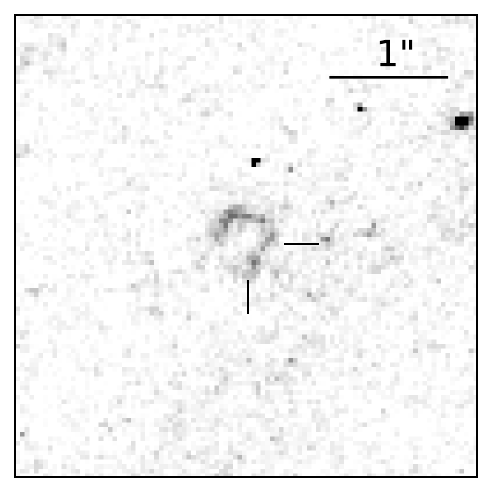}{0.25\textwidth}{(a) F438W}
          \fig{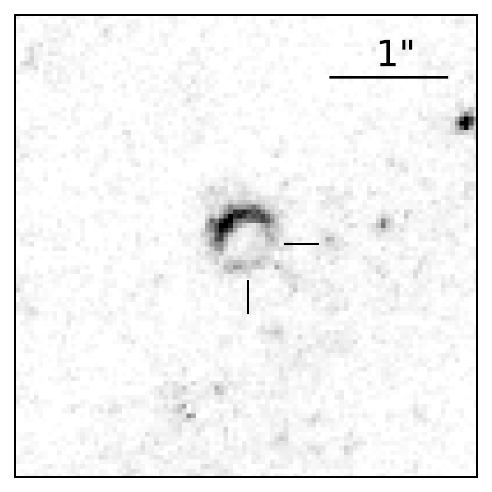}{0.25\textwidth}{(b) F625W}}
\caption{{\it Left two panels:} A portion of the WFC3 image mosaic containing SN 2000E \citep{Valentini2003}, caught serendipitously in observations of SN 2020dpw (Section~\ref{sec:2020dpw}), in (a) F555W and (b) F814W. 
The SN is no longer detectable; the site is indicated by the dashed circle.
{\it Right two panels:} A portion of the WFC3 image mosaic containing SN 2006X, caught serendipitously in observations on 2021 February 21 of SN 2019ehk (Section \ref{sec:2019ehk}), in (a) F438W and (b) F625W. Whereas the SN is no longer detectable (the site is indicated by tick marks), the light echo around it is still quite apparent. 
}
\label{fig:2000E2006X}
\end{figure*}

The site of the SN~Ia 2000E \citep{Valentini2003} was captured in our SN 2020dpw Snapshots (Section~\ref{sec:2020dpw}). However, it was no longer detectable in either band; see Figure \ref{fig:2000E2006X}.

\subsection{SN 2006X}

We managed to catch the light echo around the SN~Ia 2006X \citep{Wang2008,Crotts2008} in our SN 2019ehk Snapshots (Section \ref{sec:2019ehk}). The SN itself has disappeared, with upper limits of 26.2 and 25.8~mag in F438W and F625W, respectively; see Figure \ref{fig:2000E2006X}. An analysis of the evolution of the echo is beyond the scope of this paper.


\subsection{SN 2006ov}

\begin{figure*}[htb]
\gridline{\fig{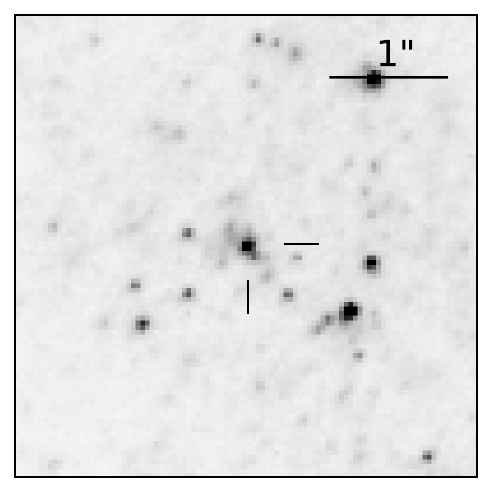}{0.25\textwidth}{(a) F555W}
          \fig{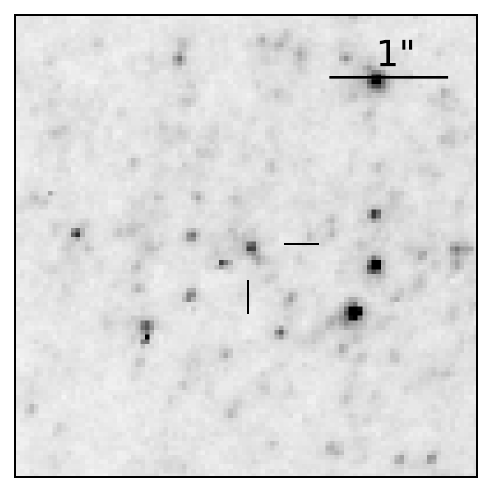}{0.25\textwidth}{(b) F814W}
          \fig{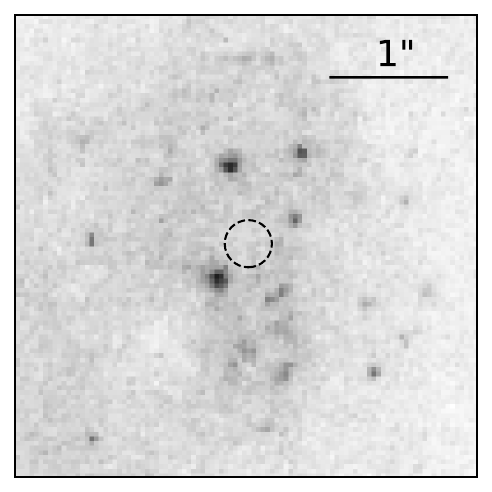}{0.25\textwidth}{(a) F555W}
          \fig{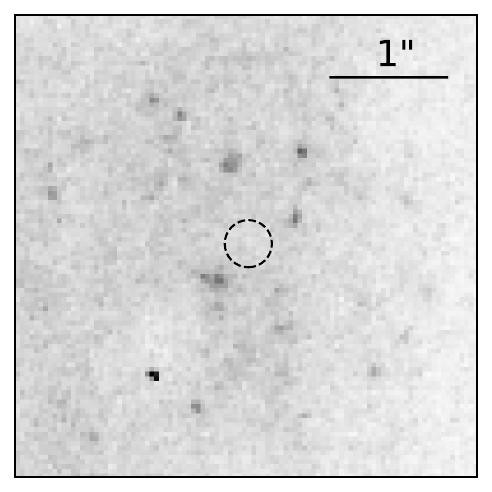}{0.25\textwidth}{(b) F814W}}
\caption{{\it Left two panels:} A portion of the WFC3 image mosaic containing SN 2006ov, caught serendipitously in observations of SN 2020jfo (Section~\ref{sec:2020jfo}), in (a) F555W and (b) F814W. The SN is still quite apparent, as indicated by the tick marks. 
{\it Right two panels:} A portion of the WFC3 image mosaic containing SN 2012bv, caught serendipitously in observations of SN 2017ixv (Section~\ref{sec:2017ixv}), in (a) F555W and (b) F814W. The SN is no longer detectable; the site is indicated by the dashed circle. 
}
\label{fig:2006ov2012bv}
\end{figure*}

The SN II-P 2006ov \citep[e.g.,][]{Spiro2014} was serendipitously captured in our SN 2020jfo Snapshots in F555W and F814W (Section~\ref{sec:2020jfo}); see Figure \ref{fig:2006ov2012bv}. We ascertained the continued presence of the SN using {\sl HST\/} ACS images from program GO-10877 (PI W.~Li) obtained around the time of discovery.

\subsection{SN 2012bv}

The SN II 2012bv was serendipitously observed in our observations of SN 2017ixv (Section~\ref{sec:2017ixv}). There was no prior {\sl HST\/} or optical ground-based imaging of the SN, and thus the absolute position was used to locate the site of the SN. The SN was no longer detectable in either band; see Figure \ref{fig:2006ov2012bv}.


\subsection{SN 2013ff}

\begin{figure*}[htb]
\gridline{\fig{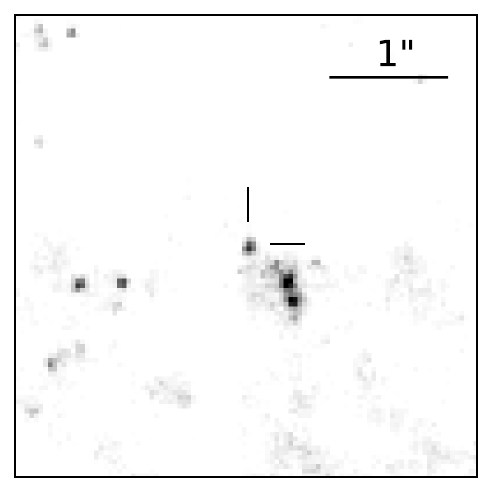}{0.25\textwidth}{(a) F555W}
          \fig{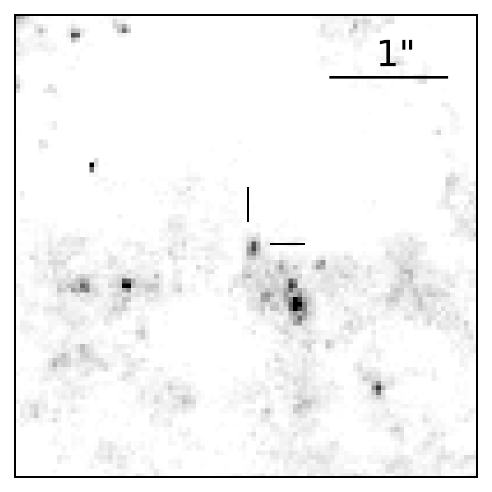}{0.25\textwidth}{(b) F814W}
          \fig{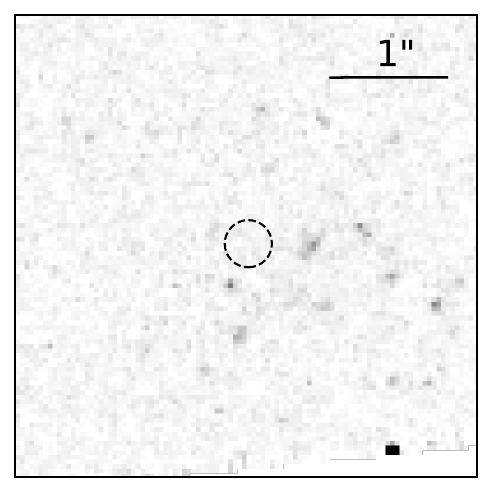}{0.25\textwidth}{(a) F555W}
          \fig{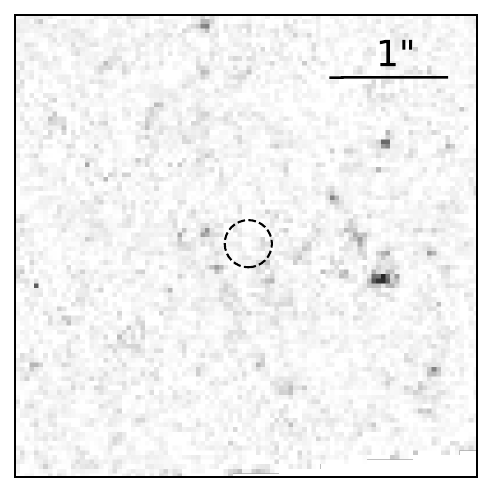}{0.25\textwidth}{(b) F814W}}
\caption{{\it Left two panels:} A portion of the WFC3 image mosaic containing SN 2013ff, caught serendipitously in observations of SN 2017gkk (Section~\ref{sec:2017gkk}), in (a) F555W and (b) F814W. The SN is detected in our Snapshots, as indicated by the tick marks. 
{\it Right two panels:} A portion of the WFC3 image mosaic containing PSN J09132750+7627410 caught serendipitously in observations of SN 2017gkk (Section~\ref{sec:2017gkk}), in (a) F555W and (b) F814W. The impostor is no longer detectable; the site is indicated by the dashed circle. 
}
\label{2013ffPSNJ09}
\end{figure*}

The SN Ic 2013ff site is in our Snapshots of SN 2017gkk (Section~\ref{sec:2017gkk}). \citet{Szalai2019a} claimed that SN 2013ff was detected by {\sl Spitzer\/} in 2014 January (at $\sim 180$~d). We isolated a probable detection of the SN in both bands by comparing the Snapshots from this program to those from our previous programs GO-14668 and GO-15166 (PI A.~Filippenko), when it appeared much fainter; see Figure \ref{2013ffPSNJ09}. This implies that, sometime between 2019 February and 2021 September, the SN shock may have encountered and was interacting with dense, or denser, CSM. This is quite unusual for an SN~Ic and would further imply that SN 2013ff may be similar to SN 2014C (Section~\ref{sec:2014C}).

\subsection{PSN J09132750+7627410}

The SN impostor PSN J09132750+7627410 was captured in our SN 2017gkk Snapshots (Section~\ref{sec:2017gkk}). Based on a direct comparison with detections shown by \citet{Tartaglia2016}, we determined that the object was no longer detectable; see Figure \ref{2013ffPSNJ09}.


\subsection{SN 2015G}

\begin{figure*}[htb]
\gridline{\fig{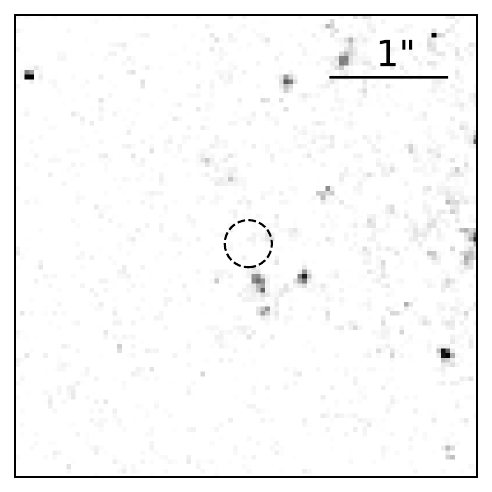}{0.25\textwidth}{(a) F555W}
          \fig{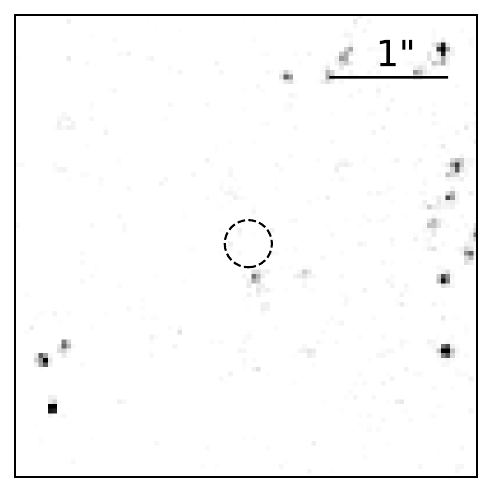}{0.25\textwidth}{(b) F814W}
          \fig{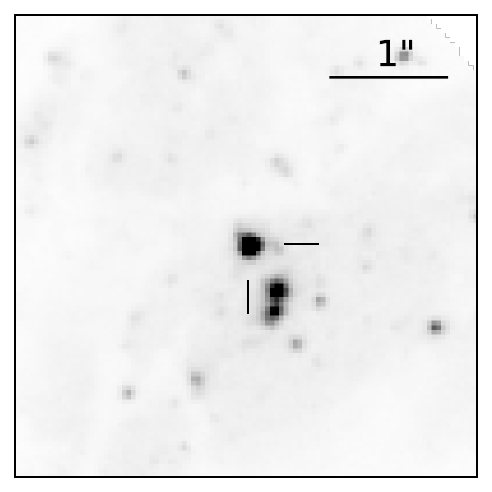}{0.25\textwidth}{(a) F438W}
          \fig{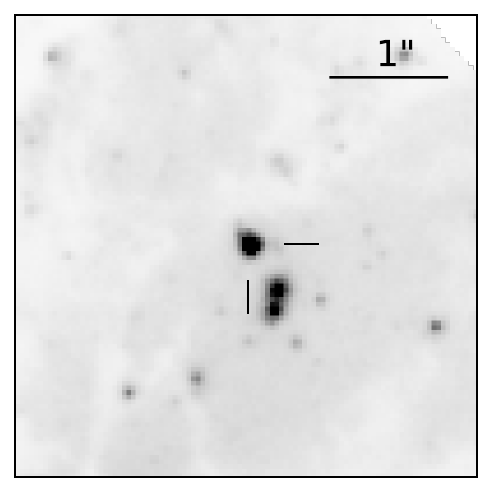}{0.25\textwidth}{(b) F625W}}
\caption{{\it Left two panels:} A portion of the WFC3 image mosaic containing SN 2015G, caught serendipitously in observations of SN 2020dpw (Section~\ref{sec:2020dpw}), in (a) F555W and (b) F814W. The SN is no longer detectable and the site is indicated by the dashed circle. 
{\it Right two panels:} A portion of the WFC3 image mosaic containing SN 2020oi, caught serendipitously in observations on 2021 February 21 of SN 2019ehk (Section \ref{sec:2019ehk}), in (a) F438W and (b) F625W. The SN is detected in our Snapshots, as indicated by the tick marks.
}
\label{2015G2020oi}
\end{figure*}

The site of the Type Ibn SN 2015G \citep{Shivvers2017} was included in our SN 2020dpw Snapshots as well (Section~\ref{sec:2020dpw}). The SN is not detected in the Snapshots, however, and had likely vanished based on comparisons with the images shown by \citet{Shivvers2017}; see Figure \ref{2015G2020oi}.

\subsection{SN 2020oi}

The SN Ic 2020oi was serendipitously detected in our SN 2019ehk Snapshots (Section \ref{sec:2019ehk}) at $m_{F438W}= 19.41 \pm 0.01$ and $m_{F625W}= 19.47 \pm 0.01$~mag; see Figure \ref{2015G2020oi}. \citet{Gagliano2022} made use of these Snapshots in their analysis of this SN.


\subsection{SN 2021J}

\begin{figure*}[htb]
\gridline{\fig{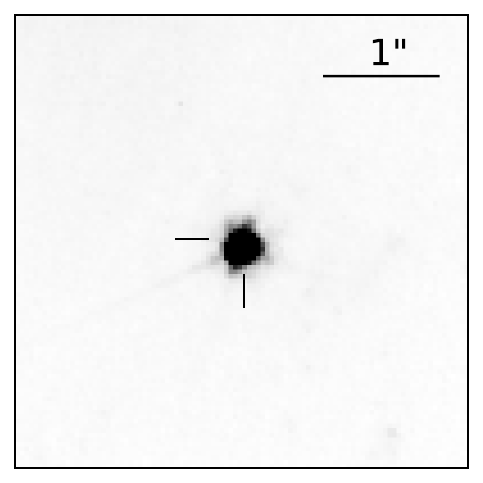}{0.25\textwidth}{(a) F336W}
          \fig{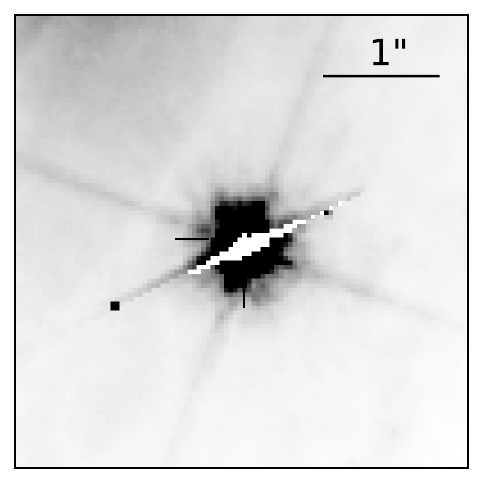}{0.25\textwidth}{(b) F555W}
          \fig{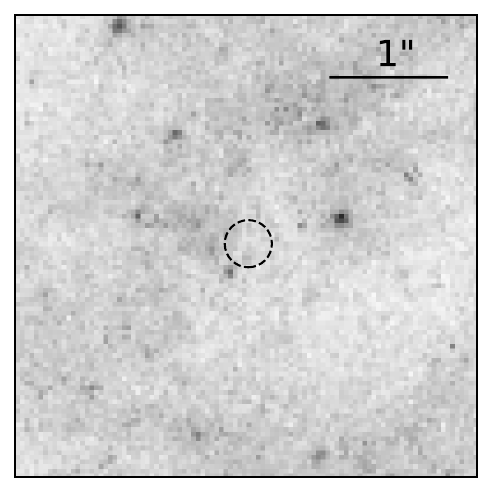}{0.25\textwidth}{(a) F555W}
          \fig{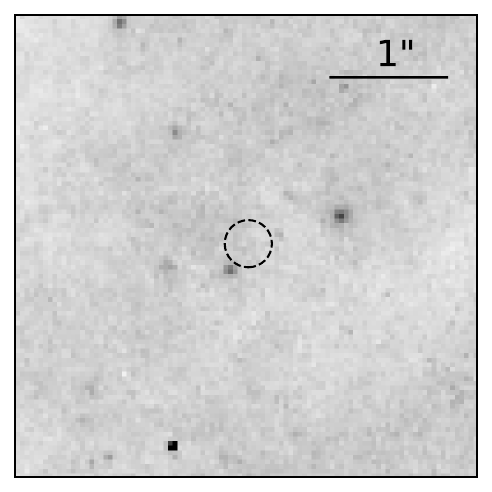}{0.25\textwidth}{(b) F814W}}
\caption{{\it Left two panels:} A portion of the WFC3 image mosaic containing SN 2021J, caught serendipitously in observations on 2021 February 15 of SN 2013df, in (a) F336W and (b) F555W. 
The SN, which was at $V = 13.63$ and $I=12.70$~mag \citep{Gallego2022}, is hopelessly saturated in both bands; hence, the Snapshots do not provide any additional photometric information.
{\it Right two panels:} A portion of the WFC3 image mosaic containing the location of SN 2021sjt, caught serendipitously in observations of SN 2020dpw, in (a) F555W and (b) F814W. A progenitor candidate is not detectable, as indicated by the dashed circle. 
}
\label{21J21sjt}
\end{figure*}

We managed to capture serendipitously the SN~Ia~2021J, which occurred in the same host galaxy (NGC~4414) as SN~2013df (Section~\ref{sec:sn2013df}). \citet{Gallego2022} undertook early-time photometric and spectroscopic monitoring of SN~2021J. Unfortunately, the SN was saturated in both bands of our Snapshot observations and therefore cannot complement the ground-based light curves; see Figure \ref{21J21sjt}.

\subsection{SN 2021sjt}

We further managed to capture the site of the SN IIb 2021sjt, pre-explosion, in the pointing for our observations of SN 2020dpw (Section \ref{sec:2020dpw}) on 2020 December 13. However, based on a spectrum posted to TNS, it appears that the extinction at the site is very large and thus any progenitor identification would likely not be possible. This lack of detection is evident from our Snapshots; see Figure \ref{21J21sjt}.


\subsection{SN 2022aau}

\begin{figure*}[htb]
\gridline{\fig{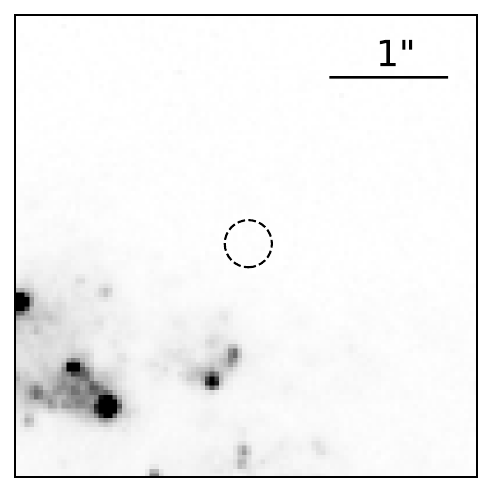}{0.25\textwidth}{(a) F336W}
          \fig{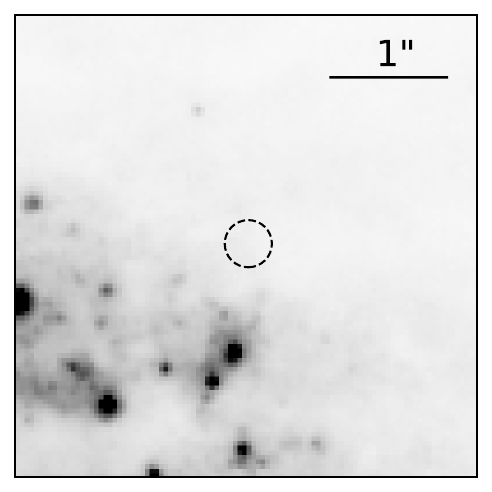}{0.25\textwidth}{(b) F814W}
}
\caption{A portion of the WFC3 image mosaic containing the location of SN 2022aau, caught serendipitously in observations of SN 2017gax, in (a) F336W and (b) F814W. The progenitor is not detectable, as indicated by the dashed circle.} 
\label{2022aau}
\end{figure*}

For the SN 2017gax observation (Section~\ref{sec:2017gax}), the site of SN~II 2022aau is in the field of view and observed somewhat over a year prior to its discovery. However, based on the classification spectrum \citep{Siebert2021}, it appears that the extinction at the site is very large and therefore any progenitor identification would likely not be possible; see Figure \ref{2022aau}. 

\begin{deluxetable}{lccccccccccc}
 \tabcolsep 0.4mm
 \tablewidth{0pt}
 \tablecaption{Photometry of Seven Supernovae with Unpublished Data (mag)$^1$}
  \tablehead{\colhead{MJD} & \colhead{$B$} & \colhead{1$\sigma$} & \colhead{$V$} & \colhead{1$\sigma$} & \colhead{$R$} & \colhead{1$\sigma$} & \colhead{$Clear$} & \colhead{1$\sigma$} & \colhead{$I$} & \colhead{1$\sigma$} & \colhead{tel}
}
\startdata
\hline
\multicolumn{12}{c}{2016bkv}\\
\hline
57470.3553 & ...    & ...   & ...    & ...   & ...    & ...   & 15.762 & 0.073  & ...    & ...   & KAIT   \\
57471.2635 & 15.412 & 0.052 & 15.463 & 0.031 & 15.399 & 0.029 & 15.243 & 0.031  & 15.402 & 0.041 & KAIT   \\
57472.3798 & 14.914 & 0.073 & 15.166 & 0.042 & 15.024 & 0.048 & 14.911 & 0.107  & 15.042 & 0.050 & KAIT   \\
57473.3044 & 14.606 & 0.066 & 14.984 & 0.032 & 14.971 & 0.032 & 14.806 & 0.036  & 15.007 & 0.038 & KAIT   \\
57474.3330 & 14.624 & 0.053 & 14.806 & 0.034 & 14.829 & 0.031 & 14.666 & 0.037  & 14.871 & 0.033 & KAIT   \\
57476.3319 & 14.685 & 0.048 & 14.754 & 0.027 & 14.751 & 0.035 & 14.635 & 0.033  & 14.756 & 0.040 & KAIT   \\
57477.3040 & 14.769 & 0.038 & 14.827 & 0.023 & 14.801 & 0.030 & 14.687 & 0.049  & 14.765 & 0.036 & KAIT   \\
57478.2828 & 14.873 & 0.042 & 14.924 & 0.025 & 14.885 & 0.030 & 14.770 & 0.035  & 14.818 & 0.036 & KAIT   \\
\enddata
\tablenotetext{}{$^1$Only a portion of the table is shown here; the full table is available in the electronic version.}
\label{PhotometryTable}
\end{deluxetable}

\bibliography{snapshot}{}
\bibliographystyle{aasjournal}

\end{document}